\documentclass[%
 reprint,
superscriptaddress,
 amsmath,amssymb,
 aps,
 prc,
 showpacs
]{revtex4-1}

\usepackage{graphicx}
\usepackage{dcolumn}
\usepackage{bm}
\usepackage{textcomp}
\usepackage{makecell}
\usepackage{subfigure}
\usepackage{longtable}
\usepackage{afterpage}
\usepackage{xcolor}
\usepackage[utf8]{inputenc}
\DeclareUnicodeCharacter{2212}{-}

\begin{document}

\preprint{APS/123-QED}

\title{Re-evaluation of the $^{22}$Ne($\alpha,\gamma$)$^{26}$Mg and $^{22}$Ne($\alpha,n$)$^{25}$Mg reaction rates}

\author{Philip Adsley}
\email{philip.adsley@wits.ac.za}
\affiliation{Institut Physique Nucl\'{e}aire d`Orsay, UMR8608, CNRS-IN2P3, Universit\'{e} Paris Sud 11, 91406 Orsay, France}
\affiliation{iThemba Laboratory for Accelerator Based Sciences, Somerset West 7129, South Africa}
\affiliation{School of Physics, University of the Witwatersrand, Johannesburg 2050, South Africa}
\author{Umberto Battino}
\affiliation{School of Physics and Astronomy, University of Edinburgh, EH9 3FD, UK}
\affiliation{The NuGrid Collaboration, http://www.nugridstars.org}
\author{Andreas Best} 
\affiliation{University of Naples \textquotedblleft Federico II\textquotedblright\, Corso Umberto I, 40, 80138 Napoli NA, Italy}
\affiliation{Istituto Nazionale di Fisica Nucleare, Sezione di Napoli, Strada Comunale Cinthia, 80126 Napoli NA, Italy}
\author{Antonio Caciolli}
\affiliation{Dipartimento di Fisica e Astronomia, Universit\`{a} degli Studi di Padova, Via F. Marzolo 8, 35131 Padova, Italy}
\affiliation{Istituto Nazionale di Fisica Nucleare, Sezione di Padova, Via F. Marzolo 8, 35131 Padova, Italy}
\author{Alessandra Guglielmetti}
\affiliation{Universit\`{a} degli Studi di Milano and INFN Milano, Via Celoria 16, I-20133 Milano}
\author{Gianluca Imbriani} 
\affiliation{University of Naples \textquotedblleft Federico II\textquotedblright\, Corso Umberto I, 40, 80138 Napoli NA, Italy}
\affiliation{Istituto Nazionale di Fisica Nucleare, Sezione di Napoli, Strada Comunale Cinthia, 80126 Napoli NA, Italy}
\author{Heshani Jayatissa}
\affiliation{Physics division, Argonne National Laboratory, Argonne IL 60439, USA}
\author{Marco La Cognata}
\affiliation{Laboratori Nazionali del Sud - Istituto Nazionale di Fisica Nucleare, Via Santa Sofia 62, 95123 Catania, Italy}
\author{Livio Lamia}
\affiliation{Universit\`{a} degli Studi di Catania, Dipartimento di Fisica e Astronomia ``E. Majorana'', via Santa Sofia 64, Italy}
\affiliation{Laboratori Nazionali del Sud - Istituto Nazionale di Fisica Nucleare, Via Santa Sofia 62, 95123 Catania, Italy}
\affiliation{CSFNSM-Centro Siciliano di Fisica Nucleare e Struttura della Materia, Via Santa Sofia 64, 95123 Catania, Italy}
\author{Eliana Masha}
\affiliation{Universit\`{a} degli Studi di Milano and INFN Milano, Via Celoria 16, I-20133 Milano}
\author{Cristian Massimi}
\affiliation{Istituto Nazionale di Fisica Nucleare, Sezione di Bologna, Bologna, Italy}
\affiliation{Dipartimento di Fisica e Astronomia, Universit\`a di Bologna, Bologna, Italy}
\author{Sara Palmerini}
\affiliation{Dipartimento di Fisica e Geologia, Universit\`a degli Studi di Perugia, Perugia, Italy}
\affiliation{Istituto Nazionale di Fisica Nucleare, Sezione di Perugia, Perugia, Italy}
\author{Ashley Tattersall}
\affiliation{School of Physics and Astronomy, University of Edinburgh, EH9 3FD, UK}
\affiliation{The NuGrid Collaboration, http://www.nugridstars.org}
\author{Raphael Hirschi}
\affiliation{School of Chemical and Physical Sciences, Keele University, Keele ST5 5BG, UK}
\affiliation{The NuGrid Collaboration, http://www.nugridstars.org}
\affiliation{Kavli IPMU (WPI), University of Tokyo, Kashiwa 277-8583, Japan}

\date{\today}

\begin{abstract}
\begin{description}
\item[Background] The competing $^{22}$Ne($\alpha,\gamma$)$^{26}$Mg and $^{22}$Ne($\alpha,n$)$^{25}$Mg reactions control the production of neutrons for the weak $s$-process in massive and AGB stars. In both systems, the ratio between the corresponding reaction rates strongly impacts the total neutron budget and strongly influences the final nucleosynthesis.
\item[Purpose] The $^{22}$Ne($\alpha,\gamma$)$^{26}$Mg and $^{22}$Ne($\alpha,n$)$^{25}$Mg reaction rates must be re-evaluated by using newly available information on $^{26}$Mg given by various recent experimental studies. Evaluations of the reaction rates following the collection of new nuclear data presently show differences of up to a factor of 500 resulting in considerable uncertainty in the resulting nucleosynthesis.
\item[Methods] The new nuclear data are evaluated and, where possible, correspondence between states observed in different experiments are made. With updated spin and parity assignments, the levels which can contribute to the reaction rates are identified. The reaction rates are computed using a Monte-Carlo method which has been used for previous evaluations of the reaction rates in order to focus solely on the changes due to modified nuclear data.
\item[Results] The evaluated $^{22}$Ne($\alpha,\gamma$)$^{26}$Mg reaction rate remains substantially similar to that of Longland {\it et al.} but, including recent results from Texas A\&M, the $^{22}$Ne($\alpha,n$)$^{25}$Mg reaction rate is lower at a range of astrophysically important temperatures. Stellar models computed with NEWTON and MESA predict decreased production of the weak branch $s$-process due to the decreased efficiency of $^{22}$Ne as a neutron source. Using the new reaction rates in the MESA model results in $^{96}$Zr/$^{94}$Zr and $^{135}$Ba/$^{136}$Ba ratios in much better agreement with the measured ratios from presolar SiC grains.
\item[Conclusion] The $^{22}$Ne$+\alpha$ reaction rates $^{22}$Ne($\alpha,\gamma$)$^{26}$Mg and $^{22}$Ne($\alpha.n$)$^{25}$Mg have been recalculated based on more recent nuclear data. The $^{22}$Ne($\alpha,\gamma$)$^{26}$Mg reaction rate remains substantially unchanged since the previous evaluation but the $^{22}$Ne($\alpha.n$)$^{25}$Mg reaction rate is substantially decreased due to updated nuclear data. This results in significant changes to the nucleosynthesis in the weak branch of the $s$-process.
\end{description}
\end{abstract}

\pacs{26.30.−k,27.30.+t,}
\maketitle


\section{\label{sec:Introduction}Introduction}

The ${^{22}}$Ne($\alpha$,$n$)${^{25}}$Mg is the main neutron source in evolved massive stars (M$>$10M$_\odot$) for the weak $s$-process (see Ref. \cite{pignatari:10}), producing most of the s-elements between iron and strontium (60$<$A$<$90). In low-mass AGB stars between 1M$_\odot$ and 4M$_\odot$ \cite{herwig:05} the reaction is activated during He-flash events, leaving its fingerprint in the final abundance of specific isotopes like ${^{87}}$Rb and ${^{96}}$Zr, which are directly observable through spectroscopy or laboratory measurements of meteoritic material \cite{herwig:05,battino:16,PALMERINI201821}. In both astrophysical scenarios, the ratio of the reaction rates of the two competing reactions $^{22}$Ne($\alpha,n$)$^{25}$Mg and $^{22}$Ne($\alpha,\gamma$)$^{26}$Mg has the largest impact on the nucleosynthesis, determining the total neutron budget.

Since the last evaluation of the $^{22}$Ne$+\alpha$ reaction rates in 2012 by Longland and collaborators \cite{PhysRevC.85.065809}, there have been a number of experimental investigations of nuclear states in $^{26}$Mg with associated re-evaluations of the $^{22}$Ne($\alpha,\gamma$)$^{26}$Mg and $^{22}$Ne($\alpha,n$)$^{25}$Mg reaction rates. The various reaction rates resulting from these studies vary by up to a factor of 500 in the astrophysically relevant region \cite{Massimi20171,PhysRevC.93.055803}. An evaluation of the nuclear data culminating in newly calculated rates is required in order to resolve or identify the main sources of these discrepancies in the $^{22}$Ne($\alpha,\gamma$)$^{26}$Mg and $^{22}$Ne($\alpha,n$)$^{25}$Mg reaction rates.

In performing the present evaluation we have chosen to use the \textsc{RatesMC} Monte-Carlo calculation code \cite{0067-0049-207-1-18}  used previously by Longland {\it et al.} \cite{PhysRevC.85.065809}. In addition we adopt as a starting point the nuclear-data evaluation of Longland {\it et al.} \cite{PhysRevC.85.065809}. This ensures that any change in the reaction rates between the current evaluation and that of Longland {\it et al.} is the result of changes in the input nuclear data rather than in the method used to evaluate the reaction rates. This is the same approach used by Talwar {\it et al.} \cite{PhysRevC.93.055803} who also adopted the \textsc{RatesMC} code to evaluate the reaction rates.

In Sec~\ref{sec:NewNuclearData} the various new sources of nuclear data are introduced along with properties about $^{26}$Mg states that may be extracted from each experimental method. The available information about $^{26}$Mg are then discussed in Section \ref{sec:SynthesisOfTheData}. Particular focus was given to spin and parity assignments of observed $^{26}$Mg levels. Since both $^{22}$Ne and $\alpha$ particle have spin-parity $J^\pi=0^+$, the $^{22}$Ne$+\alpha$ reaction can populate only natural-parity states, {\it i.~e.} $0^+$, $1^-$, $2^+...$, and so only a subset of observed nuclear levels in $^{26}$Mg can contribute to the $^{22}$Ne($\alpha$,$n$)$^{25}$Mg and $^{22}$Ne($\alpha$,$\gamma$)$^{26}$Mg reaction rates. The resulting level assignments and nuclear data on levels in $^{26}$Mg are summarised in Section \ref{sec:LevelAssignments}.  

The present characterisation of level structure in $^{26}$Mg was used for the evaluation of the reaction rates, discussed in Section~\ref{ReactionRateEvaluationSTARLIB}. Comparisons with previous reaction-rate evaluations together with the study of the contribution of individual $^{26}$Mg nuclear levels to the reaction rates are presented in Section \ref{RateComparisons}. Section \ref{sec:SuggestedFutureExperiments} presents priorities for future measurements, noting whence the major uncertainties in the reaction rates arise, and suggesting mitigating studies. In Section \ref{sec:impact} the  suggested reaction rate is used in stellar models to see the effect on AGB stars.

\section{\label{sec:NewNuclearData}New Nuclear Data}

In the following section we briefly report recent measurements performed to study the $^{26}$Mg levels. In particular, we discuss recent experiments using neutron-induced reactions on $^{25}$Mg; proton, deuteron and $\alpha$-particle inelastic-scattering reactions; the $\alpha$-particle transfer reaction $^{22}$Ne($^6$Li,$d$)$^{26}$Mg and $\gamma$-ray spectroscopy following fusion-evaporation reactions. We present and compare the different experimental work, and we discuss the  nuclear data that may be extracted from each technique.

Updated resonance parameters used in the present study are listed in Tables \ref{tab:KnownResonances} and \ref{tab:ULResonances} including the resonance strength, defined by:
\begin{equation}\label{eq:strength_alpha_gamma}
    \omega\gamma_{(\alpha,\gamma)}=\omega\frac{\Gamma_\alpha\Gamma_\gamma}{\Gamma}
\end{equation}
and 
\begin{equation}\label{eq:strength_alpha_n}
    \omega\gamma_{(\alpha,n)}=\omega\frac{\Gamma_\alpha\Gamma_n}{\Gamma},
\end{equation}
where $\Gamma$ is the total width, $\Gamma_{\alpha,n,\gamma}$ are the $\alpha$-particle, neutron and $\gamma$-ray partial widths, respectively, and $\omega=(2J+1)/[(2I+1)(2i+1)]$ the statistical spin factor, where $J$, $I$ and $i$ are the spins of the resonant level, target and projectile, respectively. Since $I=i=0$, in this case  $\omega = 2J+1$.

\subsection{$^{22}$Ne($\alpha,\gamma$)$^{26}$Mg direct measurement}

Recently, the $^{22}$Ne($\alpha,\gamma$)$^{26}$Mg resonance strength for the $E_r = 706$-keV resonance has been studied in direct kinematics at TUNL \cite{PhysRevC.99.045804}. This experiment used blister-resistant $^{22}$Ne-implanted targets, reducing the problems encountered when  using an extended gas target \cite{Wolke1989}. This experiment confirmed the resonance in $^{22}$Ne($\alpha,\gamma$)$^{26}$Mg and provided a new measurement of the resonance energy ($E_{\alpha,\mathrm{lab}} = 835.2(30)$ keV) and the resonance strength ($\omega\gamma = 0.046(12)$ meV). No corresponding $^{22}$Ne($\alpha,n$)$^{25}$Mg data have been reported.

\subsection{$^{25}$Mg$+n$ transmission and capture reactions}

The $^{25}$Mg($n,\gamma$)$^{26}$Mg reaction cross section was studied at the neutron time-of-flight (TOF) facility n\_TOF at CERN, whilst the neutron total cross section was measured at the GELINA TOF facility at JRC-GEEL \cite{Massimi20171}. These data provide useful information on nuclear levels above the neutron threshold. These recent measurements aimed at solving some inconsistencies related to neutron data in the literature~\cite{PhysRevC.66.055805, PhysRevC.85.044615}. No information about levels of $^{26}$Mg below the neutron threshold are available from these experiments. 

The n\_TOF facility at CERN generates neutrons using an high-energy proton beam incident upon a lead target. Neutrons pass down a 185-m flight path. At the end of the flight path, a detector system consisting of two C$_6$D$_6$ scintillation detectors placed on either side of the neutron beam is present. The scintillator detectors identify $\gamma$ rays resulting from $^{25}$Mg($n,\gamma$)$^{26}$Mg capture reactions. The time between the proton pulse hitting the lead target and the $\gamma$-ray detection in the scintillator detectors provides a measurement of the time of flight of the neutrons and thus the neutron energy. 

The total $n+^{25}$Mg cross section was measured on the 50-m station at the GELINA facility. At GELINA, electrons from a linear accelerator are directed onto an uranium target producing Bremsstrahlung. Photoneutrons are generated from reactions of the Bremsstrahlung $\gamma$ rays with the uranium. Neutrons are detected in a $^6$Li-glass detector. The observable in this case is the proportion of neutrons transmitted through the target to the neutron detector as a function of neutron energy.
The energy-dependent cross sections obtained with the TOF technique have sub-keV energy resolution.

Resonant neutron scattering from $^{25}$Mg can be used to simultaneously obtain the excitation energies, spin-parities, and neutron and $\gamma$-ray partial widths of $^{26}$Mg levels above the neutron threshold. This is done using an R-matrix fit to the data. For the experiments described in Ref. \cite{Massimi20171}, there is a combination of neutron-transmission data (data from GELINA) and neutron-capture data (data from n\_TOF); the R-matrix analysis is performed simultaneously on both data.

\subsection{$^{26}$Mg($\alpha,\alpha^\prime$)$^{26}$Mg}

Two studies of the $^{26}$Mg($\alpha,\alpha^\prime$)$^{26}$Mg reaction have been carried out, one by Talwar {\it et al.} using the Grand Raiden magnetic spectrometer at RCNP Osaka \cite{PhysRevC.93.055803} and the other with the K600 magnetic spectrometer at iThemba LABS by Adsley {\it et al.} \cite{PhysRevC.96.055802}. Alpha-particle inelastic scattering from even-even targets is selective to isoscalar states with natural parity. Therefore, it is well suited to probe levels in $^{26}$Mg that contribute to the $^{22}$Ne$+\alpha$ reactions, which must also be isoscalar and of natural parity. The differential cross sections of these reactions allow the spin and parity of the populated states to be determined.

In both experiments, dispersion-matched beams of $E_\alpha = 200$ MeV were incident upon metallic enriched $^{26}$Mg targets. The scattered $\alpha$ particles were momentum analysed in the spectrometers. The focal-plane detection systems of both spectrometers consisted of drift chambers which provide information on the horizontal and vertical positions and trajectories of the particles detected at the focal plane. Paddle detectors made from plastic scintillator allow the time-of-flight through the spectrometer and the total residual energy to be measured; from these quantities the particles detected at the focal plane may be identified. In both experiments, the energy resolution was around 65 keV.

Talwar {\it et al.} \cite{PhysRevC.93.055803} measured up to scattering angles of approximately 12 degrees but the number of points in the differential cross sections is only 4, making it difficult to distinguish between different spin-parities. Adsley {\it et al.} \cite{PhysRevC.96.055802} only measured up to a scattering angle of 6 degrees, and assignments are therefore limited to $J=0$ and $J=1$.

Some discrepancies were observed between the results the two $^{26}$Mg($\alpha,\alpha^\prime$)$^{26}$Mg experiments \cite{PhysRevC.93.055803,PhysRevC.96.055802}. These will be discussed in Section \ref{sec:SynthesisOfTheData}.

\subsection{$^{22}$Ne($^6$Li,$d$)$^{26}$Mg} 

In addition to studying the $^{26}$Mg($\alpha,\alpha^\prime$)$^{26}$Mg reaction, Talwar {\it et al.} also measured the $^{22}$Ne($^6$Li,$d$)$^{26}$Mg reaction with the Grand Raiden spectrometer at RCNP Osaka \cite{PhysRevC.93.055803}. This reaction was performed with $E_{^6\mathrm{Li}} = 82.3$ MeV. The resolution of this experiment was around 100 keV due to the use of a gas-cell target.

The $^{22}$Ne($^6$Li,$d$)$^{26}$Mg differential cross sections are only available at one or two angles for many of the states in the region of interest. The difficulties in these types of measurements are due to the interplay between the poor energy resolution and the high level density in the region of interest. This, in turn, makes clear identification of the resonant levels difficult, limiting the possibility of linking them to other measurements available in the literature. In addition, as discussed in Section \ref{sec:SynthesisOfTheData} there are inconsistencies between this measurement and a previous $^{22}$Ne($^6$Li,$d$)$^{26}$Mg measurement \cite{GIESEN199395} which introduce considerable uncertainty into the reaction rates.

Two additional $^{22}$Ne($^6$Li,$d$)$^{26}$Mg datasets are also available. One of these experiments used sub-Coulomb barrier $\alpha$-particle transfer with a 1 MeV/u $^{22}$Ne beam \cite{JAYATISSA2020135267}. This method allows for less model-dependence in the extraction of the $\alpha$-particle partial width compared to traditional, higher-energy transfer reactions.

The other experimental study used the $^{22}$Ne($^6$Li,$d$)$^{26}$Mg transfer reaction in inverse kinematics at a beam energy of 7 MeV/u \cite{OTA2020135256}. In this case the deuteron ejectiles were detected in an array of silicon detectors and the $^{25}$Mg and $^{26}$Mg recoils were detected at the focal plane of the MDM magnetic spectrograph. This allowed the decay branching of the populated states in $^{26}$Mg to be determined by comparing the number of deuterons associated with $^{25}$Mg and $^{26}$Mg recoils.

\subsection{$^{26}$Mg($p,p^\prime$)$^{26}$Mg and $^{26}$Mg($d,d^\prime$)$^{26}$Mg}

A study of the $^{26}$Mg($p,p^\prime$)$^{26}$Mg and $^{26}$Mg($d,d^\prime$)$^{26}$Mg reactions using 18-MeV proton and deuteron beams impinging on an enriched $^{26}$MgO target was performed using the Munich Q3D magnetic spectrograph \cite{PhysRevC.97.045807}. The proton inelastic-scattering reaction at these energies is highly unselective to the structure of the states of interest \cite{PhysRevC.89.065805,413,Moss1976429}, whilst the deuteron inelastic-scattering reaction is selective to isoscalar transitions, i.e., transitions to states with the same isospin as the ground state \cite{PhysRevC.97.045807,KAWABATA20076}. Due to the low incident energies these inelastic scattering processes are unable to give any information on the spin and parity of the resonance.

New states were observed in this experiment both above and below the neutron threshold, the existence of the states and their origin from $^{26}$Mg could be confirmed by ensuring that the kinematic shifts between angles were consistent with those expected for the $^{26}$Mg($p,p^\prime$)$^{26}$Mg reaction.

\subsection{$^{11}$B($^{16}$O,$p$)$^{26}$Mg}
Gamma-ray spectroscopy data using the Gammasphere array of high-purity germanium detectors located at Argonne National Laboratory following the $^{11}$B($^{16}$O,$p$)$^{26}$Mg reaction are available  \cite{Lotay2019}. Spin and parity assignments of levels are made using the branching ratios and angular correlations of the observed $\gamma$ rays. Due to the relatively high angular momentum imparted to the system in this fusion-evaporation reaction no $J=0$ or $J=1$ states are observed.

\section{Combining the experimental data}
\label{sec:SynthesisOfTheData}

Having  briefly introduced the updated nuclear data collected since the last evaluation of the rates, it is now necessary to build as consistent as possible a picture of all the nuclear data. In this way we are able to constrain some of the properties of possible resonances by combining data from multiple experiments.

In making new assignments updated with respect to those from Longland {\it et al.} \cite{PhysRevC.85.065809}, we adopt the following rules:

\begin{enumerate}
    \item The excitation energies of $^{26}$Mg are preferentially taken from the resonant neutron-scattering measurements of Massimi {\it et al.} \cite{Massimi20171} and the photon-scattering measurements of Longland {\it et al.} \cite{PhysRevC.80.055803} as these data have better energy resolution.
    \item For levels below the neutron threshold or for levels which were not observed in the resonant neutron-scattering data, the energies are preferentially taken from the high-resolution proton inelastic-scattering measurement of Adsley {\it et al.} \cite{PhysRevC.97.045807} or the $\gamma$-ray spectroscopy data \cite{Lotay2019}.
    \item Due to the poor energy resolutions of the experiments using $\alpha$-particle inelastic scattering \cite{PhysRevC.96.055802,PhysRevC.93.055803} and $\alpha$-particle transfer \cite{PhysRevC.93.055803}, we do not use the excitation energies resulting from these experiments.
    \item Spin-parity assignments above the neutron threshold are preferentially taken from Massimi {\it et al.} \cite{Massimi20171}.
    \item Other spin-parity assignments are taken from the $^{26}$Mg($\alpha,\alpha^\prime$)$^{26}$Mg data of Adsley {\it et al.} \cite{PhysRevC.96.055802}, Talwar {\it et al.} \cite{PhysRevC.93.055803} and the Gammasphere data of Ref. \cite{Lotay2019}.
    \item Due to the limited number of angles for the $\alpha$-particle transfer reaction of Talwar {\it et al.} \cite{PhysRevC.93.055803}, we approach the spin parities and spectroscopic factors derived from those data with some caution and, were possible, seek confirmation from another source.
    \item Levels which have not been observed in the neutron resonant-scattering data of Massimi {\it et al.} \cite{Massimi20171} but have been observed in the proton and deuteron inelastic-scattering measurements of Adsley {\it et al.} \cite{PhysRevC.97.045807} have been assigned upper limits on neutron widths based on the sensitivity of the neutron-scattering experiments.
    \item Resonance strengths for higher-lying resonances are taken from Longland {\it et al.} except for the $^{22}$Ne($\alpha,\gamma$)$^{26}$Mg resonance strength for the $E_{r}^{CM} = 706$-keV resonance, which has been recalculated using the results of a recently published direct measurement \cite{PhysRevC.99.045804}.
    \item For lower-lying resonances in $^{22}$Ne($\alpha,\gamma$)$^{26}$Mg and the recalculation of the $^{22}$Ne($\alpha,n$)$^{25}$Mg resonance strength for the $E_r = 706$-keV resonance, the results of the experiments at Texas A\&M by Jayatissa {\it et al.} \cite{JAYATISSA2020135267} and Ota {\it et al.} \cite{OTA2020135256} have been used for the calculations. When partial widths or resonance strengths are adopted from these studies is noted in the text.
\end{enumerate}

The states in $^{26}$Mg for which some re-determination must be made or some uncertainty exists are  discussed below. We do not discuss levels for which no assignment is made or changed. Table \ref{tab:LevelInformationForDiscussion} gives detailed information on the sources of excitation energies, and spins and parities for some of the levels being discussed for ease of reference.

\begin{table*}[tphb]
\caption{Excitation energies, and spins and parities for some of the levels discussed in Section \ref{sec:SynthesisOfTheData}. When there are multiple experiments of the same type (e.g. $^{25}$Mg$+n$ data from Refs. \cite{Massimi20171} and \cite{PhysRevC.85.044615}) the most up-to-date values are given. The information in the table is linked to the index for the subsection in Section \ref{sec:SynthesisOfTheData}. The adopted excitation energy, and spin and parity (if available) are given along with the excitation energies, and spins and parities available from different sources.}
\label{tab:LevelInformationForDiscussion}
\begin{ruledtabular}
\begin{tabular}{c c c c c c}
     Section & $E_x$ [MeV] & $J^\pi$ & Source 1 & Source 2 & Source 3 \\ \hline \hline \\
     \ref{statesec1} & $10.6507(4)$ & $7^-$ & \makecell{$^{26}$Mg($p,p^\prime$)$^{26}$Mg \cite{PhysRevC.96.055802} \\ $E_x = 10.650(1)$ MeV,\\ no $J^\pi$} &  \makecell{Gammasphere \cite{Lotay2019}\\$E_x = 10.6507(4)$ MeV,\\$J^\pi = 7^-$} & \\
     
      \ref{statesec2} & $10.719(1)$ & $3^+$ & \makecell{$^{26}$Mg($p,p^\prime$)$^{26}$Mg \cite{PhysRevC.96.055802} \\ $E_x = 10.719(1)$ MeV\\ no $J^\pi$} & \makecell{Gammasphere \cite{Lotay2019} \\ $E_x = 10.7227(22)$ MeV\\ $J^\pi = (3^+)$} \\
      
      \ref{statesec3} & $10.746(1)$ & $3^+$ & \makecell{$^{26}$Mg($p,p^\prime$)$^{26}$Mg \cite{PhysRevC.96.055802} \\ $E_x = 10.746(3)$ MeV\\ no $J^\pi$} & \makecell{Gammasphere \cite{Lotay2019} \\ $E_x = 10741.7(30)$ MeV\\ $J^\pi = 3^+$} \\
      
      \ref{statesec4} & $10.771(1)$ & $(3-7)$ & \makecell{$^{26}$Mg($p,p^\prime$)$^{26}$Mg \cite{PhysRevC.96.055802} \\ $E_x = 10.771(1)$ MeV\\ no $J^\pi$} & \makecell{Gammasphere \cite{Lotay2019} \\ $E_x = 10765.1(30)$\\ J$^\pi = (3-7)$} \\
      
      \ref{statesec5} & $10.8057(7)$ & $1^-$ & \makecell{$^{26}$Mg($\gamma,\gamma^\prime$)$^{26}$Mg \cite{PhysRevC.80.055803}\\$E_x = 10.8057(7)$ MeV\\$J^\pi = 1^-$} & \makecell{$^{26}$Mg($\alpha,\alpha^\prime$)$^{26}$Mg \cite{PhysRevC.96.055802}\\$E_x = 10.806(10)$ MeV\\$J^\pi = 1^-$} & \makecell{$^{26}$Mg($p,p^\prime$)$^{26}$Mg \cite{PhysRevC.96.055802}\\
     $E_x = 10.806(1)$ MeV\\no $J^\pi$}\\
     
      \ref{statesec5} & $10.818(1)$ & $0^+$ & \makecell{$^{26}$Mg($\alpha,\alpha^\prime$)$^{26}$Mg \cite{PhysRevC.93.055803}/\cite{PhysRevC.96.055802} \\$E_x = 10.822(10)/10.824(10)$,\\$J^\pi = 0^+$} & \makecell{$^{26}$Mg($p,p^\prime$)$^{26}$Mg \cite{PhysRevC.96.055802}\\$E_x = 10.818(1)$ MeV,\\no $J^\pi$} \\
      
      \ref{statesec5} & $10.826(1)$ & $(2^+)$ & \makecell{$^{26}$Mg($p,p^\prime$)$^{26}$Mg \cite{PhysRevC.96.055802}\\$E_x = 10.826(1)$ MeV\\ no $J^\pi$} & \makecell{Gammasphere \cite{Lotay2019}\\$E_x = 10.8226(30)$ MeV\\$J^\pi = 2^+$} \\
      
      \ref{statesec8} & $10.915(1)$ & $2^+$ or $6^+$ & \makecell{$^{26}$Mg($p,p^\prime$)$^{26}$Mg \cite{PhysRevC.96.055802}\\
     $E_x = 10.915(1)$ MeV\\
     no $J^\pi$} & \makecell{Gammasphere \cite{Lotay2019}\\$E_x = 10.9128(30)$ MeV,\\ $J^\pi = 2^+$ or $6^+$} \\
     
      \ref{statesec9} & $11.017(1)$ & $2^+, 3, 4, 5^+$ & \makecell{$^{26}$Mg($p,p^\prime$)$^{26}$Mg \cite{PhysRevC.96.055802} \\ $E_x = 11.017(1)$ MeV\\ no $J^\pi$} & \makecell{Gammasphere \cite{Lotay2019} \\ $E_x = 11.0173(36)$ MeV\\ $J^\pi = 2^+, 3, 4, 5^+$} \\
      
      \ref{statesec10} & $11.084(1)$ & $2^+$ & \makecell{$^{26}$Mg($p,p^\prime$)$^{26}$Mg \cite{PhysRevC.96.055802} \\ $E_x = 11.084(1)$ MeV\\ no $J^\pi$} & \makecell{Gammasphere \cite{Lotay2019}\\$E_x = 11.0809(40)$ MeV,\\$J^\pi = 2^+$} \\

      \ref{statesec11} & $11.16310(4)$ & $2^+$ & \makecell{$^{25}$Mg$+n$ \cite{Massimi20171}\\$E_x = 11.16310(4)$ MeV\\$J^\pi=2^+$} & & \\
      \ref{statesec11} & $11.16926(4)$ & $3^-$ & \makecell{$^{25}$Mg$+n$ \cite{Massimi20171}\\$E_x = 11.16926(4)$ MeV\\$J^\pi = 3^-$} & \\
      \ref{statesec11} & $11.1707(4)$ & $(2^+)$ & \makecell{Gammasphere \cite{Lotay2019}\\$E_x = 11.1717(30)$\\$J^\pi = (2^+)$} & \makecell{$^{25}$Mg$+n$ \cite{Massimi20171}\\$E_x = 11.1707(4)$ MeV\\no $J^\pi$} & \\

      \ref{statesec12} & $11.321(1)$ & $0^+$ or $1^-$ & See text & \\
\end{tabular}
\end{ruledtabular}
\end{table*}

\subsubsection{$E_x = 10.6507$ MeV; $E_r = 36$ keV}
 \label{statesec1}
The level at $E_x = 10.6507(4)$ MeV was assigned $J^\pi = 7^-$ from the Gammasphere data \cite{Lotay2019}. Due to the high spin of the state it has negligible astrophysical impact. Note that we do not assume that this state is the same as the $E_x = 10.650(1)$-MeV, $J^\pi = 1^+$ state assigned by Adsley {\it et al.} \cite{PhysRevC.97.045807} based on the $^{26}$Mg($p,p^\prime$)$^{26}$Mg data of Crawley {\it et al.} \cite{PhysRevC.39.311} as a $J=1$ state would not have been observed in the Gammasphere data \cite{Lotay2019}.

\subsubsection{$E_x = 10.719$ MeV; $E_r = 104$ keV}
 \label{statesec2}
 The level at $E_x = 10.719(2)$ MeV observed in the $^{26}$Mg($p,p^\prime$)$^{26}$Mg inelastic-scattering reaction at low energies \cite{PhysRevC.97.045807} was assigned as $J^\pi = 3^+$ under the assumption that it is the $E_x = 10.7227(22)$-MeV state observed in the Gammasphere experiment \cite{Lotay2019}.
 
\subsubsection{$E_x = 10.746$ MeV; $E_r = 131$ keV}
 \label{statesec3}
The level at $E_x = 10.746(1)$ MeV observed in the $^{26}$Mg($p,p^\prime$)$^{26}$Mg inelastic-scattering reaction at low energies \cite{PhysRevC.97.045807} was assigned to $J^\pi = 3^+$ under the assumption that it is the $E_x = 10.7417(30)$-MeV state observed in the Gammasphere experiment \cite{Lotay2019}.

\subsubsection{$E_x = 10.771$ MeV; $E_r = 149$ keV}
 \label{statesec4}
The level at $E_x = 10.7651(30)$~MeV was assigned to $J^\pi = (3-7)$ from the Gammasphere data \cite{Lotay2019}. This state is assumed to be the {$E_x = 10.771(1)$-MeV} state in the $^{26}$Mg($p,p^\prime$)$^{26}$Mg inelastic-scattering reaction at low energies \cite{PhysRevC.97.045807}.

\subsubsection{$E_x = 10.806$, $E_x = 10.818$ and $E_x = 10.826$ MeV; $E_r = 191$, $E_r = 203$ and $E_r = 211$ keV}
 \label{statesec5}

The $\alpha$-particle inelastic-scattering measurement of Talwar {\it et al.} \cite{PhysRevC.93.055803} gave an assignment of $J^\pi = 1^-$ to the state observed at $E_x = 10.82$ MeV assuming it to be the $J^\pi = 1^-$ level observed by Longland {\it et al.} in a $\gamma$-ray inelastic-scattering measurement \cite{PhysRevC.80.055803}. However, Adsley {\it et al.} \cite{PhysRevC.96.055802}, also using $\alpha$-particle inelastic scattering, showed that there is a $J^\pi = 0^+$ level at around $E_x = 10.82$ MeV. Multiple levels were confirmed by the high-resolution proton and deuteron inelastic-scattering data of Adsley {\it et al.} \cite{PhysRevC.97.045807}. Adsley {\it et al.} argue that the third state at around this excitation energy observed in the high-resolution proton scattering experiment at Munich \cite{PhysRevC.97.045807} corresponds to a $J^\pi = 1^+$ level at $E_x = 10.81$ MeV observed in high-energy $^{26}$Mg($p,p^\prime$)$^{26}$Mg scattering data \cite{PhysRevC.39.311}.

Lotay {\it et al.} \cite{Lotay2019} give a tentative $J^\pi = 2^+$ assignment for the state at $E_x = 10.8226(30)$ MeV combining a $J = 2-6$ assignment from the Gammasphere data with the $^{26}$Mg($e,e^\prime$)$^{26}$Mg reaction study of Lees {\it et al.}, which finds only a $J^\pi = 2^+$ state in this excitation-energy region though with poor energy resolution. 

The summary of the experimental studies is this: there are three levels at $E_x = 10.8057(7)$, $10.818(1)$ and $10.826(1)$ MeV based on the $^{26}$Mg($p,p^\prime$)$^{26}$Mg data \cite{PhysRevC.97.045807}. The level at $E_x = 10.8057(7)$ MeV has $J^\pi = 1^-$ \cite{PhysRevC.80.055803}. The level at $E_x = 10.826(1)$ MeV must have $J>1$ as it was observed in the Gammasphere study \cite{Lotay2019}. The result is therefore that the previous assignment of $J^\pi = 0^+$ to the $E_x = 10.826(1)$-MeV level was incorrect, and the $J^\pi = 0^+$ assignment should instead be associated with the $E_x = 10.818(1)$ MeV state.

Finally, for computing the reaction rates, the $\alpha$-particle partial widths must be estimated. The previous evaluation of Longland {\it et al.} used the $^{22}$Ne($^6$Li,$d$)$^{26}$Mg data of Ugalde \cite{PhysRevC.76.025802} to calculate the $\alpha$-particle width for the $J^\pi = 1^-$ state at $E_x = 10.806$ MeV. Adsley {\it et al.} \cite{PhysRevC.96.055802} relaxed the constraints on the $\alpha$-particle partial widths since the states at these excitation energies are not resolved and it is not clear to which $^{26}$Mg state or states the transfer reaction proceeds. The sub-Coulomb transfer data of Jayatissa {\it et al.} \cite{JAYATISSA2020135267} show that a state is populated at around $E_x = 10.8$ MeV but it is not clear which of the three states in $^{26}$Mg is the one populated. We adopt the widths from Jayatissa {\it et al.} \cite{JAYATISSA2020135267} for the computation of the recommended reaction rate.

\subsubsection{$E_x = 10.915$ MeV; $E_r = 300$ keV}
 \label{statesec8}
The level observed at $E_x = 10.9128(30)$ MeV in the Gammasphere data \cite{Lotay2019} was constrained as $J^\pi = 2^+$ or $6^+$. This state is assumed to be the level at $E_x = 10.915(1)$~MeV observed in the low-energy $^{26}$Mg($p,p^\prime$)$^{26}$Mg measurement of Adsley {\it et al.} \cite{PhysRevC.97.045807}.

As this state does not have a confirmed spin-parity assignment it has not been included in the current evaluation. However, this state does have a natural parity and could potentially contribute to the $^{22}$Ne($\alpha,\gamma$)$^{26}$Mg reaction rate.

\subsubsection{$E_x = 11.017$ MeV; $E_r = 402$ keV}
 \label{statesec9}
The level observed at $E_x = 11.0173(36)$ MeV in the Gammasphere data \cite{Lotay2019} was constrained to $J^\pi = 2^+, 3, 4, 5^+$. This state is assumed to be the level observed at $E_x = 11.017(1)$~MeV in the low-energy $^{26}$Mg($p,p^\prime$)$^{26}$Mg measurement of Adsley {\it et al.} \cite{PhysRevC.97.045807}.

This state does not have a confirmed spin-parity assignment and has therefore not been included in the current evaluation. It is not clear if this state has a natural parity and so it is unclear if it may contribute to the $^{22}$Ne($\alpha,\gamma$)$^{26}$Mg reaction rate.

\subsubsection{$E_x = 11.084$ MeV; $E_r = 469$ keV}
 \label{statesec10}

This state was observed in $^{26}$Mg($p,p^\prime$)$^{26}$Mg ($E_x = 11.084(1)$ MeV) reactions \cite{PhysRevC.97.045807,Moss1976429} and in the Gammasphere data \cite{Lotay2019} ($E_x = 11.809(40)$ MeV) in which an assignment of $J^\pi = 2^+$ is made. The $\alpha$-particle partial width was determined for this resonance to be: $\Gamma_\alpha = 5.7(15) \times 10^{-11}$ eV from sub-Coulomb $\alpha$-particle transfer \cite{JAYATISSA2020135267}.

\subsubsection{$E_x = 11.102$ MeV; $E_r = 487$ keV}

A state at $E_x = 11.102(1)$ MeV was observed in the $^{26}$Mg($p,p^\prime$)$^{26}$Mg data of Adsley {\it et al.} \cite{PhysRevC.97.045807}. This state lies $8$ keV above the neutron threshold but was not observed in the $^{25}$Mg$+n$ experiments of Massimi {\it et al.} \cite{Massimi20171}. Based on this non-observation it is possible to estimate the maximum neutron width for this state. Depending on the $\gamma$-ray partial width used in the calculation, the neutron partial width, $\Gamma_n$, must be below $2-4$ eV. No spin-parity assignment is available for this state.

\subsubsection{$E_x = 11.119$ MeV; $E_r = 504$ keV}

This state was observed in the $^{26}$Mg($p,p^\prime$)$^{26}$Mg data of Adsley {\it et al.} \cite{PhysRevC.97.045807}. This state lies $26$ keV above the neutron threshold but was not observed in the $^{25}$Mg$+n$ experiments of Massimi {\it et al.} \cite{Massimi20171}. Using the same methodology as for the $E_x = 11.102$-MeV state, the upper limit of the neutron partial width is found to be $\Gamma_n = 6-8$ eV. No spin-parity assignment is available for this state.

\subsubsection{$E_x = 11.163$, $E_x = 11.169$ and $E_x = 11.171$ MeV; $E_r = 548$, $E_r = 554$, $E_r = 556$ keV}
 \label{statesec11}
\label{sec:11167keVState}

The region from $E_x = 11.163$ to $11.171$ MeV is extremely important, due to the possible observation of an $\alpha$-particle cluster state in the $^{22}$Ne($^6$Li,$d$)$^{26}$Mg study of Talwar {\it et al.} \cite{PhysRevC.93.055803}. However, as we discuss below, there are outstanding problems in the nuclear data in this region and further confirmation of the proposed state of Talwar {\it et al.} is required.

First let us consider the assignments in this region which are simple to make. Two states have been observed in the $^{25}$Mg$+n$ experiments of Massimi {\it et al.} at $E_x = 11.163$ and $11.167$ MeV with $J^\pi  = 2^+$ and $3^-$, respectively \cite{Massimi20171}.

A third state, at $E_x = 11.171$ MeV was also observed by Massimi {\it et al.} \cite{Massimi20171}, but no $J^\pi$ could be assigned. This state has comparable $\gamma$-ray and neutron partial widths, and is thus likely to have an observable $\gamma$-ray decay. This is therefore likely the state observed by Lotay {\it et al.} at $E_x = 11.1717(30)$ MeV and given a tentative $J^\pi = 2^+$ assignment \cite{Lotay2019}.

This third state has been associated with the strong $\alpha$-cluster candidate at $E_x = 11.17$ MeV observed by Talwar {\it et al.} in $^{22}$Ne($^6$Li,$d$)$^{26}$Mg reactions \cite{PhysRevC.93.055803}. This association was made because no corresponding strong $^{22}$Ne($\alpha,n$)$^{25}$Mg resonance was observed in the direct measurements of Jaeger {\it et al.} \cite{PhysRevLett.87.202501}, meaning that the state cannot have a large $^{22}$Ne($\alpha,n$)$^{25}$Mg resonance strength. This, in turn, means that the $\gamma$-ray partial width is larger than the neutron partial width. Talwar {\it et al.} also reported observing the $E_x = 11.31$-MeV state corresponding to the $E_r = 706$-keV resonance of Jaeger \cite{PhysRevLett.87.202501}, Wolke \cite{Wolke1989} and Giesen \cite{GIESEN199395}. In summary, Talwar {\it et al.} report two $\alpha$-cluster states at $E_x = 11.17$ and $11.31$ MeV, separated by around 140 keV.

There are older $^{22}$Ne($^6$Li,$d$)$^{26}$Mg data, those of Giesen {\it et al.} \cite{GIESEN199395} in which no $E_x = 11.17$-MeV $\alpha$-cluster state is observed. Rather two states are observed at $E_x = 11.31$ and $11.45$ MeV, separated by around 140 keV. This $E_x = 11.45$ MeV $\alpha$-cluster state matches a resonance observed in the direct measurement of Jaeger {\it et al.} at $E_x = 11.441$ MeV \cite{PhysRevLett.87.202501}. Unfortunately, Talwar {\it et al.} \cite{PhysRevC.93.055803} report that the $E_x = 11.44$-MeV state does not fall on the focal plane in their measurement, though it should not be entirely resolved from the $E_x = 11.31$-MeV state and so some indication of its population may be observable in their data.

In addition to the Giesen and Talwar $^{22}$Ne($^6$Li,$d$)$^{26}$Mg data, the more recent $^{22}$Ne($^6$Li,$d$)$^{26}$Mg results from Texas A\&M at 7 MeV/u (corresponding to a $^6$Li beam energy of 42 MeV) \cite{OTA2020135256} and the sub-Coulomb barrier transfer results at 1 MeV/u \cite{JAYATISSA2020135267} both fail to observe a strongly populated state at around $E_x = 11.17$ MeV. In the case of the sub-Coulomb transfer measurement, the non-observation of the state could be due to the state of interest having a high spin \cite{JAYATISSA2020135267}.

The higher-energy measurement from Texas A\&M (at 7 MeV/u) is at an energy between the Giesen and Talwar measurements, and a state at $E_x = 11.17$ MeV is observed. The change of beam energy between the various $^{22}$Ne($^6$Li,$d$)$^{26}$Mg measurements may be the cause of why the $E_x = 11.17$-MeV state is not seen in the experiments of Giesen {\it et al.} \cite{GIESEN199395} and Jayatissa {\it et al.} \cite{JAYATISSA2020135267}, though there is no clear evidence of the change in the population ratio of e.g. the $E_x = 11.31$-MeV and the $E_x = 11.45$-MeV states between Ota measurement \cite{OTA2020135256} and that of Giesen {\it et al.} \cite{OTA2020135256} which may be expected if the beam-energy changes were causing marked changes in the population of levels with different spins.

It may be coincidental that the two $^{22}$Ne($^6$Li,$d$)$^{26}$Mg measurements using magnetic spectrographs (the data of Talwar {\it et al.} and Giesen {\it et al.}) observe two $\alpha$-cluster states separated by approximately the same energy. However, it could be that one of the measurements has a problem in the excitation-energy calibration which may cause this discrepancy.

Given the discrepancies between the Talwar {\it et al.} \cite{PhysRevC.93.055803} result and the results of two above-barrier $^{22}$Ne($^6$Li,$d$)$^{26}$Mg measurements \cite{GIESEN199395,OTA2020135256}, one sub-barrier $^{22}$Ne($^6$Li,$d$)$^{26}$Mg measurement \cite{JAYATISSA2020135267}, and the direct $^{22}$Ne($\alpha,n$)$^{25}$Mg measurement of Jaeger \cite{PhysRevLett.87.202501}, we have decided not to use the spectroscopic factors derived from the $^{22}$Ne($^6$Li,$d$)$^{26}$Mg measurement of Talwar {\it et al.} in our computation of the reaction rates. We have instead used the $J^\pi = 2^+$ assignment for the $E_x = 11.171$-MeV state and have used the experimental upper limits on the $\alpha$-particle partial widths from the sub-Coulomb barrier transfer measurement of Jayatissa {\it et al.} \cite{JAYATISSA2020135267} and the transfer-reaction of Ota {\it et al.} \cite{OTA2020135256} to constrain the resonance strengths for the states at $E_x = 11.163$, $11.169$ and $11.171$ MeV.

\subsubsection{$E_x = 11.27963(4)$ MeV; $E_r = 665$ keV}

The state at $E_x = 11.27963(4)$ MeV observed in resonant neutron scattering at $E_n = 194.01(2)$ keV has been reassigned, on the basis of the neutron-scattering data, from $J^\pi = 4^{(-)}$ to $J^\pi = 3^-$, a change from unnatural to natural parity. This state now potentially contributes to the reaction rates \cite{Massimi20171}.

\subsubsection{$E_x = 11.321$ MeV; $E_r = 706$ keV} 
 \label{statesec12}

This is the location of the lowest-energy directly measured resonances. Wolke {\it et al.} \cite{Wolke1989} measured a resonance in the $^{22}$Ne($\alpha,\gamma$)$^{26}$Mg reaction at $E_{\alpha,\mathrm{lab}} = 828(5)$ keV ($E_x = 11.315(4)$ MeV) with $\omega\gamma_{(\alpha,\gamma)} = 0.036(4)$ meV.

Hunt {\it et al.} measured a resonance in the $^{22}$Ne($\alpha,\gamma$)$^{26}$Mg reaction at $E_{\alpha,\mathrm{lab}} = 835.2(30)$ keV ($E_r = 705(3)$ keV, $E_x = 11.3195(25)$ MeV) with $\omega\gamma_{(\alpha,\gamma)}=0.046(12)$ meV. Based on the dipole-or-E2 rule of Endt \cite{ENDT19901}, a spin-parity assignment of $J^\pi = 0^+$, $1^-$, $2^+$ or $3^-$ is given.

Using the results of the measurements above, the weighted average of the resonance strength is $\omega\gamma = 0.037(4)$ meV. As the results of the two $^{22}$Ne($\alpha,\gamma$)$^{26}$Mg measurements \cite{Wolke1989,PhysRevC.99.045804} agree within their uncertainties, we do not perform the procedure described in Ref. \cite{PhysRevC.85.065809} to account for unknown systematic biases within the direct measurements.

Jaeger {\it et al.} \cite{PhysRevLett.87.202501} measured a resonance in the $^{22}$Ne($\alpha,n$)$^{25}$Mg reaction at $E_{\alpha,\mathrm{lab}} = 832(2)$ keV ($E_x = 11.319(2) $ MeV) with $\omega\gamma_{(\alpha,n)} = 0.118(11)$ meV with a total width of $\Gamma = \Gamma_n+\Gamma_\gamma+\Gamma_\alpha =0.25(17)$ keV.

The $^{22}$Ne($\alpha,n$)$^{25}$Mg reaction has also been measured by Giesen {\it et al.} \cite{GIESEN199395}, Drotleff {\it et al.} \cite{1993ApJ...414..735D} and Harms {\it et al.} \cite{PhysRevC.43.2849}. Giesen {\it et al.} determined a resonance energy $E_r = 701(5)$ keV and a resonance strength $\omega\gamma = 0.234(77)$ meV. Drotleff {\it et al.} found $E_r = 703(3)$ keV and $\omega\gamma = 0.18(3)$ meV. Harms {\it et al.} measured $E_r = 702(3)$ keV and $\omega\gamma = 0.083(24)$ meV. Longland {\it et al.} \cite{PhysRevC.85.065809} used a method of estimating parameters and uncertainties including potential systematic effects, which resulted in $\omega\gamma = 0.14(3)$ meV.

From the weighted averages of the direct measurements described above, the $\alpha$-particle partial width may be computed resulting in $\Gamma_\alpha = 180(30) \mu$eV ($60(10) \mu$eV) assuming a $J^\pi = 0^+$ ($J^\pi = 1^-$) assignment.

The $\alpha$-particle partial width for this resonance has been measured with sub-Coulonb barrier $\alpha$-particle transfer at Texas A\&M \cite{JAYATISSA2020135267}, and found to be smaller than expected based on the weighted averages of the direct $^{22}$Ne($\alpha,\gamma$)$^{26}$Mg and $^{22}$Ne($\alpha,n$)$^{25}$Mg measurements. This, coupled with a smaller $\Gamma_n/\Gamma_\gamma$ ratio determined by Ota {\it et al.}, detecting $^{25}$Mg and $^{26}$Mg recoils using the MDM magnetic spectrograph at Texas A\&M following the $^{22}$Ne($^6$Li,$d$)$^{26}$Mg reaction in inverse kinematics leads to a reduced $^{22}$Ne($\alpha,n$)$^{25}$Mg resonance strength. Jayatissa {\it et al.} \cite{JAYATISSA2020135267} give a resonance strength of $\omega\gamma_{(\alpha,n)} = 42(11)$ $\mu$eV combining the weighted average of the direct $^{22}$Ne($\alpha,\gamma$)$^{26}$Mg measurements of the resonance strength with the neutron/gamma branching ratio measured by Ota {\it et al.} \cite{OTA2020135256}. The resonance strength for the $^{22}$Ne($\alpha,\gamma$)$^{26}$Mg channel remains unchanged.

The measurements of Jayatissa {\it et al.} \cite{JAYATISSA2020135267} and Ota {\it et al.} \cite{OTA2020135256} both favour $J^\pi = 0^+$ or $J^\pi = 1^-$ assignments, and we adopt $J^\pi = 0^+$ for our calculations; as the resonance is narrow the choice of the $J^\pi$ assignment does not change the calculated reaction rate.

A weighed average of the $^{22}$Ne($\alpha,n$)$^{25}$Mg resonance strengths from the Jaeger \cite{PhysRevLett.87.202501}, Giesen \cite{GIESEN199395}, Drotleff \cite{1993ApJ...414..735D}, Jayatissa \cite{JAYATISSA2020135267} and Ota \cite{OTA2020135256} measurements gives a resonance strength of $\omega\gamma = 71(22)$ $\mu$eV, falling between the result for the TAMU experiments and the previous weighted average with a rate that falls between the rates calculated using the TAMU and Longland resonance strengths. However, as the two TAMU results are consistent with one another and differ from the Longland evaluation of the resonance strength by $3.1\sigma$ we instead evaluate the rate separately, using the results from the Texas A\&M experiments for one evaluation and the Longland weighted average for the other rather than using the weighted average of the global data.

There remains some dispute as to whether there is one or two resonances at this energy. Koehler \cite{PhysRevC.66.055805} argued that the $^{22}$Ne($\alpha,\gamma$)$^{26}$Mg and $^{22}$Ne($\alpha,n$)$^{25}$Mg resonances cannot be the same since, using the total width determined by Jaeger and the measured resonance strengths, the $\gamma$-ray partial width would be $\Gamma_\gamma = 56$ eV, much larger than a typical $\gamma$-ray partial width. However, Longland {\it et al.} \cite{PhysRevC.85.065809} noted that, due to the high uncertainty on the total width from Jaeger {\it et al.}, the uncertainty on the calculated $\Gamma_\gamma$ is extremely high, meaning that this argument cannot be used to conclude that the resonances are not the same.

Talwar {\it et al.} \cite{PhysRevC.93.055803} argued that the resonances observed in $^{22}$Ne($\alpha,\gamma$)$^{26}$Mg and $^{22}$Ne($\alpha,n$)$^{25}$Mg are the same. The basis of their argument is that the experimental energy resolution of the study of Jaeger {\it et al.} leads to an upper limit of 0.42 keV for the total width. In addition, they cite the unpublished $^{22}$Ne($\alpha,\gamma$)$^{26}$Mg data in the PhD thesis of Jaeger \cite{JaegerThesis} which uses the same resonance parameters as those of the $^{22}$Ne($\alpha,n$)$^{25}$Mg study \cite{PhysRevLett.87.202501}.

We note two additional pieces of evidence that the $^{22}$Ne($\alpha,\gamma$)$^{26}$Mg and $^{22}$Ne($\alpha,n$)$^{25}$Mg resonances originate from the same state. First, in the high-resolution $^{26}$Mg($p,p^\prime$)$^{26}$Mg data of Adsley {\it et al.} \cite{PhysRevC.97.045807} a state is observed at $E_x = 11.321(1)$ MeV, matching the energies of $^{22}$Ne($\alpha,\gamma$)$^{26}$Mg and $^{22}$Ne($\alpha,n$)$^{25}$Mg resonances within uncertainties. No other state is observed at around this excitation energy.

Second, the non-observation of a level at around this energy in the $^{25}$Mg$+n$ scattering study  by Massimi {\it et al.} \cite{Massimi20171} implies that the neutron partial width of the resonance is below the experimental sensitivity of the  $^{25}$Mg$+n$ scattering study, which is of the order of 20 eV at $E_n = 234$ keV. As a consequence, the neutron partial width is much smaller than might be expected from the total width taken from Jaeger {\it et al.} and the $\gamma$-ray partial width need not to be abnormally high to reproduce the relative resonance strengths. This is consistent with the argument of Talwar {\it et al.} \cite{PhysRevC.93.055803} that the total width determined by Jaeger {\it et al.} \cite{PhysRevLett.87.202501} should instead be interpreted as an upper limit, and the real total width could be much smaller.

In conclusion, the $^{22}$Ne($\alpha,\gamma$)$^{26}$Mg and $^{22}$Ne($\alpha,n$)$^{25}$Mg resonances correspond to the same state in $^{26}$Mg and the reaction rate calculations are performed under this assumption in the present paper. The resonance energy is taken from the proton inelastic scattering data of Adsley {\it et al.} \cite{PhysRevC.97.045807}. The $^{22}$Ne($\alpha,\gamma$)$^{26}$Mg resonance strength is taken from the weighted average of the direct measurements, and the $^{22}$Ne($\alpha,n$)$^{25}$Mg resonance strength is taken from the weighted average of the direct measurements and the indirect Texas A\&M $^{22}$Ne($^6$Li,$d$)$^{26}$Mg measurements given by Jayatissa {\it et al.} \cite{JAYATISSA2020135267}.

\section{Tables of level assignments \label{sec:LevelAssignments}}

The available nuclear data and the properties of states in $^{26}$Mg are summarised  in Tables \ref{tab:KnownResonances}-\ref{tab:UnknownResonances}.

Table \ref{tab:KnownResonances} contains the properties of resonances which are known to be natural parity and for which estimates or measurements of the $\alpha$-particle partial width exist. These include states which have been directly measured in $^{22}$Ne($\alpha,\gamma$)$^{26}$Mg and $^{22}$Ne($\alpha,n$)$^{25}$Mg reactions, and states which have been observed in $\alpha$-particle transfer reactions. In contrast to the evaluation of Longland {\it et al.} \cite{PhysRevC.85.065809}, when resonances have been observed in both $^{22}$Ne($\alpha,\gamma$)$^{26}$Mg and $^{22}$Ne($\alpha,n$)$^{25}$Mg reactions at the same energy we use the weighted average of the resonance energy, though for all resonances below the resonance at $E_{r,\mathrm{lab}} = 1433.7(12)$ keV, the resonance energies for the $^{22}$Ne($\alpha,\gamma$)$^{26}$Mg and $^{22}$Ne($\alpha,n$)$^{25}$Mg reactions are identical.

\begin{table*}
  \caption{Relevant data for resonances in $^{22}$Ne$+\alpha$ with known partial widths or resonance strengths. Resonance strengths and partial widths are taken from the evaluation of Longland {\it et al.} \cite{PhysRevC.85.065809} except for the $E_{r}^{CM} = 706$-keV resonance which has been recalculated using the results of a recently published study  \cite{PhysRevC.99.045804} (see Section \ref{statesec12}). Excitation and resonance energies ($E_{r}^{CM}$ is the resonance energy in the $\alpha+^{22}$Ne centre of mass system) have been re-calculated where appropriate (see Section \ref{sec:SynthesisOfTheData}). The final column indicates whether the contribution of the resonance is numerically integrated to account for the energy dependence of the partial widths.\label{tab:KnownResonances}}
  
   \begin{ruledtabular}
      \begin{tabular}{ c  c  c  c  c  c  c  c  c }
	\makecell{$E_x$\\{[MeV]}} & \makecell{$E_{r}^{CM}$\\{[keV]}} & $J^\pi$ & \makecell{$\omega\gamma_{(\alpha,\gamma)}$\\{[eV]}} & \makecell{$\omega\gamma_{(\alpha,n)}$\\{[eV]}} & \makecell{$\Gamma_\alpha$\\{[eV]}}  & \makecell{$\Gamma_\gamma$\\{[eV]}} & \makecell{$\Gamma_n$\\{[eV]}} & \makecell{Integrate\\resonance?} \\ \\
	\hline \\
	$10.6963(4)$ & $81.6(4)$ & $4^+$ & & & $3.5(18) \times 10^{-46}$ & $3.0(15)$ & $0$ & No \\
    $11.084(1)$ & $469(1)$ & $2^+$ & & & $5.7(1.5) \times 10^{-11}$ & $3.0(15)$& 0 & No \\
	$11.321(1)$& $706(1)$\footnote{Resonance energy is taken using the state observed in Ref. \cite{PhysRevC.97.045807} and assuming a single resonance.}                & $0^+/1^-$& $3.7(4) \times 10^{-5}$ & $4.2(11) \times 10^{-5}$ & & & & No \\
	$11.44120(4)$& $826.46(5)$ & $3^-$ & &$3.9(10)\times10^{-5}$ & $5.50(14) \times 10^{-6}$ & $3.0(15)$ & $1.47(8) \times 10^{3}$ & Yes \\
	$11.46574(6)$ & $851.00(6)$& $3^-$ & & $5.5(17)\times10^{-5}$ & $7.9(2.4) \times 10^{-6}$ & $3.0(15)$ & $6.55(9) \times 10^3$ & Yes\\
	$11.5080(9)$ & $893.3(9)$  & $1^-$ & & $3.5(6)\times10^{-4}$& $1.2(4)\times10^{-4}$&$3.0(15)$ & $1.27(25)\times10^3$& Yes\\
	$11.5260(15)$ & $911.3(15)$&$1^-$  & & $1.3(4)\times10^{-3}$ &$4.3(11)\times10^{-4}$ &$3.0(15)$ &$1.80(25)\times10^3$ & Yes\\
	$11.630(1)$ & $1015.3(14)$  & $1^-$ & &$7.1(15)\times10^{-3}$ &$2.4(5)\times10^{-3}$ &$3.0(15)$ & $13.5(17)\times10^3$&Yes \\
	$11.749(5)$ & $1133(6)$     & $1^-$ & & $5.9(8)\times10^{-2}$ &$2.0(3)\times 10^{-2}$ & $3.0(15)$& $64(9)\times10^3$&Yes \\
	$11.787(3)$ & $1172(3)$     &$1^-$  & &$2.5(9)\times10^{-2}$ &$8(3)\times10^{-3}$ &$3.0(15)$ &$24.5(24)\times10^3$ &Yes \\
	$11.828(1)$ & $1213(1)$     &$2^+$  & &$2.5(3)\times10^{-4}$ & $1.8(1)\times10^{-1}$ &$3.0(15)$ &$1.10(25)\times10^3$ &Yes \\
	$11.863(3)$ & $1248(3)$ & $1^-$ & & & $1.5(10) \times 10^{-2}$ & $3.0(15)$ & $2.45(34) \times 10^4$ & Yes \\
	$11.880(3)$ & $1265(3)$ & $1^-$ & & $1.9(19) \times 10^{-1}$ & $6.30(63) \times 10^{-2}$ & $3.0(15)$ & $3.0(15) \times 10^3$ & No \\
	$11.895(4)$ & $1280(4)$     &$1^-$  & $2.0(2)\times10^{-3}$& $4.1(4)\times10^{-1}$& & &  & No\\
	$11.911(1)$ & $1297(3)$     &$1^-$  & $3.4(4)\times10^{-3}$&$1.4(1)$ &$1.9(9.8)$ &$3.0(15)$ &$5(2)\times10^3$ & Yes\\
	$11.953(3)$ & $1338(3)$     &$2^+$  & $3.4(4)\times10^{-3}$& $1.60(13)$ & $3.2(1.7)\times 10^{-1}$ & $3.0(15)$ & $2(1) \times 10^3$ & Yes\\
	$12.050(1)$ & $1436(3)$     &$2^+$  & $6.0(7)\times10^{-3}$&$4.7(3)$ &$1.1(3)\times10^{-1}$ &$3.0(15)$ &$4(1)\times10^3$ & Yes\\
	$12.141(1)$ & $1526(3)$     &$1^-$  & $1.0(2)\times10^{-3}$&$2.4(2)$ & $1.7(5)$& $3.0(15)$& $1.5(2)\times10^{4}$& Yes\\
	$12.184(5)$ & $1569(7)$     &$0^+$  & $1.1(2)\times10{-3}$& $1.21(29)\times10^1$& $0.90(11)$&$3.0(15)$ & $3.3(5)\times10^4$& Yes\\
	$12.270(5)$ & $1658(7)$     &$0^+$  &$8.9(1)\times10^{-3}$ &$2.1(2)\times10^1$ &$2.2(4)\times10^2$ &$3.0(15)$ & $7.3(9)\times10^{4}$& Yes\\
	$12.344(2)$ & $1728(4)$     &$0^+$  &$5.4(7)\times10^{-2}$ &$1.57(10)\times10^{2}$&$6.30(12)\times10^{2}$ &$3.0(15)$ & $3.5(5)\times10^{4}$ & Yes \\
      \end{tabular}
  \end{ruledtabular}
\end{table*}

Tables \ref{tab:ULResonances} contains properties of resonances which are known to have natural parity and for which the spin of the state is known. Therefore, these states can in principle contribute to the reaction rates but no estimate of the $\alpha$-particle partial width is available. The upper limits for the $\alpha$-particle partial widths for these states come from the Wigner limits or, in some cases, from $\alpha$-particle transfer-reaction data.

\begin{table*}
\caption{Nuclear data inputs for unobserved resonances in the $^{22}$Ne$+\alpha$ system with known natural parity. Alpha-particle partial widths are computed from the Wigner limits unless a lower estimate is available. Excitation and resonance energies ($E_{r}^{CM}$ is the resonance energy in the $\alpha-^{22}$Ne centre of mass system) have been re-calculated where required.\label{tab:ULResonances}}

  \begin{ruledtabular}
    \begin{tabular}{ c  c  c  c  c  c  c }
      \makecell{$E_x$\\{[MeV]}} & \makecell{$E_r^{CM}$\\{[keV]}} & $J^\pi$ & \makecell{$\Gamma_{\alpha,UL}$\\{[eV]}} & \makecell{$\Gamma_\gamma$\\{[eV]}} & \makecell{$\Gamma_n$\\{[eV]}} & \makecell{Integrate\\resonance?} \\ \\
      \hline \\
      $10.6507(4)$ & $36.0(4)$ & $7^-$ & $1.60 \times 10^{-76}$ & $3.0(15)$ & 0 & No\\
      $10.8057(7)$ & $191.0(7)$ & $1^-$ & $3.2 \times 10^{-23}$ & $0.72(18)$ & 0 & No \\
      $10.818(1)$  & $203(1)$ & $0^+$ & $1.29\times 10^{-20}$ & $3.0(15)$ & 0 & No \\
      $10.826(1)$ & $211(1)$ & $(2^+)$ & $6.65\times10^{-21}$ & $3.0(15)$ & 0 & No \\
      $10.8976(47)$ & $278(1)$ & $(4^+)$ & $1.41\times10^{-18}$ & $3.0(15)$ & 0 & No \\
      $10.9491(1)$ & $334.4(8)$ & $1^-$ & $2.90 \times 10^{-15}$ &$1.9(3)$ & 0 & No \\
      $11.11223(4)$ & $497.49(5)$ & $2^+$ & $4.3 \times 10^{-10}$ & $1.37(6)\times10^{-2}$ &$2.095(5)\times10^3$ & Yes \\
      $11.16310(4)$ & $548.36(5)$ & $2^+$ & $5.2\times10^{-9}$ & $2.8(2)$ & $5.31(5)\times10^3$ & Yes  \\
      $11.16926(4)$ & $554.52(5)$ &$3^-$ & $4.4\times10^{-10}$& $3.3(2)$ &$1.94(2)\times10^{3}$ & Yes \\
      $11.17107(4)$ &  $556.33(5)$           &$2^+$ & $1.3\times10^{-11}$& $3(2)$   & $0.8(7)$             & No \\
      $11.27380(4)$ & $659.06(5)$ & $2^+$& $1.00\times10^{-6}$&$2.2(2)$ &$4.1(1)\times10^2$& Yes \\
      $11.27963(4)$ & $664.89(5)$ & $3^-$&$9.20\times10^{-8}$ &$3(1)\times10^{-1}$ &$1.81(2)\times10^{3}$ & Yes \\
      $11.30100(9)$ & $686.26(9)$ & $(2^+)$&$1.53\times10^{-5}$ &$<3$ & $<2.0\times10^1$& No \\
      $11.32768(4)$ & $712.94(5)$ & $(1^-)$&$1.80\times10^{-6}$ &$2.2(3)$ &$1.71(6)\times10^2$ & Yes \\
      $11.33696(4)$ & $722.22(5)$ &$(1^-)$ &$1.74\times10^{-4}$ & $<3$& $<2.0\times10^1$& No \\
      $11.34389(9)$ & $729.15(9)$ & $(2^+)$& $1.10\times10^{-6}$&$1.0(2)$ &$<1.95\times10^3$ & Yes \\
      $11.50022(4)$ & $885.48(5)$ & $1^-$&$1.95\times10^{-1}$ &$3.0(15)$ & $3.0(15)\times10^3$& Yes \\
    \end{tabular}
  \end{ruledtabular}
\end{table*}

States with unknown or uncertain spin-parities which could potentially contribute to the reaction rates are given in Table \ref{tab:UnknownResonances} along with potentially useful nuclear data such as constraints 
on the partial widths for the state. Where available, known limits on the spin-parity of the states are also included. 

\begin{table*}
  \caption{States in $^{26}$Mg without confirmed spin-parity (i.e. unknown or tentative). Any known limit on the spin and parity of a state is included in the table. Note that this list is limited to levels between the $\alpha$-particle threshold and the lowest directly measured resonance at $E_{r}^{CM} =~706$-keV ($E_x = 11.321$ MeV). We assume that all important resonances at higher energy would have been observed in Refs. \cite{PhysRevLett.87.202501,Wolke1989}.\label{tab:UnknownResonances}}
  
    \begin{ruledtabular}
      \begin{tabular}{ c  c  c  c }
      \makecell{$E_x$\\{[MeV]}} & \makecell{$E_{r}^{CM}$\\{[keV]}} & \makecell{Possible\\$J^\pi$} & Comments\\ \\
      \hline \\
      $10.684(1)$ & $69(3)$& & \\
      $10.7037(22)$ & $89(2)$ & $2^+,4$ & \\
      $10.7089(22)$ & $94(2)$ & $3-6$ & \\
      $10.730(1)$ & $115(1)$ & $3^+$ & Tentative assignment from the Gammasphere data \cite{Lotay2019} \\
      $10.882(1)$ &$267(1)$ & & \\
      $10.915(1)$ &$300(1)$ & $2^+/6^+$ &  \\
      $10.928(1)$ &$313(1)$ & & \\
      $10.978(1)$ &$363(1)$ & & \\
      $11.017(1)$ &$402(1)$ & $2-5$ & \\
      $11.0386(20)$ & $424(3)$ & $2-6$ & \\
      $11.047(1)$ &$432(1)$ & & \\
      $11.074(1)$ &$459(1)$ &  & \\
      $11.102(1)$ &$487(1)$ & & $E_n = 9$ keV, $\Gamma_n < 3(1)$ eV \\
      $11.1818(20)$ & $567(2)$ & $2-6$ & \\
      $11.119(1)$ &$504(1)$ & & $E_n = 27$ keV, $\Gamma_n < 7(1)$ eV \\
      \end{tabular}
    \end{ruledtabular}
\end{table*}

\section{Reaction-rate evaluation using RatesMC}
\label{ReactionRateEvaluationSTARLIB}

The re-evaluation of the reaction rates was performed using the STARLIB Monte-Carlo code \textsc{RatesMC} \cite{0067-0049-207-1-18,PhysRevC.85.065809}, accessible online \cite{STARLIBWebsite}. This was done to ensure that the only variations observed in the reaction rates were due to changes in the nuclear data while the calculation methodology remained identical to that of Longland {\it et al.} \cite{PhysRevC.85.065809}. The resulting reaction rates  are given in Tables \ref{tab:Ne22_alphagamma_rate} and \ref{tab:Ne22_alphan_rate}. The rates were evaluated using the resonances given in Table \ref{tab:KnownResonances} as measured resonance values and the resonances listed in Table \ref{tab:ULResonances} as upper limits for resonances.

To show the effect of the changes in the nuclear data for the $E_r = 706$-keV resonance, the reaction rates for $^{22}$Ne($\alpha,\gamma$)$^{26}$Mg and $^{22}$Ne($\alpha,n$)$^{25}$Mg without the inclusion of the results from the Texas A\&M experiments were also computed. These reaction rate are given in Tables \ref{tab:Ne22_alphagamma_rate_no_TAMU} and \ref{tab:Ne22_alphan_rate_no_TAMU}. For details on how \textsc{RatesMC} estimates the reaction rates, refer to Refs. \cite{0067-0049-207-1-18,PhysRevC.85.065809}.

For the avoidance of doubt: the rates in the present paper which are recommended for use are those in Tables \ref{tab:Ne22_alphagamma_rate} and \ref{tab:Ne22_alphan_rate} which incorporate the results from the Texas A\&M experiments. We recommend the rates with the inclusion of data from the Texas A\&M measurements as these two experiments show consistent results, suggesting a reduction of the ${22}$Ne($\alpha,n$)$^{25}$Mg resonance strength for the $E_r = 706$-keV resonance.

\begin{table*}
\caption{Recommended $^{22}$Ne($\alpha,\gamma$)$^{26}$Mg rate from the present evaluation incorporating the TAMU results. The low, median and high rates correspond to the 16\%, 50\% and 84\% values of the cumulative probability distribution. The $\mu$ and $\sigma$ parameters are resulting from fitting the distribution of rates at that temperature to a log-normal distribution. The Anderson-Darling statistic is a measure of how the data are well described by a log-normal distribution - see Refs. \cite{0067-0049-207-1-18,PhysRevC.85.065809} for details. All rates and the $\mu$ and $\sigma$ parameters are given in cm$^3$ mol$^{-1}$ s$^{-1}$. At temperatures above 1.33 GK, the rates should be taken from Hauser-Feshbach models - see Ref. \cite{PhysRevC.85.065809}.}
\begin{ruledtabular}
\begin{tabular}{c c c c c c c }
T [GK] & Low rate & Median rate & High rate & Log-normal $\mu$ & Log-normal $\sigma$ & Anderson-Darling statistic \\
\hline \\
0.010 & 3.38$\times$10$^{-79}$ & 5.52$\times$10$^{-79}$ &
      9.04$\times$10$^{-79}$ &  -1.802$\times$10$^{+02}$ &
       5.01$\times$10$^{-01}$  & 2.73$\times$10$^{-01}$ \\ 
0.011 & 1.59$\times$10$^{-75}$ & 2.58$\times$10$^{-75}$ &
      4.21$\times$10$^{-75}$ &  -1.717$\times$10$^{+02}$ &
       4.95$\times$10$^{-01}$  & 3.22$\times$10$^{-01}$ \\ 
0.012 & 1.82$\times$10$^{-72}$ & 2.93$\times$10$^{-72}$ &
      4.75$\times$10$^{-72}$ &  -1.647$\times$10$^{+02}$ &
       4.89$\times$10$^{-01}$  & 3.22$\times$10$^{-01}$ \\ 
0.013 & 6.96$\times$10$^{-70}$ & 1.12$\times$10$^{-69}$ &
      1.81$\times$10$^{-69}$ &  -1.588$\times$10$^{+02}$ &
       4.85$\times$10$^{-01}$  & 3.26$\times$10$^{-01}$ \\ 
0.014 & 1.13$\times$10$^{-67}$ & 1.81$\times$10$^{-67}$ &
      2.93$\times$10$^{-67}$ &  -1.537$\times$10$^{+02}$ &
       4.83$\times$10$^{-01}$  & 3.25$\times$10$^{-01}$ \\ 
0.015 & 9.26$\times$10$^{-66}$ & 1.48$\times$10$^{-65}$ &
      2.40$\times$10$^{-65}$ &  -1.493$\times$10$^{+02}$ &
       4.83$\times$10$^{-01}$  & 3.23$\times$10$^{-01}$ \\ 
0.016 & 4.34$\times$10$^{-64}$ & 6.93$\times$10$^{-64}$ &
      1.13$\times$10$^{-63}$ &  -1.454$\times$10$^{+02}$ &
       4.83$\times$10$^{-01}$  & 3.10$\times$10$^{-01}$ \\ 
0.018 & 2.60$\times$10$^{-61}$ & 4.16$\times$10$^{-61}$ &
      6.75$\times$10$^{-61}$ &  -1.390$\times$10$^{+02}$ &
       4.85$\times$10$^{-01}$  & 2.84$\times$10$^{-01}$ \\ 
0.020 & 4.27$\times$10$^{-59}$ & 6.83$\times$10$^{-59}$ &
      1.12$\times$10$^{-58}$ &  -1.339$\times$10$^{+02}$ &
       4.87$\times$10$^{-01}$  & 2.55$\times$10$^{-01}$ \\ 
0.025 & 5.79$\times$10$^{-55}$ & 9.65$\times$10$^{-55}$ &
      1.59$\times$10$^{-54}$ &  -1.244$\times$10$^{+02}$ &
       5.03$\times$10$^{-01}$  & 9.62$\times$10$^{-01}$ \\ 
0.030 & 8.22$\times$10$^{-50}$ & 3.94$\times$10$^{-49}$ &
      1.40$\times$10$^{-48}$ &  -1.116$\times$10$^{+02}$ &
       1.37$\times$10$^{+00}$  & 2.50$\times$10$^{+01}$ \\ 
0.040 & 1.18$\times$10$^{-41}$ & 4.00$\times$10$^{-41}$ &
      1.10$\times$10$^{-40}$ &  -9.314$\times$10$^{+01}$ &
       1.13$\times$10$^{+00}$  & 2.34$\times$10$^{+01}$ \\ 
0.050 & 9.31$\times$10$^{-37}$ & 2.76$\times$10$^{-36}$ &
      6.55$\times$10$^{-36}$ &  -8.200$\times$10$^{+01}$ &
       1.01$\times$10$^{+00}$  & 3.40$\times$10$^{+01}$ \\ 
0.060 & 1.70$\times$10$^{-33}$ & 4.78$\times$10$^{-33}$ &
      1.07$\times$10$^{-32}$ &  -7.455$\times$10$^{+01}$ &
       9.68$\times$10$^{-01}$  & 4.07$\times$10$^{+01}$ \\ 
0.070 & 3.65$\times$10$^{-31}$ & 9.84$\times$10$^{-31}$ &
      2.19$\times$10$^{-30}$ &  -6.921$\times$10$^{+01}$ &
       9.44$\times$10$^{-01}$  & 3.71$\times$10$^{+01}$ \\ 
0.080 & 2.20$\times$10$^{-29}$ & 5.64$\times$10$^{-29}$ &
      1.22$\times$10$^{-28}$ &  -6.513$\times$10$^{+01}$ &
       8.60$\times$10$^{-01}$  & 1.85$\times$10$^{+01}$ \\ 
0.090 & 8.06$\times$10$^{-28}$ & 1.72$\times$10$^{-27}$ &
      3.24$\times$10$^{-27}$ &  -6.169$\times$10$^{+01}$ &
       7.24$\times$10$^{-01}$  & 2.00$\times$10$^{+01}$ \\ 
0.100 & 1.68$\times$10$^{-26}$ & 4.26$\times$10$^{-26}$ &
      8.66$\times$10$^{-26}$ &  -5.852$\times$10$^{+01}$ &
       7.83$\times$10$^{-01}$  & 4.51$\times$10$^{+01}$ \\ 
0.110 & 2.73$\times$10$^{-25}$ & 8.38$\times$10$^{-25}$ &
      2.12$\times$10$^{-24}$ &  -5.553$\times$10$^{+01}$ &
       9.06$\times$10$^{-01}$  & 5.08$\times$10$^{+01}$ \\ 
0.120 & 5.43$\times$10$^{-24}$ & 1.44$\times$10$^{-23}$ &
      3.57$\times$10$^{-23}$ &  -5.263$\times$10$^{+01}$ &
       8.12$\times$10$^{-01}$  & 6.23$\times$10$^{+01}$ \\ 
0.130 & 1.11$\times$10$^{-22}$ & 2.11$\times$10$^{-22}$ &
      4.36$\times$10$^{-22}$ &  -4.989$\times$10$^{+01}$ &
       5.97$\times$10$^{-01}$  & 4.85$\times$10$^{+01}$ \\ 
0.140 & 1.71$\times$10$^{-21}$ & 2.62$\times$10$^{-21}$ &
      4.30$\times$10$^{-21}$ &  -4.738$\times$10$^{+01}$ &
       4.15$\times$10$^{-01}$  & 2.27$\times$10$^{+01}$ \\ 
0.150 & 1.90$\times$10$^{-20}$ & 2.63$\times$10$^{-20}$ &
      3.62$\times$10$^{-20}$ &  -4.509$\times$10$^{+01}$ &
       3.09$\times$10$^{-01}$  & 6.46$\times$10$^{+00}$ \\ 
0.160 & 1.58$\times$10$^{-19}$ & 2.07$\times$10$^{-19}$ &
      2.68$\times$10$^{-19}$ &  -4.303$\times$10$^{+01}$ &
       2.63$\times$10$^{-01}$  & 1.91$\times$10$^{+00}$ \\ 
0.180 & 5.56$\times$10$^{-18}$ & 7.02$\times$10$^{-18}$ &
      8.96$\times$10$^{-18}$ &  -3.950$\times$10$^{+01}$ &
       2.41$\times$10$^{-01}$  & 1.76$\times$10$^{-01}$ \\ 
0.200 & 1.08$\times$10$^{-16}$ & 1.33$\times$10$^{-16}$ &
      1.67$\times$10$^{-16}$ &  -3.655$\times$10$^{+01}$ &
       2.19$\times$10$^{-01}$  & 9.84$\times$10$^{-01}$ \\ 
0.250 & 5.68$\times$10$^{-14}$ & 6.42$\times$10$^{-14}$ &
      7.25$\times$10$^{-14}$ &  -3.038$\times$10$^{+01}$ &
       1.22$\times$10$^{-01}$  & 3.76$\times$10$^{-01}$ \\ 
0.300 & 7.63$\times$10$^{-12}$ & 8.60$\times$10$^{-12}$ &
      9.78$\times$10$^{-12}$ &  -2.547$\times$10$^{+01}$ &
       1.28$\times$10$^{-01}$  & 4.29$\times$10$^{+00}$ \\ 
0.350 & 2.85$\times$10$^{-10}$ & 3.23$\times$10$^{-10}$ &
      3.71$\times$10$^{-10}$ &  -2.185$\times$10$^{+01}$ &
       1.36$\times$10$^{-01}$  & 6.78$\times$10$^{+00}$ \\ 
0.400 & 4.32$\times$10$^{-09}$ & 4.89$\times$10$^{-09}$ &
      5.64$\times$10$^{-09}$ &  -1.913$\times$10$^{+01}$ &
       1.40$\times$10$^{-01}$  & 9.12$\times$10$^{+00}$ \\ 
0.450 & 3.52$\times$10$^{-08}$ & 3.99$\times$10$^{-08}$ &
      4.61$\times$10$^{-08}$ &  -1.703$\times$10$^{+01}$ &
       1.42$\times$10$^{-01}$  & 1.13$\times$10$^{+01}$ \\ 
0.500 & 1.86$\times$10$^{-07}$ & 2.11$\times$10$^{-07}$ &
      2.44$\times$10$^{-07}$ &  -1.536$\times$10$^{+01}$ &
       1.44$\times$10$^{-01}$  & 1.33$\times$10$^{+01}$ \\ 
0.600 & 2.19$\times$10$^{-06}$ & 2.48$\times$10$^{-06}$ &
      2.89$\times$10$^{-06}$ &  -1.289$\times$10$^{+01}$ &
       1.46$\times$10$^{-01}$  & 1.68$\times$10$^{+01}$ \\ 
0.700 & 1.25$\times$10$^{-05}$ & 1.42$\times$10$^{-05}$ &
      1.65$\times$10$^{-05}$ &  -1.115$\times$10$^{+01}$ &
       1.46$\times$10$^{-01}$  & 1.94$\times$10$^{+01}$ \\ 
0.800 & 4.70$\times$10$^{-05}$ & 5.31$\times$10$^{-05}$ &
      6.15$\times$10$^{-05}$ &  -9.829$\times$10$^{+00}$ &
       1.44$\times$10$^{-01}$  & 1.95$\times$10$^{+01}$ \\ 
0.900 & 1.36$\times$10$^{-04}$ & 1.54$\times$10$^{-04}$ &
      1.80$\times$10$^{-04}$ &  -8.762$\times$10$^{+00}$ &
       1.46$\times$10$^{-01}$  & 1.71$\times$10$^{+01}$ \\ 
1.000 & 3.36$\times$10$^{-04}$ & 3.85$\times$10$^{-04}$ &
      4.56$\times$10$^{-04}$ &  -7.846$\times$10$^{+00}$ &
       1.60$\times$10$^{-01}$  & 1.89$\times$10$^{+01}$ \\ 
1.250 & 2.11$\times$10$^{-03}$ & 2.55$\times$10$^{-03}$ &
      3.23$\times$10$^{-03}$ &  -5.946$\times$10$^{+00}$ &
       2.23$\times$10$^{-01}$  & 2.58$\times$10$^{+01}$ \\ 
\end{tabular}
\end{ruledtabular}
\label{tab:Ne22_alphagamma_rate}
\end{table*}

\begin{table*}
\caption{The $^{22}$Ne($\alpha,\gamma$)$^{26}$Mg reaction rate evaluated without the incorporation of the TAMU results. The low, median and high rates correspond to the 16\%, 50\% and 84\% values of the cumulative probability distribution. The $\mu$ and $\sigma$ parameters are resulting from fitting the distribution of rates at that temperature to a log-normal distribution. The Anderson-Darling statistic is a measure of how the data are well described by a log-normal distribution - see Refs. \cite{0067-0049-207-1-18,PhysRevC.85.065809} for details. All rates and the $\mu$ and $\sigma$ parameters are given in cm$^3$ mol$^{-1}$ s$^{-1}$. At temperatures above 1.33 GK, the rates should be taken from Hauser-Feshbach models - see Ref. \cite{PhysRevC.85.065809}.}
\begin{ruledtabular}
\begin{tabular}{c c c c c c c }
T [GK] & Low rate & Median rate & High rate & Log-normal $\mu$ & Log-normal $\sigma$ & Anderson-Darling statistic \\
\hline \\
0.010 & 3.38$\times$10$^{-79}$ & 5.55$\times$10$^{-79}$ &
      9.05$\times$10$^{-79}$ &  -1.802$\times$10$^{+02}$ &
       4.96$\times$10$^{-01}$  & 1.42$\times$10$^{-01}$ \\ 
0.011 & 1.59$\times$10$^{-75}$ & 2.58$\times$10$^{-75}$ &
      4.20$\times$10$^{-75}$ &  -1.717$\times$10$^{+02}$ &
       4.91$\times$10$^{-01}$  & 1.88$\times$10$^{-01}$ \\ 
0.012 & 1.82$\times$10$^{-72}$ & 2.94$\times$10$^{-72}$ &
      4.79$\times$10$^{-72}$ &  -1.647$\times$10$^{+02}$ &
       4.86$\times$10$^{-01}$  & 2.32$\times$10$^{-01}$ \\ 
0.013 & 6.96$\times$10$^{-70}$ & 1.13$\times$10$^{-69}$ &
      1.82$\times$10$^{-69}$ &  -1.588$\times$10$^{+02}$ &
       4.83$\times$10$^{-01}$  & 2.77$\times$10$^{-01}$ \\ 
0.014 & 1.13$\times$10$^{-67}$ & 1.83$\times$10$^{-67}$ &
      2.95$\times$10$^{-67}$ &  -1.537$\times$10$^{+02}$ &
       4.81$\times$10$^{-01}$  & 3.15$\times$10$^{-01}$ \\ 
0.015 & 9.23$\times$10$^{-66}$ & 1.49$\times$10$^{-65}$ &
      2.42$\times$10$^{-65}$ &  -1.493$\times$10$^{+02}$ &
       4.81$\times$10$^{-01}$  & 3.50$\times$10$^{-01}$ \\ 
0.016 & 4.30$\times$10$^{-64}$ & 6.99$\times$10$^{-64}$ &
      1.14$\times$10$^{-63}$ &  -1.454$\times$10$^{+02}$ &
       4.82$\times$10$^{-01}$  & 3.65$\times$10$^{-01}$ \\ 
0.018 & 2.57$\times$10$^{-61}$ & 4.21$\times$10$^{-61}$ &
      6.84$\times$10$^{-61}$ &  -1.390$\times$10$^{+02}$ &
       4.84$\times$10$^{-01}$  & 2.76$\times$10$^{-01}$ \\ 
0.020 & 4.22$\times$10$^{-59}$ & 6.93$\times$10$^{-59}$ &
      1.13$\times$10$^{-58}$ &  -1.339$\times$10$^{+02}$ &
       4.87$\times$10$^{-01}$  & 2.27$\times$10$^{-01}$ \\ 
0.025 & 5.83$\times$10$^{-55}$ & 9.83$\times$10$^{-55}$ &
      1.62$\times$10$^{-54}$ &  -1.244$\times$10$^{+02}$ &
       5.05$\times$10$^{-01}$  & 1.44$\times$10$^{+00}$ \\ 
0.030 & 8.41$\times$10$^{-50}$ & 4.14$\times$10$^{-49}$ &
      1.48$\times$10$^{-48}$ &  -1.116$\times$10$^{+02}$ &
       1.41$\times$10$^{+00}$  & 3.25$\times$10$^{+01}$ \\ 
0.040 & 1.19$\times$10$^{-41}$ & 4.25$\times$10$^{-41}$ &
      1.18$\times$10$^{-40}$ &  -9.310$\times$10$^{+01}$ &
       1.16$\times$10$^{+00}$  & 3.31$\times$10$^{+01}$ \\ 
0.050 & 9.66$\times$10$^{-37}$ & 2.96$\times$10$^{-36}$ &
      7.00$\times$10$^{-36}$ &  -8.195$\times$10$^{+01}$ &
       1.03$\times$10$^{+00}$  & 4.45$\times$10$^{+01}$ \\ 
0.060 & 1.78$\times$10$^{-33}$ & 5.14$\times$10$^{-33}$ &
      1.11$\times$10$^{-32}$ &  -7.450$\times$10$^{+01}$ &
       9.83$\times$10$^{-01}$  & 5.10$\times$10$^{+01}$ \\ 
0.070 & 3.77$\times$10$^{-31}$ & 1.06$\times$10$^{-30}$ &
      2.21$\times$10$^{-30}$ &  -6.916$\times$10$^{+01}$ &
       9.54$\times$10$^{-01}$  & 4.60$\times$10$^{+01}$ \\ 
0.080 & 2.28$\times$10$^{-29}$ & 5.95$\times$10$^{-29}$ &
      1.23$\times$10$^{-28}$ &  -6.509$\times$10$^{+01}$ &
       8.67$\times$10$^{-01}$  & 2.45$\times$10$^{+01}$ \\ 
0.090 & 8.31$\times$10$^{-28}$ & 1.76$\times$10$^{-27}$ &
      3.28$\times$10$^{-27}$ &  -6.167$\times$10$^{+01}$ &
       7.28$\times$10$^{-01}$  & 2.19$\times$10$^{+01}$ \\ 
0.100 & 1.75$\times$10$^{-26}$ & 4.25$\times$10$^{-26}$ &
      8.54$\times$10$^{-26}$ &  -5.852$\times$10$^{+01}$ &
       7.87$\times$10$^{-01}$  & 4.46$\times$10$^{+01}$ \\ 
0.110 & 2.42$\times$10$^{-25}$ & 7.97$\times$10$^{-25}$ &
      2.02$\times$10$^{-24}$ &  -5.561$\times$10$^{+01}$ &
       9.86$\times$10$^{-01}$  & 5.22$\times$10$^{+01}$ \\ 
0.120 & 3.02$\times$10$^{-24}$ & 1.22$\times$10$^{-23}$ &
      3.24$\times$10$^{-23}$ &  -5.294$\times$10$^{+01}$ &
       1.10$\times$10$^{+00}$  & 6.52$\times$10$^{+01}$ \\ 
0.130 & 3.78$\times$10$^{-23}$ & 1.47$\times$10$^{-22}$ &
      3.61$\times$10$^{-22}$ &  -5.048$\times$10$^{+01}$ &
       1.09$\times$10$^{+00}$  & 7.52$\times$10$^{+01}$ \\ 
0.140 & 3.97$\times$10$^{-22}$ & 1.40$\times$10$^{-21}$ &
      3.09$\times$10$^{-21}$ &  -4.823$\times$10$^{+01}$ &
       1.04$\times$10$^{+00}$  & 8.16$\times$10$^{+01}$ \\ 
0.150 & 3.47$\times$10$^{-21}$ & 1.09$\times$10$^{-20}$ &
      2.19$\times$10$^{-20}$ &  -4.616$\times$10$^{+01}$ &
       9.84$\times$10$^{-01}$  & 6.05$\times$10$^{+01}$ \\ 
0.160 & 2.45$\times$10$^{-20}$ & 6.86$\times$10$^{-20}$ &
      1.47$\times$10$^{-19}$ &  -4.424$\times$10$^{+01}$ &
       9.44$\times$10$^{-01}$  & 2.59$\times$10$^{+01}$ \\ 
0.180 & 8.92$\times$10$^{-19}$ & 1.99$\times$10$^{-18}$ &
      4.96$\times$10$^{-18}$ &  -4.073$\times$10$^{+01}$ &
       8.47$\times$10$^{-01}$  & 1.92$\times$10$^{+00}$ \\ 
0.200 & 2.77$\times$10$^{-17}$ & 5.00$\times$10$^{-17}$ &
      1.06$\times$10$^{-16}$ &  -3.747$\times$10$^{+01}$ &
       6.40$\times$10$^{-01}$  & 2.56$\times$10$^{+01}$ \\ 
0.250 & 4.52$\times$10$^{-14}$ & 5.44$\times$10$^{-14}$ &
      6.84$\times$10$^{-14}$ &  -3.052$\times$10$^{+01}$ &
       2.13$\times$10$^{-01}$  & 1.87$\times$10$^{+01}$ \\ 
0.300 & 7.45$\times$10$^{-12}$ & 8.54$\times$10$^{-12}$ &
      9.85$\times$10$^{-12}$ &  -2.548$\times$10$^{+01}$ &
       1.42$\times$10$^{-01}$  & 2.44$\times$10$^{+00}$ \\ 
0.350 & 2.86$\times$10$^{-10}$ & 3.25$\times$10$^{-10}$ &
      3.73$\times$10$^{-10}$ &  -2.184$\times$10$^{+01}$ &
       1.39$\times$10$^{-01}$  & 5.49$\times$10$^{+00}$ \\ 
0.400 & 4.33$\times$10$^{-09}$ & 4.94$\times$10$^{-09}$ &
      5.66$\times$10$^{-09}$ &  -1.912$\times$10$^{+01}$ &
       1.42$\times$10$^{-01}$  & 8.52$\times$10$^{+00}$ \\ 
0.450 & 3.52$\times$10$^{-08}$ & 4.02$\times$10$^{-08}$ &
      4.63$\times$10$^{-08}$ &  -1.702$\times$10$^{+01}$ &
       1.44$\times$10$^{-01}$  & 1.10$\times$10$^{+01}$ \\ 
0.500 & 1.86$\times$10$^{-07}$ & 2.12$\times$10$^{-07}$ &
      2.45$\times$10$^{-07}$ &  -1.536$\times$10$^{+01}$ &
       1.46$\times$10$^{-01}$  & 1.32$\times$10$^{+01}$ \\ 
0.600 & 2.19$\times$10$^{-06}$ & 2.50$\times$10$^{-06}$ &
      2.89$\times$10$^{-06}$ &  -1.289$\times$10$^{+01}$ &
       1.49$\times$10$^{-01}$  & 1.69$\times$10$^{+01}$ \\ 
0.700 & 1.25$\times$10$^{-05}$ & 1.43$\times$10$^{-05}$ &
      1.66$\times$10$^{-05}$ &  -1.115$\times$10$^{+01}$ &
       1.49$\times$10$^{-01}$  & 1.99$\times$10$^{+01}$ \\ 
0.800 & 4.71$\times$10$^{-05}$ & 5.33$\times$10$^{-05}$ &
      6.19$\times$10$^{-05}$ &  -9.826$\times$10$^{+00}$ &
       1.47$\times$10$^{-01}$  & 2.06$\times$10$^{+01}$ \\ 
0.900 & 1.36$\times$10$^{-04}$ & 1.55$\times$10$^{-04}$ &
      1.81$\times$10$^{-04}$ &  -8.758$\times$10$^{+00}$ &
       1.50$\times$10$^{-01}$  & 1.86$\times$10$^{+01}$ \\ 
1.000 & 3.36$\times$10$^{-04}$ & 3.87$\times$10$^{-04}$ &
      4.58$\times$10$^{-04}$ &  -7.843$\times$10$^{+00}$ &
       1.65$\times$10$^{-01}$  & 1.83$\times$10$^{+01}$ \\ 
1.250 & 2.10$\times$10$^{-03}$ & 2.56$\times$10$^{-03}$ &
      3.25$\times$10$^{-03}$ &  -5.944$\times$10$^{+00}$ &
       2.29$\times$10$^{-01}$  & 2.58$\times$10$^{+01}$ \\ 
\end{tabular}
\end{ruledtabular}
\label{tab:Ne22_alphagamma_rate_no_TAMU}
\end{table*}

\begin{table*}
\caption{Recommended $^{22}$Ne($\alpha,n$)$^{25}$Mg rate from the present evaluation incorporating the TAMU results. For a description of the table content, refer to the caption of Table \ref{tab:Ne22_alphagamma_rate}. At temperatures below 0.03 GK the rate is $<10^{-99}$ and is omitted from the table. At temperatures above 1.33 GK, the rates should be taken from Hauser-Feshbach models - see Ref. \cite{PhysRevC.85.065809}.}
\begin{ruledtabular}
\begin{tabular}{c c c c c c c }
T [GK] & Low rate & Median rate & High rate  & Log-normal $\mu$  & Log-normal $\sigma$ & Anderson-Darling statistic \\
\hline \\
0.030 & 6.98$\times$10$^{-88}$ & 4.57$\times$10$^{-87}$ &
      1.82$\times$10$^{-86}$ &  -1.990$\times$10$^{+02}$ &
       1.60$\times$10$^{+00}$  & 4.18$\times$10$^{+01}$ \\ 
0.040 & 1.67$\times$10$^{-67}$ & 1.38$\times$10$^{-66}$ &
      5.65$\times$10$^{-66}$ &  -1.519$\times$10$^{+02}$ &
       1.72$\times$10$^{+00}$  & 5.50$\times$10$^{+01}$ \\ 
0.050 & 3.11$\times$10$^{-55}$ & 2.89$\times$10$^{-54}$ &
      1.19$\times$10$^{-53}$ &  -1.236$\times$10$^{+02}$ &
       1.77$\times$10$^{+00}$  & 6.28$\times$10$^{+01}$ \\ 
0.060 & 4.86$\times$10$^{-47}$ & 4.69$\times$10$^{-46}$ &
      1.94$\times$10$^{-45}$ &  -1.047$\times$10$^{+02}$ &
       1.79$\times$10$^{+00}$  & 6.49$\times$10$^{+01}$ \\ 
0.070 & 3.73$\times$10$^{-41}$ & 3.38$\times$10$^{-40}$ &
      1.39$\times$10$^{-39}$ &  -9.121$\times$10$^{+01}$ &
       1.76$\times$10$^{+00}$  & 6.13$\times$10$^{+01}$ \\ 
0.080 & 1.04$\times$10$^{-36}$ & 8.27$\times$10$^{-36}$ &
      3.38$\times$10$^{-35}$ &  -8.107$\times$10$^{+01}$ &
       1.70$\times$10$^{+00}$  & 5.31$\times$10$^{+01}$ \\ 
0.090 & 3.13$\times$10$^{-33}$ & 2.13$\times$10$^{-32}$ &
      8.56$\times$10$^{-32}$ &  -7.317$\times$10$^{+01}$ &
       1.62$\times$10$^{+00}$  & 4.35$\times$10$^{+01}$ \\ 
0.100 & 1.96$\times$10$^{-30}$ & 1.16$\times$10$^{-29}$ &
      4.50$\times$10$^{-29}$ &  -6.683$\times$10$^{+01}$ &
       1.54$\times$10$^{+00}$  & 3.52$\times$10$^{+01}$ \\ 
0.110 & 3.97$\times$10$^{-28}$ & 2.01$\times$10$^{-27}$ &
      7.48$\times$10$^{-27}$ &  -6.164$\times$10$^{+01}$ &
       1.45$\times$10$^{+00}$  & 2.92$\times$10$^{+01}$ \\ 
0.120 & 3.35$\times$10$^{-26}$ & 1.51$\times$10$^{-25}$ &
      5.32$\times$10$^{-25}$ &  -5.730$\times$10$^{+01}$ &
       1.37$\times$10$^{+00}$  & 2.56$\times$10$^{+01}$ \\ 
0.130 & 1.43$\times$10$^{-24}$ & 5.97$\times$10$^{-24}$ &
      1.95$\times$10$^{-23}$ &  -5.361$\times$10$^{+01}$ &
       1.30$\times$10$^{+00}$  & 2.37$\times$10$^{+01}$ \\ 
0.140 & 3.65$\times$10$^{-23}$ & 1.42$\times$10$^{-22}$ &
      4.24$\times$10$^{-22}$ &  -5.044$\times$10$^{+01}$ &
       1.22$\times$10$^{+00}$  & 2.19$\times$10$^{+01}$ \\ 
0.150 & 6.24$\times$10$^{-22}$ & 2.24$\times$10$^{-21}$ &
      6.19$\times$10$^{-21}$ &  -4.767$\times$10$^{+01}$ &
       1.13$\times$10$^{+00}$  & 1.91$\times$10$^{+01}$ \\ 
0.160 & 7.87$\times$10$^{-21}$ & 2.54$\times$10$^{-20}$ &
      6.58$\times$10$^{-20}$ &  -4.521$\times$10$^{+01}$ &
       1.02$\times$10$^{+00}$  & 1.51$\times$10$^{+01}$ \\ 
0.180 & 6.91$\times$10$^{-19}$ & 1.63$\times$10$^{-18}$ &
      3.61$\times$10$^{-18}$ &  -4.098$\times$10$^{+01}$ &
       7.82$\times$10$^{-01}$  & 8.65$\times$10$^{+00}$ \\ 
0.200 & 3.27$\times$10$^{-17}$ & 5.61$\times$10$^{-17}$ &
      1.03$\times$10$^{-16}$ &  -3.740$\times$10$^{+01}$ &
       5.43$\times$10$^{-01}$  & 9.32$\times$10$^{+00}$ \\ 
0.250 & 5.27$\times$10$^{-14}$ & 6.80$\times$10$^{-14}$ &
      8.70$\times$10$^{-14}$ &  -3.032$\times$10$^{+01}$ &
       2.52$\times$10$^{-01}$  & 2.42$\times$10$^{-01}$ \\ 
0.300 & 8.33$\times$10$^{-12}$ & 1.05$\times$10$^{-11}$ &
      1.30$\times$10$^{-11}$ &  -2.528$\times$10$^{+01}$ &
       2.26$\times$10$^{-01}$  & 4.96$\times$10$^{-01}$ \\ 
0.350 & 3.31$\times$10$^{-10}$ & 4.12$\times$10$^{-10}$ &
      5.09$\times$10$^{-10}$ &  -2.161$\times$10$^{+01}$ &
       2.18$\times$10$^{-01}$  & 4.74$\times$10$^{-01}$ \\ 
0.400 & 5.60$\times$10$^{-09}$ & 6.83$\times$10$^{-09}$ &
      8.28$\times$10$^{-09}$ &  -1.880$\times$10$^{+01}$ &
       2.00$\times$10$^{-01}$  & 7.07$\times$10$^{-01}$ \\ 
0.450 & 5.63$\times$10$^{-08}$ & 6.64$\times$10$^{-08}$ &
      7.86$\times$10$^{-08}$ &  -1.652$\times$10$^{+01}$ &
       1.70$\times$10$^{-01}$  & 1.48$\times$10$^{+00}$ \\ 
0.500 & 4.21$\times$10$^{-07}$ & 4.81$\times$10$^{-07}$ &
      5.52$\times$10$^{-07}$ &  -1.454$\times$10$^{+01}$ &
       1.37$\times$10$^{-01}$  & 1.76$\times$10$^{+00}$ \\ 
0.600 & 1.39$\times$10$^{-05}$ & 1.59$\times$10$^{-05}$ &
      1.87$\times$10$^{-05}$ &  -1.104$\times$10$^{+01}$ &
       1.50$\times$10$^{-01}$  & 9.31$\times$10$^{+00}$ \\ 
0.700 & 2.60$\times$10$^{-04}$ & 3.09$\times$10$^{-04}$ &
      3.76$\times$10$^{-04}$ &  -8.071$\times$10$^{+00}$ &
       1.86$\times$10$^{-01}$  & 9.47$\times$10$^{+00}$ \\ 
0.800 & 2.92$\times$10$^{-03}$ & 3.48$\times$10$^{-03}$ &
      4.25$\times$10$^{-03}$ &  -5.649$\times$10$^{+00}$ &
       1.86$\times$10$^{-01}$  & 9.15$\times$10$^{+00}$ \\ 
0.900 & 2.16$\times$10$^{-02}$ & 2.54$\times$10$^{-02}$ &
      3.05$\times$10$^{-02}$ &  -3.664$\times$10$^{+00}$ &
       1.70$\times$10$^{-01}$  & 9.47$\times$10$^{+00}$ \\ 
1.000 & 1.15$\times$10$^{-01}$ & 1.33$\times$10$^{-01}$ &
      1.57$\times$10$^{-01}$ &  -2.008$\times$10$^{+00}$ &
       1.50$\times$10$^{-01}$  & 7.96$\times$10$^{+00}$ \\ 
1.250 & 2.72$\times$10$^{+00}$ & 3.06$\times$10$^{+00}$ &
      3.46$\times$10$^{+00}$ &  1.122$\times$10$^{+00}$ &
       1.20$\times$10$^{-01}$  & 1.19$\times$10$^{+00}$ \\ 
\end{tabular}
\end{ruledtabular}
\label{tab:Ne22_alphan_rate}
\end{table*}

\begin{table*}
\caption{The $^{22}$Ne($\alpha,n$)$^{25}$Mg reaction rate evaluated without the incorporation of the TAMU results. For a description of the table contents, refer to the caption of Table \ref{tab:Ne22_alphagamma_rate}. At temperatures below 0.03 GK the rate is $<10^{-99}$ and is omitted from the table. At temperatures above 1.33 GK, the rates should be taken from Hauser-Feshbach models - see Ref. \cite{PhysRevC.85.065809}.}
\begin{ruledtabular}
\begin{tabular}{c c c c c c c }
T [GK] & Low rate & Median rate & High rate  & Log-normal $\mu$  & Log-normal $\sigma$ & Anderson-Darling statistic \\
\hline \\
0.030 & 7.71$\times$10$^{-88}$ & 4.87$\times$10$^{-87}$ &
      2.01$\times$10$^{-86}$ &  -1.989$\times$10$^{+02}$ &
       1.62$\times$10$^{+00}$  & 2.84$\times$10$^{+01}$ \\ 
0.040 & 1.90$\times$10$^{-67}$ & 1.47$\times$10$^{-66}$ &
      6.28$\times$10$^{-66}$ &  -1.518$\times$10$^{+02}$ &
       1.74$\times$10$^{+00}$  & 4.00$\times$10$^{+01}$ \\ 
0.050 & 3.62$\times$10$^{-55}$ & 3.05$\times$10$^{-54}$ &
      1.32$\times$10$^{-53}$ &  -1.235$\times$10$^{+02}$ &
       1.79$\times$10$^{+00}$  & 4.71$\times$10$^{+01}$ \\ 
0.060 & 5.76$\times$10$^{-47}$ & 4.96$\times$10$^{-46}$ &
      2.15$\times$10$^{-45}$ &  -1.046$\times$10$^{+02}$ &
       1.81$\times$10$^{+00}$  & 4.88$\times$10$^{+01}$ \\ 
0.070 & 4.29$\times$10$^{-41}$ & 3.58$\times$10$^{-40}$ &
      1.55$\times$10$^{-39}$ &  -9.110$\times$10$^{+01}$ &
       1.77$\times$10$^{+00}$  & 4.44$\times$10$^{+01}$ \\ 
0.080 & 1.16$\times$10$^{-36}$ & 8.80$\times$10$^{-36}$ &
      3.76$\times$10$^{-35}$ &  -8.095$\times$10$^{+01}$ &
       1.69$\times$10$^{+00}$  & 3.56$\times$10$^{+01}$ \\ 
0.090 & 3.54$\times$10$^{-33}$ & 2.29$\times$10$^{-32}$ &
      9.52$\times$10$^{-32}$ &  -7.304$\times$10$^{+01}$ &
       1.60$\times$10$^{+00}$  & 2.59$\times$10$^{+01}$ \\ 
0.100 & 2.34$\times$10$^{-30}$ & 1.25$\times$10$^{-29}$ &
      4.98$\times$10$^{-29}$ &  -6.669$\times$10$^{+01}$ &
       1.50$\times$10$^{+00}$  & 1.79$\times$10$^{+01}$ \\ 
0.110 & 4.78$\times$10$^{-28}$ & 2.21$\times$10$^{-27}$ &
      8.33$\times$10$^{-27}$ &  -6.148$\times$10$^{+01}$ &
       1.40$\times$10$^{+00}$  & 1.23$\times$10$^{+01}$ \\ 
0.120 & 4.22$\times$10$^{-26}$ & 1.68$\times$10$^{-25}$ &
      5.92$\times$10$^{-25}$ &  -5.713$\times$10$^{+01}$ &
       1.31$\times$10$^{+00}$  & 8.87$\times$10$^{+00}$ \\ 
0.130 & 1.89$\times$10$^{-24}$ & 6.73$\times$10$^{-24}$ &
      2.21$\times$10$^{-23}$ &  -5.342$\times$10$^{+01}$ &
       1.22$\times$10$^{+00}$  & 6.64$\times$10$^{+00}$ \\ 
0.140 & 5.06$\times$10$^{-23}$ & 1.62$\times$10$^{-22}$ &
      4.89$\times$10$^{-22}$ &  -5.022$\times$10$^{+01}$ &
       1.12$\times$10$^{+00}$  & 4.77$\times$10$^{+00}$ \\ 
0.150 & 9.10$\times$10$^{-22}$ & 2.62$\times$10$^{-21}$ &
      7.27$\times$10$^{-21}$ &  -4.741$\times$10$^{+01}$ &
       1.01$\times$10$^{+00}$  & 3.09$\times$10$^{+00}$ \\ 
0.160 & 1.25$\times$10$^{-20}$ & 3.13$\times$10$^{-20}$ &
      7.94$\times$10$^{-20}$ &  -4.490$\times$10$^{+01}$ &
       8.88$\times$10$^{-01}$  & 3.23$\times$10$^{+00}$ \\ 
0.180 & 1.37$\times$10$^{-18}$ & 2.42$\times$10$^{-18}$ &
      4.83$\times$10$^{-18}$ &  -4.051$\times$10$^{+01}$ &
       6.09$\times$10$^{-01}$  & 1.47$\times$10$^{+01}$ \\ 
0.200 & 7.79$\times$10$^{-17}$ & 1.08$\times$10$^{-16}$ &
      1.65$\times$10$^{-16}$ &  -3.672$\times$10$^{+01}$ &
       3.80$\times$10$^{-01}$  & 1.81$\times$10$^{+01}$ \\ 
0.250 & 1.49$\times$10$^{-13}$ & 1.81$\times$10$^{-13}$ &
      2.24$\times$10$^{-13}$ &  -2.934$\times$10$^{+01}$ &
       2.08$\times$10$^{-01}$  & 3.87$\times$10$^{-01}$ \\ 
0.300 & 2.48$\times$10$^{-11}$ & 3.03$\times$10$^{-11}$ &
      3.72$\times$10$^{-11}$ &  -2.422$\times$10$^{+01}$ &
       2.04$\times$10$^{-01}$  & 7.57$\times$10$^{-01}$ \\ 
0.350 & 9.76$\times$10$^{-10}$ & 1.19$\times$10$^{-09}$ &
      1.46$\times$10$^{-09}$ &  -2.055$\times$10$^{+01}$ &
       2.02$\times$10$^{-01}$  & 7.98$\times$10$^{-01}$ \\ 
0.400 & 1.54$\times$10$^{-08}$ & 1.87$\times$10$^{-08}$ &
      2.27$\times$10$^{-08}$ &  -1.780$\times$10$^{+01}$ &
       1.96$\times$10$^{-01}$  & 9.26$\times$10$^{-01}$ \\ 
0.450 & 1.37$\times$10$^{-07}$ & 1.63$\times$10$^{-07}$ &
      1.96$\times$10$^{-07}$ &  -1.563$\times$10$^{+01}$ &
       1.83$\times$10$^{-01}$  & 1.27$\times$10$^{+00}$ \\ 
0.500 & 8.52$\times$10$^{-07}$ & 9.89$\times$10$^{-07}$ &
      1.17$\times$10$^{-06}$ &  -1.382$\times$10$^{+01}$ &
       1.61$\times$10$^{-01}$  & 1.77$\times$10$^{+00}$ \\ 
0.600 & 1.94$\times$10$^{-05}$ & 2.21$\times$10$^{-05}$ &
      2.53$\times$10$^{-05}$ &  -1.072$\times$10$^{+01}$ &
       1.35$\times$10$^{-01}$  & 2.20$\times$10$^{+00}$ \\ 
0.700 & 2.94$\times$10$^{-04}$ & 3.43$\times$10$^{-04}$ &
      4.11$\times$10$^{-04}$ &  -7.965$\times$10$^{+00}$ &
       1.71$\times$10$^{-01}$  & 8.99$\times$10$^{+00}$ \\ 
0.800 & 3.04$\times$10$^{-03}$ & 3.59$\times$10$^{-03}$ &
      4.37$\times$10$^{-03}$ &  -5.614$\times$10$^{+00}$ &
       1.82$\times$10$^{-01}$  & 9.35$\times$10$^{+00}$ \\ 
0.900 & 2.19$\times$10$^{-02}$ & 2.56$\times$10$^{-02}$ &
      3.07$\times$10$^{-02}$ &  -3.651$\times$10$^{+00}$ &
       1.69$\times$10$^{-01}$  & 9.09$\times$10$^{+00}$ \\ 
1.000 & 1.16$\times$10$^{-01}$ & 1.34$\times$10$^{-01}$ &
      1.57$\times$10$^{-01}$ &  -2.003$\times$10$^{+00}$ &
       1.51$\times$10$^{-01}$  & 7.67$\times$10$^{+00}$ \\ 
1.250 & 2.72$\times$10$^{+00}$ & 3.05$\times$10$^{+00}$ &
      3.45$\times$10$^{+00}$ &  1.121$\times$10$^{+00}$ &
       1.20$\times$10$^{-01}$  & 1.84$\times$10$^{+00}$ \\ 
\end{tabular}
\end{ruledtabular}
\label{tab:Ne22_alphan_rate_no_TAMU}
\end{table*}

\begin{figure}[htbp]
    \centering
   \subfigure{\includegraphics[angle=0,width=0.95\columnwidth]{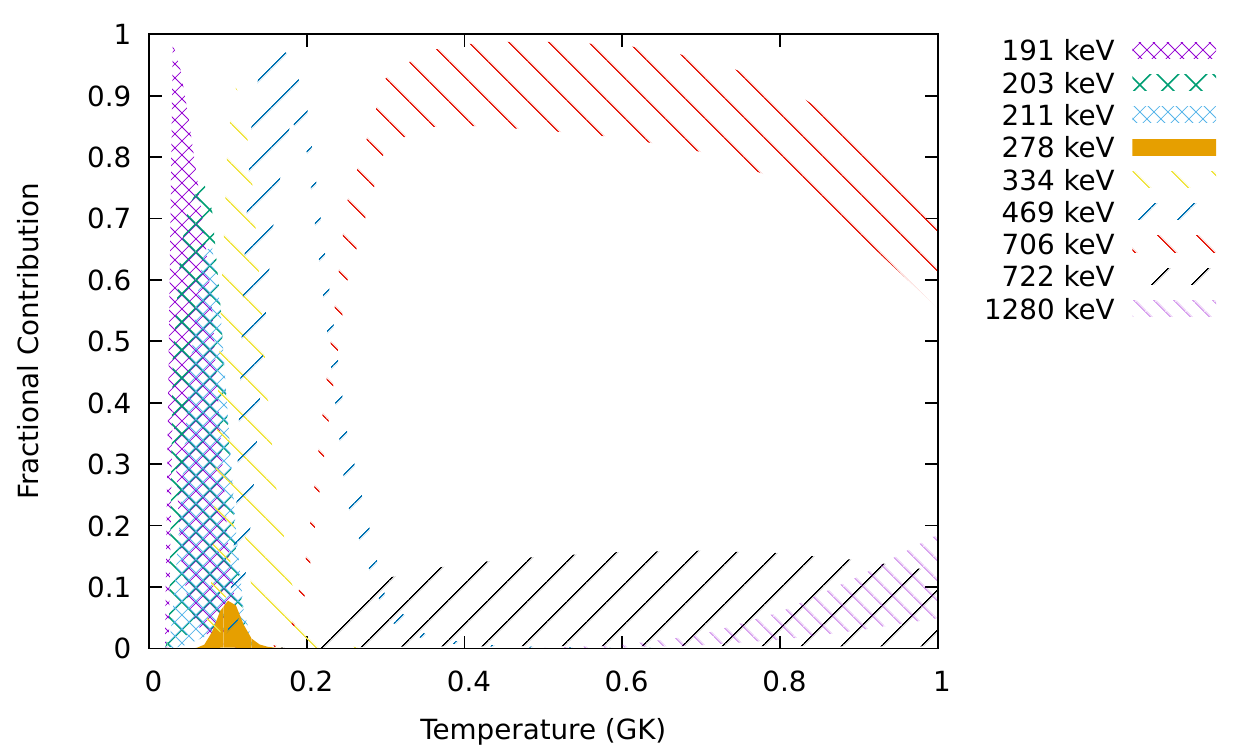}}
   \subfigure{\includegraphics[angle=0,width=0.95\columnwidth]{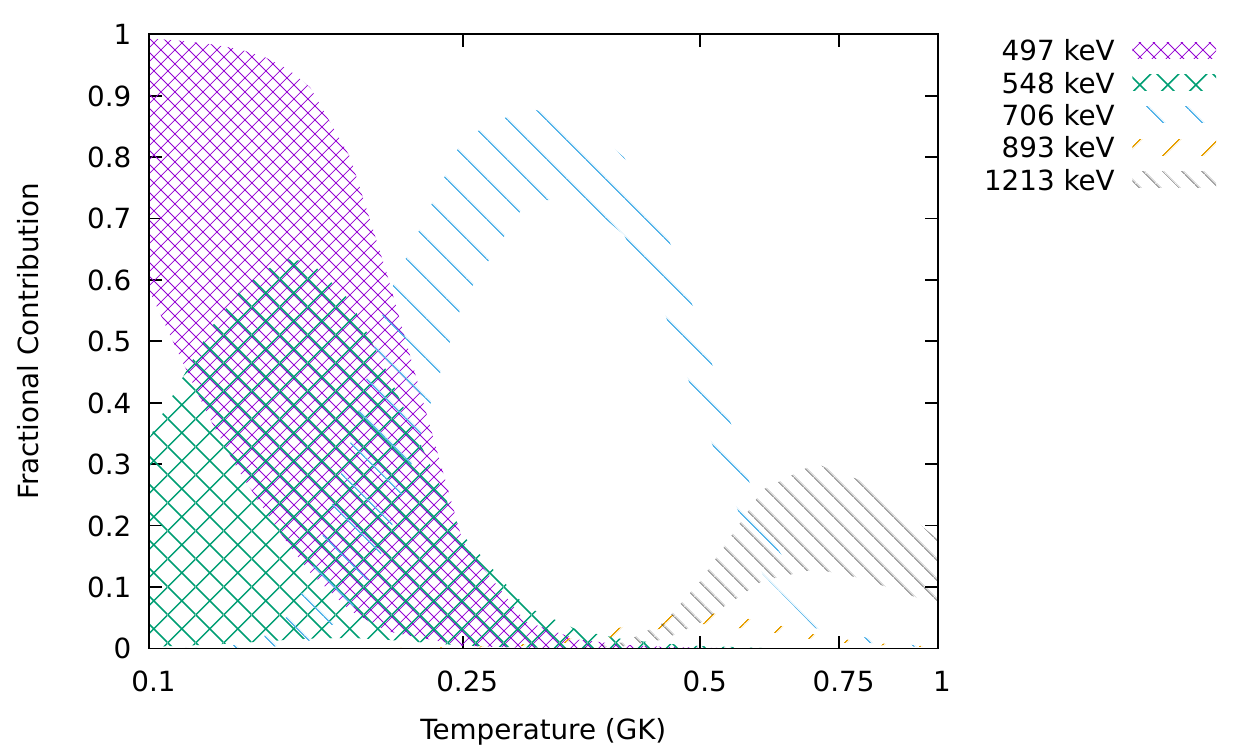}}
    \caption{Fractional contributions of selected resonances to the (top) $^{22}$Ne($\alpha,\gamma$)$^{26}$Mg and (bottom) $^{22}$Ne($\alpha,n$)$^{25}$Mg reaction rates. These fractional contributions are for the recommended reaction rates, which incorporate the Texas A\&M results. The shaded region gives the 68\% coverage limit for the contribution of each resonance. Note that only the most significant resonances are included in the figure; the sum of the contributions may not reach 100\% due to contributions from omitted resonances.}
    \label{fig:FractionalRateContributions}
\end{figure}

Figure \ref{fig:FractionalRateContributions} shows the fractional contributions of individual resonances to the recommended reaction rates for the $^{22}$Ne($\alpha,\gamma$)$^{26}$Mg and $^{22}$Ne($\alpha,n$)$^{25}$Mg reactions. Above 0.2 GK, both reaction rates are dominated by the $E_r = 706$-keV resonance and other higher-lying directly measured resonances. In this region, we expect little deviations from the $^{22}$Ne($\alpha,\gamma$)$^{26}$Mg rate of Longland {\it et al.} \cite{PhysRevC.85.065809} as both evaluations are based on the similar nuclear data, with only the updated (and consistent) measurement of Hunt {\it et al.} \cite{PhysRevC.99.045804}. The updated, reduced, $^{22}$Ne($\alpha,n$)$^{25}$Mg resonance strength for this state is expected to cause a concomitant reduction in the reaction rate above 0.2 GK.

Below 0.2 GK, lower-energy resonances in $^{26}$Mg are predicted to dominate the reaction rate. Some of these resonances have been observed in the $^{25}$Mg$+n$ reactions of Massimi {\it et al.} \cite{Massimi20171,PhysRevC.85.044615} and neutron and $\gamma$-ray partial widths have been determined. No $\alpha$-particle partial widths have been directly measured for the resonances below $E_{cm} = 706$ keV but an estimate of the partial width for the $E_r = 469$-keV resonance is available from the transfer data of Jayatissa {\it et al.} \cite{JAYATISSA2020135267}. Evaluations of the reaction rates are based on the potential upper limits for the other resonances resulting in larger uncertainties for the rates.

The states below the $E_r = 706$-keV resonance which have been identified as potentially strongly contributing  to the $^{22}$Ne($\alpha,\gamma$)$^{26}$Mg reaction rate are those at $E_r = 191$, $203$, $211$, $278$, $334$, and $469$ keV ($E_x = 10.806$, $10.818$, $10.826$, $10.893$, $10.949$, and $11.084$ MeV. The states at around $E_x = 11.17$ MeV are predicted to have negligible impact due to the weak cross section observed in the sub-barrier transfer reaction of Jayatissa {\it et al.} \cite{JAYATISSA2020135267}.

The states below the $E_r = 706$-keV resonance which have been identified as potentially strongly contributing to the $^{22}$Ne($\alpha,n$)$^{25}$Mg reaction rate are those at $E_r = 497$ and $548$ keV ($E_x = 11.112$ and $11.163$ MeV).

\section{Comparison to previous reaction-rate evaluations}
\label{RateComparisons}

We now compare the presently computed reaction rates with those of Longland {\it et al.} \cite{PhysRevC.85.065809}, Talwar {\it et al.} \cite{PhysRevC.93.055803} and Massimi {\it et al.} \cite{Massimi20171}. The Longland \cite{PhysRevC.85.065809} and Talwar \cite{PhysRevC.93.055803} evaluations both use the \textsc{RatesMC} code whereas the Massimi evaluation was differently performed and reported upper limits rather than median reaction rates.

\begin{figure}[htbp]
    \centering
    \subfigure{
    \includegraphics[angle=0, width=0.95\columnwidth]{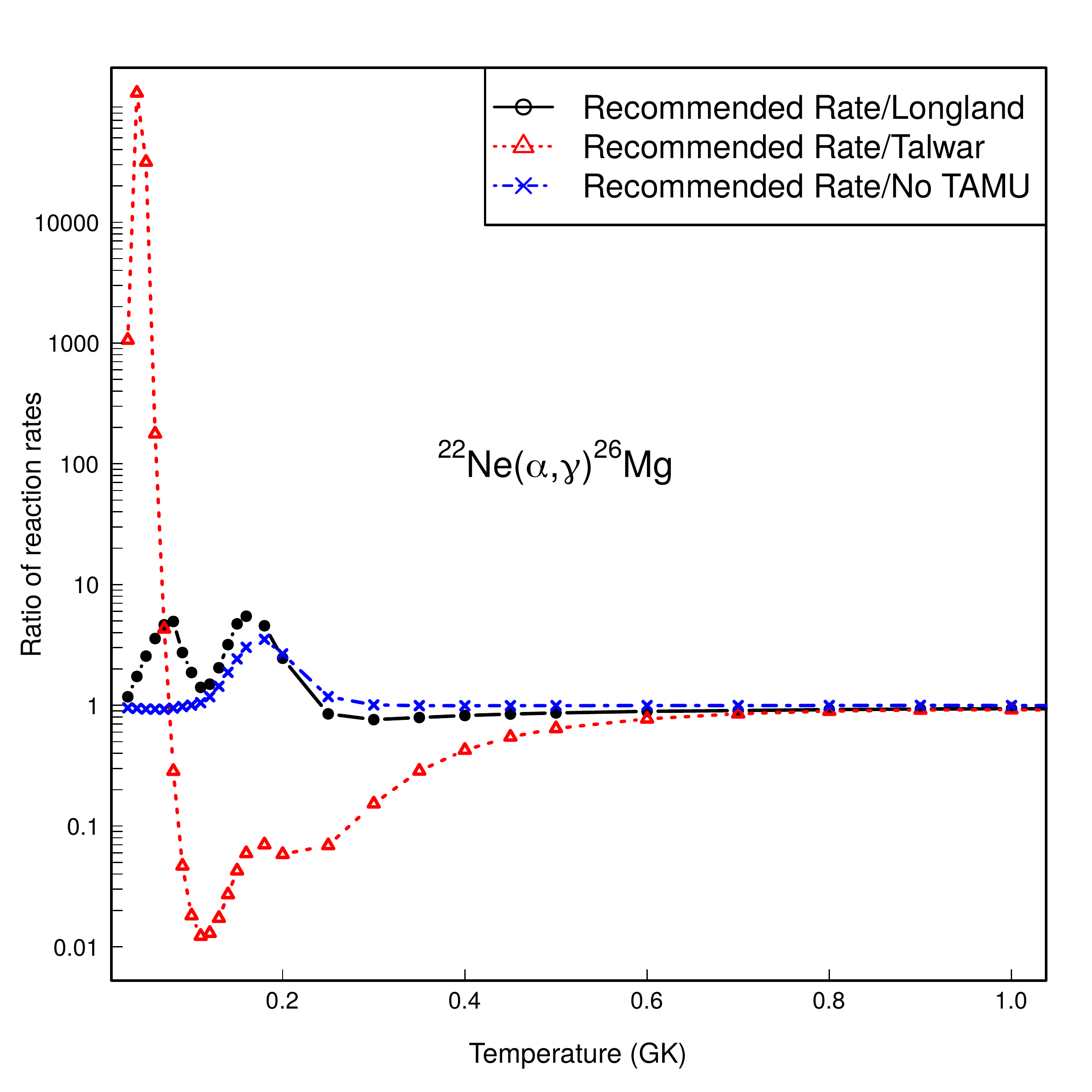}}
    \subfigure{
    \includegraphics[angle=0, width=0.95\columnwidth]{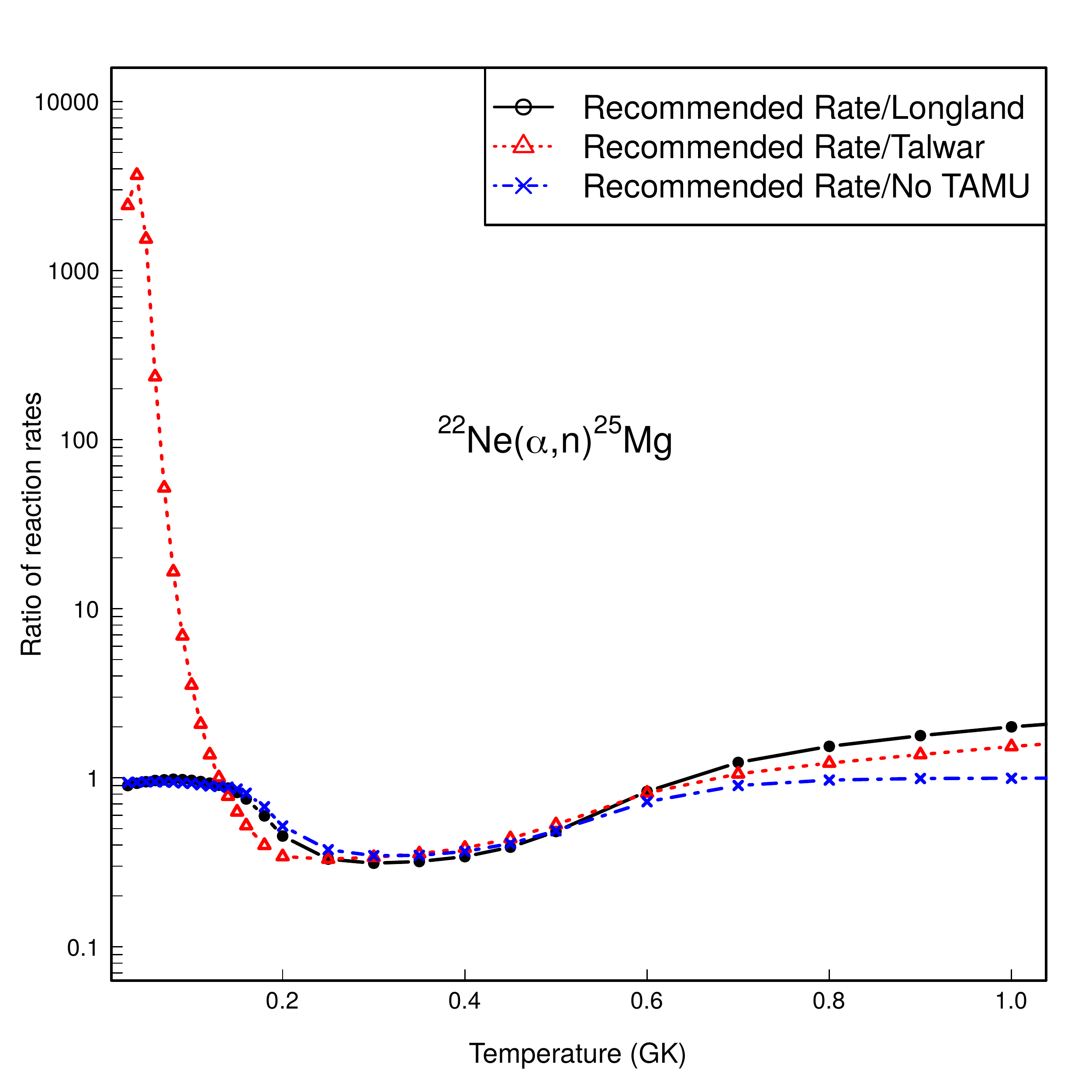}}
    \caption{Ratio of the (top) $^{22}$Ne($\alpha,\gamma$)$^{26}$Mg and (bottom) $^{22}$Ne($\alpha,n$)$^{25}$Mg recommended reaction rates of the present calculation to the rates from (black, circles, solid line) Longland {\it et al.} \cite{PhysRevC.85.065809}, (red, triangles, broken line) Talwar {\it et al.} \cite{PhysRevC.93.055803} and (blue, crosses, broken line) the calculations performed for this measurement without using the Texas A\&M data. The solid points are the calculated ratios of median reaction rates; lines are drawn to guide the eye.}
    \label{fig:RatioToHistorics}
\end{figure}

The ratios of the $^{22}$Ne($\alpha,\gamma$)$^{26}$Mg and $^{22}$Ne($\alpha,n$)$^{25}$Mg reaction rates calculated in the present study and the reaction rates from Longland {\it et al.} \cite{PhysRevC.85.065809} and Talwar {\it et al.} \cite{PhysRevC.93.055803} are shown in Figure \ref{fig:RatioToHistorics}.

Above 0.2 GK, the $^{22}$Ne($\alpha,\gamma$)$^{26}$Mg and $^{22}$Ne($\alpha,n$)$^{25}$Mg reaction rates are dominated by resonances for which direct measurements exist, in particular by the $706$-keV resonance discussed in Section \ref{statesec12}. The consistency in the $^{22}$Ne($\alpha,\gamma$)$^{26}$Mg reaction rate (upper panel of Figure \ref{fig:RatioToHistorics}) above 0.2 GK is due to the fact that the $^{22}$Ne($\alpha,\gamma$)$^{26}$Mg resonance strength for the $E_r = 706$-keV resonance is unchanged between the different evaluations. There is a reduction in the recommended $^{22}$Ne($\alpha,n$)$^{25}$Mg reaction rate relative to the Longland and Talwar rates due to the results from the two transfer reactions \cite{OTA2020135256,JAYATISSA2020135267} carried out at Texas A\&M.. A small reduction in the $^{22}$Ne($\alpha,\gamma$)$^{26}$Mg reaction rate is observed due to the conclusion that the resonances observed in $^{22}$Ne($\alpha,\gamma$)$^{26}$Mg and $^{22}$Ne($\alpha,n$)$^{25}$Mg reactions are the same and the subsequent shift in the resonance energy. This reduction is within the uncertainty on the reaction rate computed by Longland {\it et al.} \cite{PhysRevC.85.065809}.

There is a small increase in the recommended $^{22}$Ne($\alpha,\gamma$)$^{26}$Mg reaction rate relative to the Longland rate below 0.1 GK. This is due to the additional contributions from the $J^\pi = 0^+$ state at $E_x = 10.818$ MeV from the $^{26}$Mg($\alpha,\alpha^\prime$)$^{26}$Mg data of Adsley {\it et al.} \cite{PhysRevC.96.055802}. 

The increase in the $^{22}$Ne($\alpha,\gamma$)$^{26}$Mg reaction rate between 0.1 and 0.2 GK seen when comparing the recommended rate to the rate computed without the Texas A\& M results and the Longland evaluation is due to there now being an estimated resonance strength rather than an upper limit for the the  $J^\pi = 2^+$ state at $E_x = 11.084$ MeV ($E_r = 469$ keV) \cite{OTA2020135256,JAYATISSA2020135267}.

The present $^{22}$Ne($\alpha,\gamma$)$^{26}$Mg reaction rate is much smaller than that of Talwar {\it et al.} \cite{PhysRevC.93.055803} between 0.1 and 0.4 GK due to the increased contribution from the $E_x = 11.171$-MeV state in the Talwar evaluation. As discussed in Section \ref{sec:11167keVState}, there is some uncertainty about the existence and properties of the states around $E_x = 11.17$ MeV; for this reason it has not been included in the present evaluation.

At the lowest temperatures, the significant increase in the reaction rate with respect to that of  Talwar {\it et al.} \cite{PhysRevC.93.055803} is due to the treatment of the low-lying resonances at $E_r = 191$ and $E_r = 211$ keV for which the limits on the spectroscopic factors have been relaxed. However, these resonances are at low energy and are unlikely to have any astrophysical impact.

At temperatures above 0.2 GK, Massimi {\it et al.} \cite{Massimi20171} calculated an upper limit for the $^{22}$Ne($\alpha,\gamma$)$^{26}$Mg reaction rate which is much smaller than the rate of Longland {\it et al.} \cite{PhysRevC.85.065809} and the current rate. This is because the methodology used to calculate the reaction rates is based on the ratio of the rates which are inferred from the ratios of the neutron and $\gamma$-ray partial widths for the observed resonances. As a consequence, the $^{22}$Ne($\alpha,\gamma$)$^{26}$Mg reaction rate is underestimated by Massimi {\it et al.} \cite{Massimi20171}.

\subsection{Reaction-rate ratio}

The neutron yield depends on the $^{22}$Ne abundance and on the competition between the $^{22}$Ne($\alpha,\gamma$)$^{26}$Mg and $^{22}$Ne($\alpha,n$)$^{25}$Mg reactions at various temperatures. Therefore, it is necessary to  evaluate both the $^{22}$Ne($\alpha,\gamma$)$^{26}$Mg and the $^{22}$Ne($\alpha,n$)$^{25}$Mg reaction rates, and their ratio. 

\begin{figure}[htbp]
    \centering
    \includegraphics[width=0.95\columnwidth]{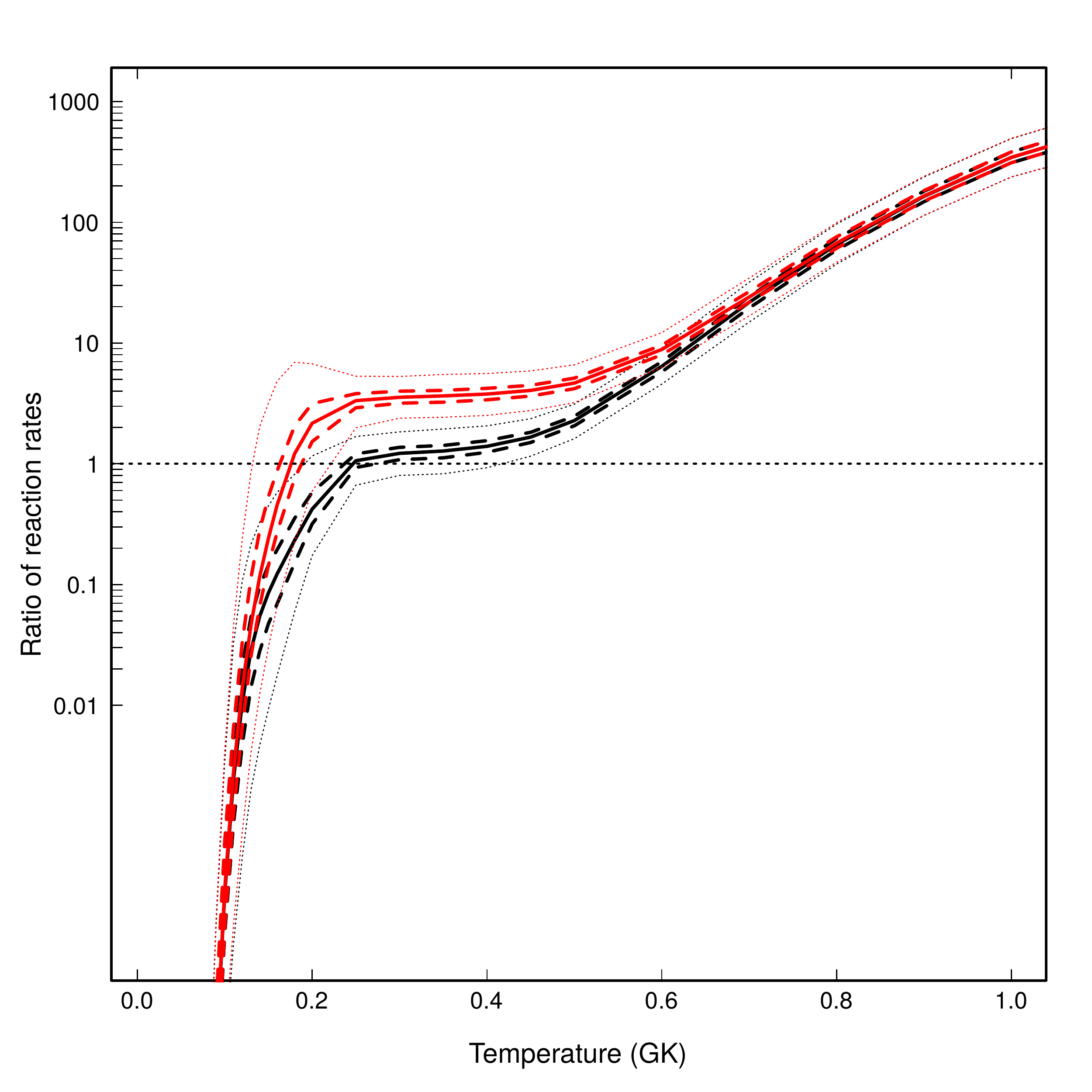}
    \caption{Ratio of $^{22}$Ne($\alpha,n$)$^{25}$Mg and $^{22}$Ne($\alpha,\gamma$)$^{26}$Mg reaction rates for the recommended rate (black solid line) and the rate computed without the Texas A\& M results (red solid line). Values above unity imply that neutron production is more likely than radiative capture. The broken (dotted) lines are the 68\% (95\%) confidence limits for the ratio of the recommended rates, the colours represent the same evaluations.}
    \label{fig:RRRs}
\end{figure}

The ratio of the two rates as a function of temperature is shown in Figure \ref{fig:RRRs}. At lower temperatures the $^{22}$Ne($\alpha,\gamma$)$^{26}$Mg reaction rate is expected to dominate as the $^{22}$Ne($\alpha,n$)$^{25}$Mg reaction is endothermic. As the temperature increases, the $^{22}$Ne($\alpha,n$)$^{25}$Mg reaction rate becomes stronger relative to the $^{22}$Ne($\alpha,\gamma$)$^{26}$Mg until it eventually exceeds the $^{22}$Ne($\alpha,\gamma$)$^{26}$Mg reaction rate at around 0.2 GK. The dominant states in the temperature region around 0.2 GK for $^{22}$Ne($\alpha,\gamma$)$^{26}$Mg reaction are those at $E_x = 10.949$ and $11.084$ MeV, while the dominant states for the $^{22}$Ne($\alpha,n$)$^{25}$Mg reaction are those at $E_x = 11.112$ and $11.163$ MeV. It is likely that the strengths of these resonances determine at which temperature the $^{22}$Ne($\alpha,n$)$^{25}$Mg reaction becomes stronger than the $^{22}$Ne($\alpha,\gamma$)$^{26}$Mg one.

The Monte-Carlo calculations performed in this evaluation and in the evaluations of e.g. Refs. \cite{PhysRevC.85.065809,PhysRevC.93.055803} are independent calculations of the $^{22}$Ne($\alpha,\gamma$)$^{26}$Mg and $^{22}$Ne($\alpha,n$)$^{25}$Mg reaction rates. However, the properties of resonances in the $^{22}$Ne($\alpha,\gamma$)$^{26}$Mg and $^{22}$Ne($\alpha,n$)$^{25}$Mg reactions are correlated which can lead to overestimation of the uncertainties in the ratio of the reaction rates. This is not a problem for the reaction rate calculated without the inclusion of the Texas A\&M results as all of these measurements are independent.

Combining the $\Gamma_n/\Gamma_\gamma$ ratio determined from the measurement of Ota {\it et al.} \cite{OTA2020135256} with the weighted average of the the direct $^{22}$Ne($\alpha,\gamma$)$^{26}$Mg measurements means that the $^{22}$Ne($\alpha,\gamma$)$^{26}$Mg and $^{22}$Ne($\alpha,n$)$^{25}$Mg resonance strengths are no longer independent.
However, in the present case the uncertainty is dominated by the uncertainty in the ratio of $\Gamma_n$ and $\Gamma_\gamma$ (23\%) not the uncertainty in the $^{22}$Ne($\alpha,\gamma$)$^{26}$Mg resonance strength (11\%). This means that the overestimation in the uncertainty from the correlations arising from the $^{22}$Ne($\alpha,\gamma$)$^{26}$Mg resonance strength are a small contribution to the total uncertainty and, therefore, that the uncertainty in the ratio is not significantly overestimated.


Bearing in mind the above arguments, the 68\% and 95\% confidence limits on the reaction-rate ratio have been computed from the Monte-Carlo samples for each rate at each temperature. The probability distribution function of the ratio of the reaction rates has been constructed from these samples and the confidence limits extracted.

\section{Priorities for future measurements\label{sec:SuggestedFutureExperiments}}

As discussed in the previous section, the comparison to previous reaction-rate estimations shows sizeable uncertainties at low temperatures and significant disagreement with the rates of Talwar {\it et al.} \cite{PhysRevC.93.055803}. The fair agreement between our results and those of Longland {\it et al.} \cite{PhysRevC.85.065809} for the $^{22}$Ne($\alpha,\gamma$)$^{26}$Mg reaction rate may also be attributable to the common method (i.e. the \textsc{RatesMC} Monte-Carlo code \cite{0067-0049-207-1-18}) adopted for the evaluation of the $\alpha+^{22}$Ne reaction rates. Although a full R-Matrix evaluation of the reaction rate would be preferable, the lack of experimental information on several $^{26}$Mg states as well as on the interference patterns between distant levels prevented this kind of approach.

As a consequence, future measurements should aim at determining the ($\alpha$,$n$), ($\alpha,\gamma$) and elastic scattering cross sections as well as characterising $^{26}$Mg levels in terms of spin and parity, eventually leading to an estimation of the reaction rate based on a full R-matrix analysis of available experimental data, including the effect of interference also due to sub-threshold resonances. Hereafter we report a high-priority request list based on the impact on the reaction rate and astrophysical implications.

\subsubsection{$^{22}$Ne($\alpha,\gamma$)$^{26}$Mg}
The $^{22}$Ne($\alpha,\gamma$)$^{26}$Mg reaction rate at  temperature of interest to the $s$-process is linked to a few resonances. These levels could be studied in inverse kinematics using a recoil separator, as well as with direct measurements. In particular:
\begin{itemize}
    \item The resonance at $E_x=11.321$~MeV ($E_r=706$~keV) is crucial at temperatures relevant to massive stars (see Figure~\ref{fig:FractionalRateContributions}). Two consistent direct measurements of the $^{22}$Ne($\alpha,\gamma$)$^{26}$Mg resonance strength are reported in the literature~\cite{Wolke1989,PhysRevC.99.045804}. Further direct measurements of this resonance with high beam intensities and improved resonance-energy resolution are required in order to determine any possible interference effects. The blister-proof neon-implanted targets of Hunt {\it et al.} \cite{PhysRevC.99.045804} would appear to be a productive approach.
    \item The tentative $E_r = 557$-keV resonance reported by Talwar {\it et al.} \cite{PhysRevC.93.055803} is a source of great uncertainty in the $^{22}$Ne($\alpha,\gamma$)$^{26}$Mg reaction rate, increasing it by up to a factor of 100 relative to our evaluation. Based on the estimate of Talwar {\it et al.} \cite{PhysRevC.93.055803}, should this resonance exist (which we consider an open question) then it should be within reach of direct measurements. Confirmation or refutation of this state with additional measurements of the $^{22}$Ne($^6$Li,$d$)$^{26}$Mg and $^{22}$Ne($^7$Li,$t$)$^{26}$Mg reactions would help to reduce the uncertainties in the rate considerably. The results of the Texas A\&M experiments appear to suggest that this state does not exist but higher-resolution measurements would be preferred in order to render a definitive verdict.
    \item An associated important piece of experimental data would be the confirmation of the $E_x = 11.44$-MeV state observed by Giesen {\it et al.} \cite{GIESEN199395} in $^{22}$Ne($^6$Li,$d$)$^{26}$Mg and Jaeger {\it et al.} \cite{PhysRevLett.87.202501} in $^{22}$Ne($\alpha,n$)$^{25}$Mg reactions. This state falls at the edge of the focal-plane acceptance of the experiment of Talwar {\it et al.} \cite{PhysRevC.93.055803}, being around 140 keV above the $E_x = 11.31$-MeV state.
    \item A re-examination of the focal-plane spectra of Refs. \cite{PhysRevC.93.055803,TalwarThesis} and re-measurements of the $^{22}$Ne($^6$Li,$d$)$^{26}$Mg reaction or a measurement of the $^{22}$Ne($^7$Li,$t$)$^{26}$Mg reaction may help to clarify which and how many $\alpha$-cluster states are populated in this region. As explained in Section \ref{sec:SynthesisOfTheData}, both Giesen {\it et al.} \cite{GIESEN199395} and Talwar {\it et al.} \cite{PhysRevC.93.055803} observed two $\alpha$-particle cluster states separated by $\sim150$ keV using the $^{22}$Ne($^6$Li,$d$)$^{26}$Mg reaction. This may be coincidental but one possible cause is that the energy calibration in one of the experiments is shifted. A valuable step which may take place without requiring any further experiments is the re-analysis of the data of Talwar {\it et al.} \cite{PhysRevC.93.055803}, confirming that the two states observed in Talwar {\it et al.} \cite{PhysRevC.93.055803} at $E_x = 11.17$ and $11.31$ MeV are not, in fact, the $E_x = 11.31$- and $11.44$-MeV levels observed by Giesen {\it et al.} \cite{GIESEN199395} and Jaeger {\it et al.} \cite{PhysRevLett.87.202501}.  \color{black} Re-analyses of the data of Giesen {\it et al.} \cite{GIESEN199395} and Jaeger {\it et al.} \cite{PhysRevLett.87.202501} would also be valuable but those data may no longer be available. This reconsideration of the results of the transfer reactions is supported by the non-observation of the $E_x = 11.17$-MeV state of Talwar {\it et al.} in the results from the Texas A\&M measurements of Jayatissa {\it et al.} \cite{JAYATISSA2020135267}. We note that the low-energy transfer reaction of Jayatissa {\it et al.} \cite{JAYATISSA2020135267} cannot populate high-spin states which may support the argument of Ota {\it et al.} \cite{OTA2020135256} that this state has high spin, as it was observed in that experiment. Additionally, the relative population of states of different spins in the various transfer measurements (Refs. \cite{GIESEN199395,PhysRevC.93.055803,JAYATISSA2020135267,OTA2020135256}) could also be considered as this would provide a practical method of testing the high-spin hypothesis of Ota {\it et al.} \cite{OTA2020135256}.
    \item Resonances at $E_x=10.949$ MeV ($E_r=334$~keV) and $E_x=11.084$ MeV ($E_r=469$~keV) dominate the $^{22}$Ne($\alpha,\gamma$)$^{26}$Mg reaction rate at temperature between 0.1 and 0.2~GK. No direct measurement of the $E_x = 10.949$-MeV state are available yet though some progress has been made \cite{10.1007/978-3-030-13876-9_50} but improved estimates with transfer reactions would also help to reduce uncertainties.
\end{itemize}

\subsubsection{$^{22}$Ne($\alpha$,$n$)$^{25}$Mg}
For an accurate determination of the $^{22}$Ne($\alpha$,$n$)$^{26}$Mg reaction rate and its uncertainty, several new experimental data are required. In particular: 
\begin{itemize}
\item The partial widths or resonance strengths of the $E_x=11.112$ MeV ($E_r=497$ keV), $E_x=11.163$ MeV ($E_r=548$ keV), $E_x=11.321$ MeV ($E_r=706$ keV) and $E_x=11.911$ MeV ($E_r=1297$ keV) levels in $^{26}$Mg. Estimates of some of these quantities are available from the decay branching measurements of Ota {\it et al.} \cite{OTA2020135256} but higher-resolution data obtained with a less selective reaction may help in quantifying the branching of these states.
\item As for the $^{22}$Ne($\alpha,\gamma$)$^{26}$Mg reaction, a firm spin and parity assignment of the potential cluster state observed in the vicinity of $E_x = 11.167$ MeV and a clear connection between this state and the states observed in the $n+^{25}$Mg experiments \cite{PhysRevC.85.044615,Massimi20171} are required. Confirmation of the $E_x = 11.44$-MeV state would also be useful in clarifying the properties of the levels in this region.
\item The interference pattern between distant levels, including sub-threshold states. The implanted targets of Hunt {\it et al.} \cite{PhysRevC.99.045804} provide an excellent opportunity to constrain the interference pattern.
\end{itemize}

\section{Impact for $s$-process calculations}\label{sec:impact}
\subsection{Main s-process from low-mass AGB star}

 We re-computed the $s$-process nucleosynthesis for stellar models of 2 and 3M$_\odot$
 adopting our new evaluations for ${^{22}}$Ne($\alpha$,$n$)${^{25}}$Mg and ${^{22}}$Ne($\alpha$,$\gamma$)${^{26}}$Mg reaction rates using two different codes:  the stellar models computed with the MESA code \cite{paxton:11} presented by Battino {\it et al.} \cite{battino:19} and the NEWTON code \cite{PALMERINI201821} presented by Trippella {\it et al.} \cite{trippella}. 
 In both cases, we include the results obtained adopting the reaction rates of Longland {\it et al.} \cite{PhysRevC.85.065809} and the rates evaluated in the present work with and without the Texas A\&M results.
 
\begin{figure}[htbp]
    \centering
    \includegraphics[width=0.95\columnwidth]{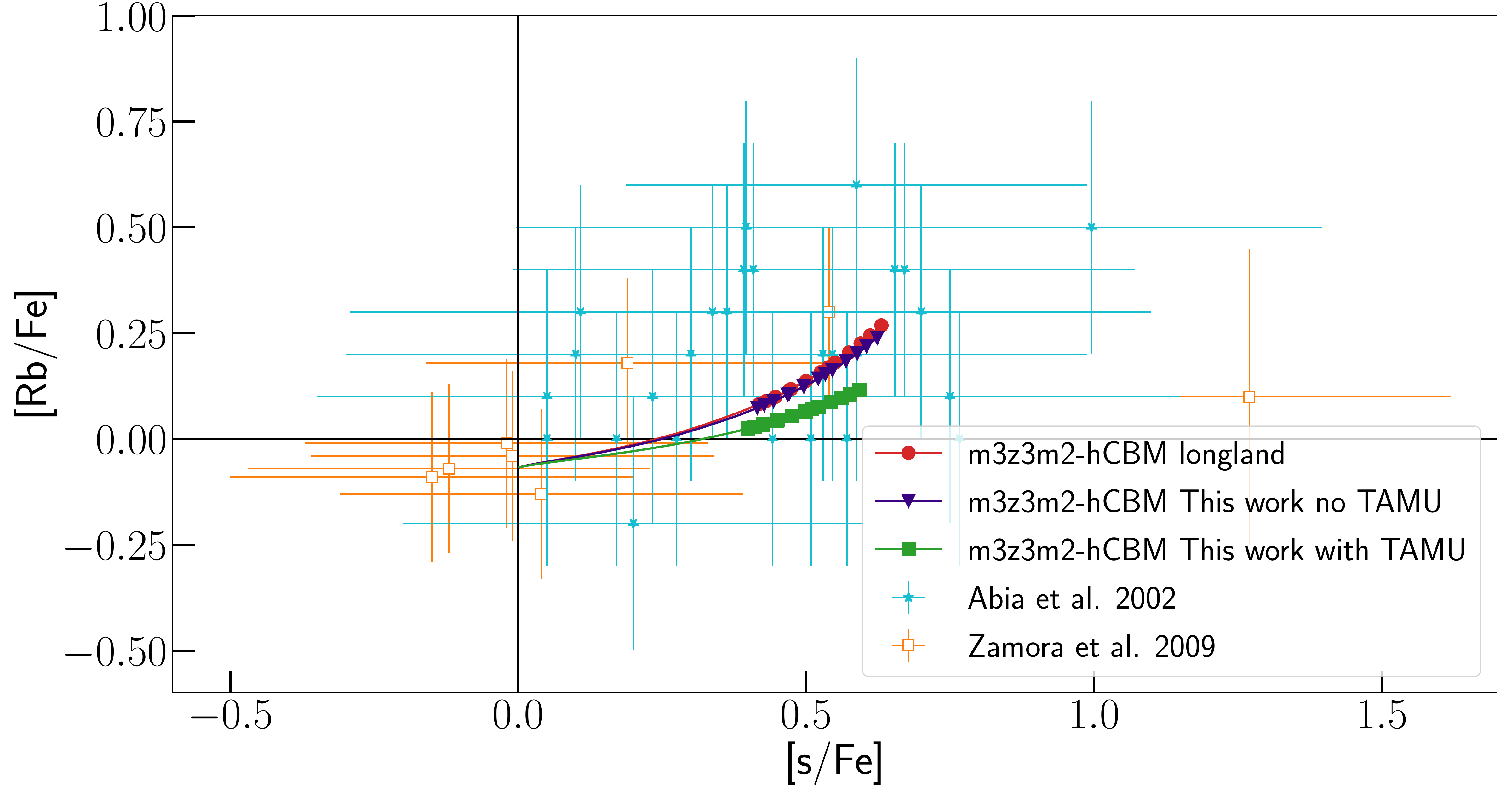}
    \caption{Rb abundances as a function of $s$-process abundances - results obtained using the MESA models of Battino {\it et al.} \cite{battino:19} compared to spectroscopy.}
    \label{fig:rb_cstars}
\end{figure}

\begin{figure}[htbp]
   \centering
    \subfigure{
     \includegraphics[width=0.95\columnwidth]{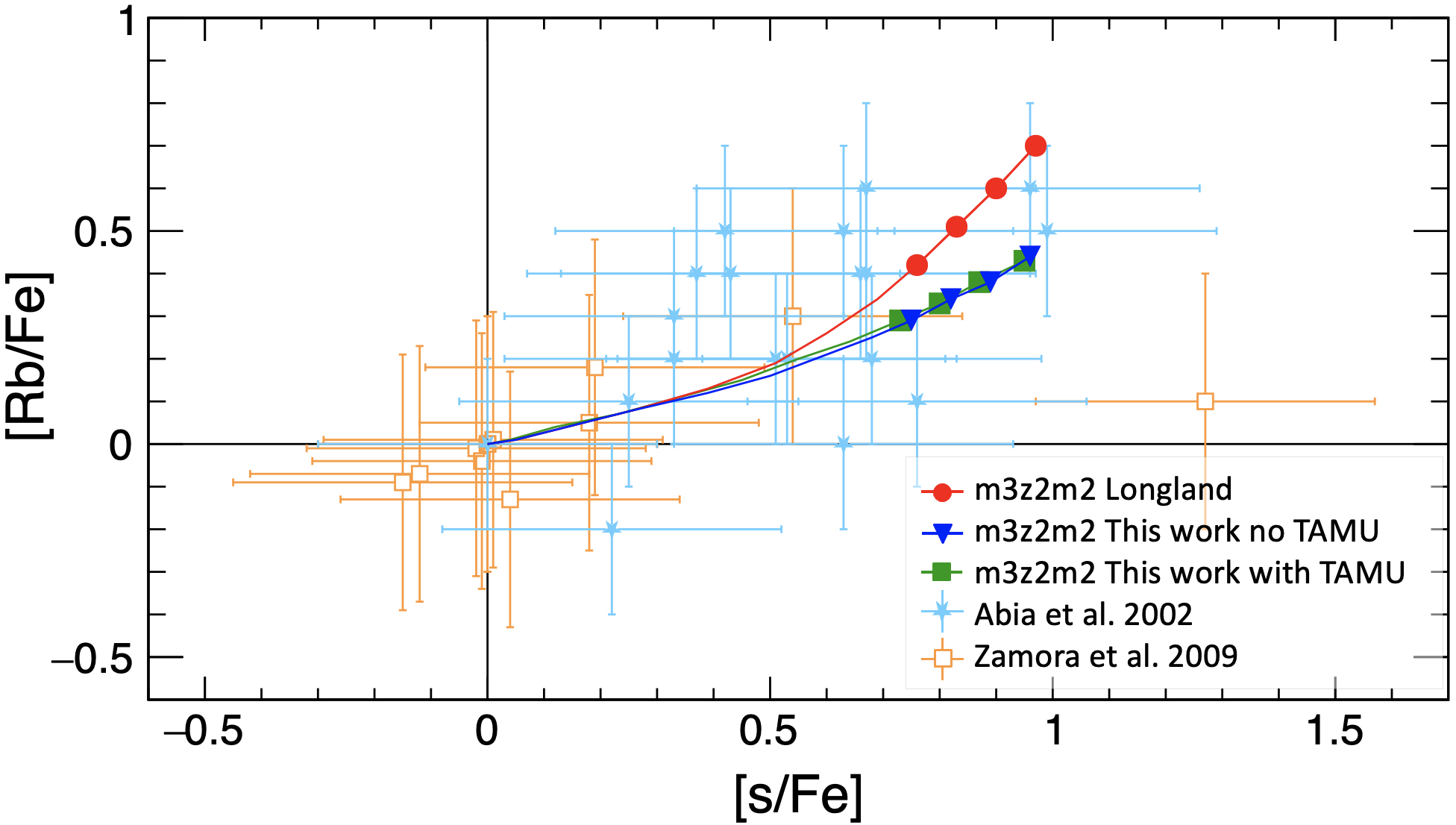}}
      \subfigure{
    \includegraphics[width=0.95\columnwidth]{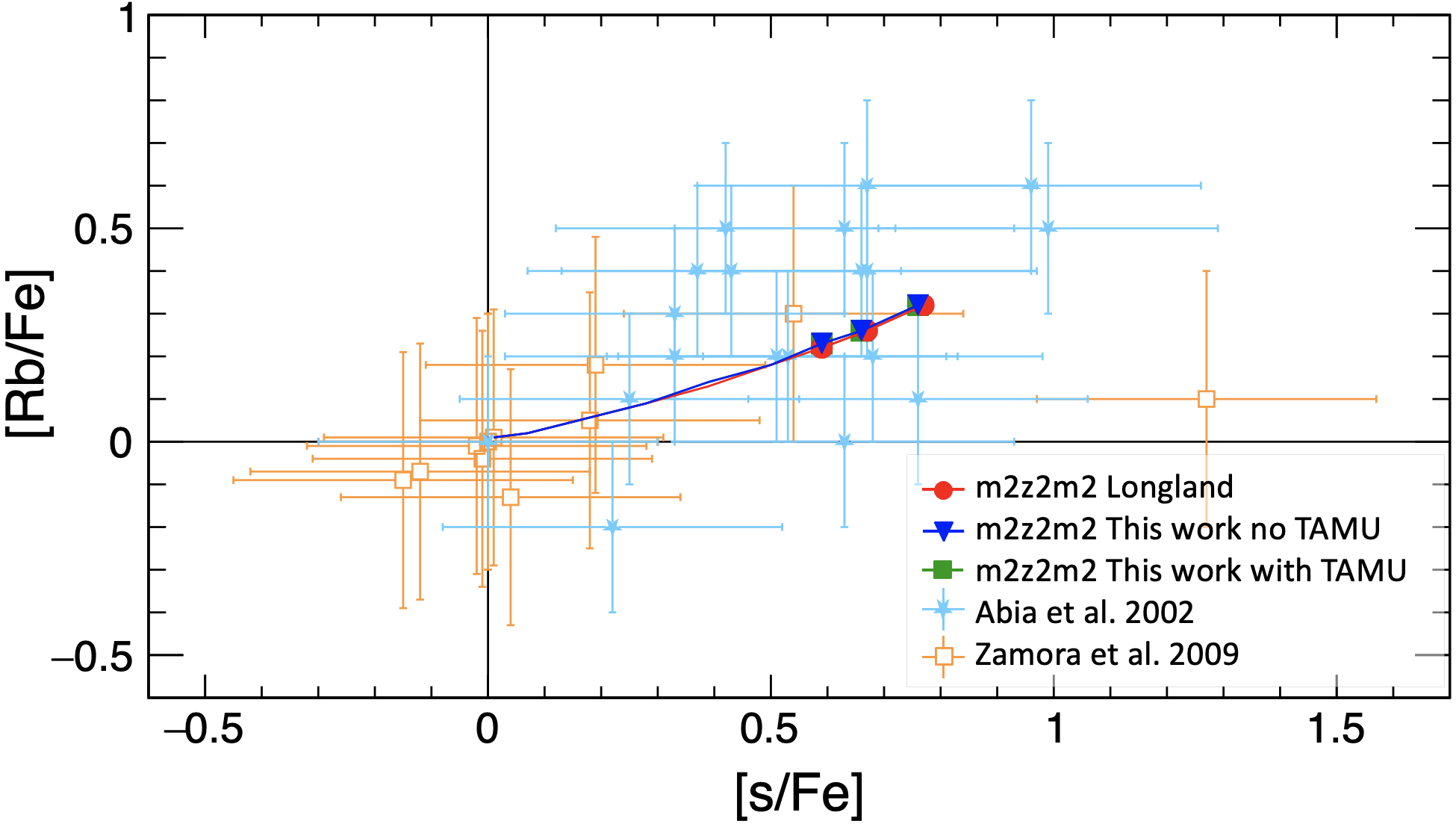}
    }
    \caption{As in Figure \ref{fig:rb_cstars}, but showing the theoretical results obtained with the NEWTON code for a 3$M_{\odot}$ model (upper panel) and  2$M_{\odot}$ model (lower panel).}
    \label{fig:Rb_cstars_sara}
\end{figure}

 In Figures \ref{fig:rb_cstars} and \ref{fig:Rb_cstars_sara} we compare our calculated Rb abundances as a function of the total $s$-process abundances observed through the spectroscopy of carbon stars \citep{abia:02,zamora:09} of initial metal content similar to our models (Z = 0.02 and Z=0.03). The M3.z3m2-hCBM model computed by Battino {\it et al.} \cite{battino:19} are reported in Figure \ref{fig:rb_cstars} while Figure \ref{fig:Rb_cstars_sara} reports the comparison of the NEWTON code output with observations. The variations in the predictions from the models using the Longland rates and the rates in the current paper without the Texas A\&M results included is small. However, the variations when including the Texas A\&M results is much more significant due to the reduced $^{22}$Ne($\alpha,n$)$^{25}$Mg reaction rate.

\begin{figure}[htbp]
    \centering
\includegraphics[width=\columnwidth]{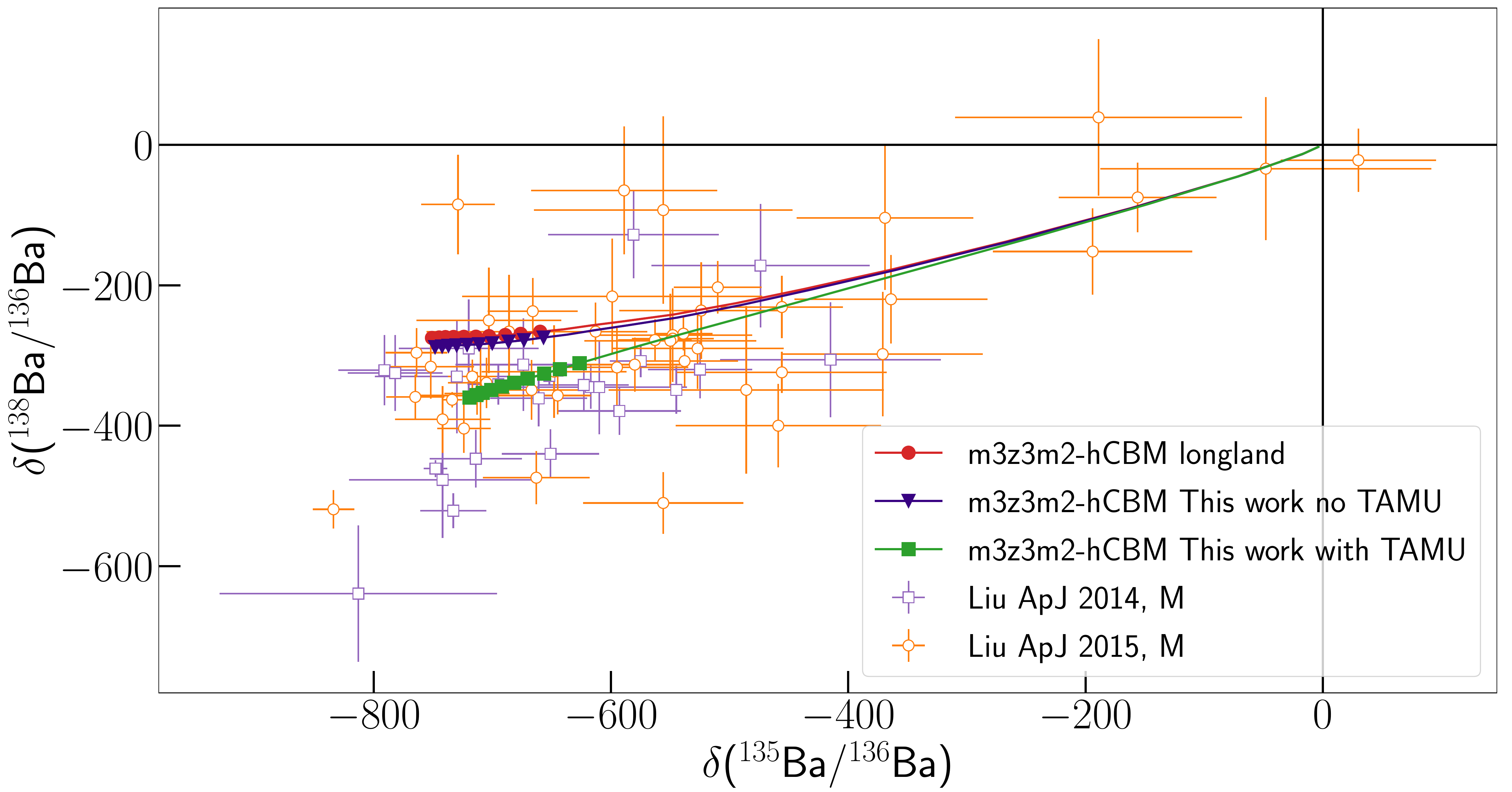}
    \caption{Comparison of measured Ba isotopic rations from presolar SiC grains with the results of MESA stellar models using the reaction rates from the present work.}
    \label{fig:Ba_isoratios}
\end{figure}

\begin{figure}[htbp]
    \centering
    \subfigure{
     \includegraphics[width=0.95\columnwidth]{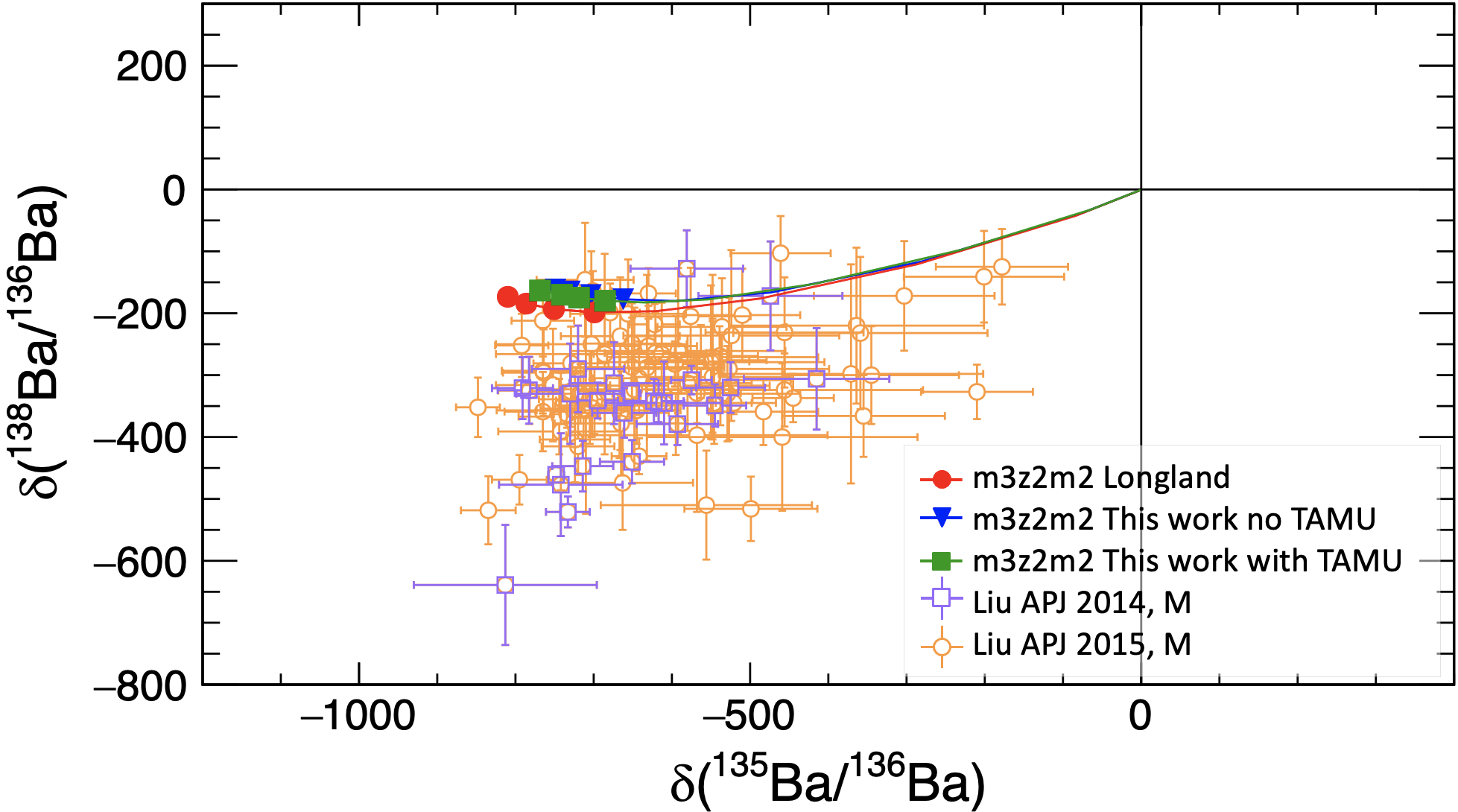}
     }
      \subfigure{
    \includegraphics[width=0.95\columnwidth]{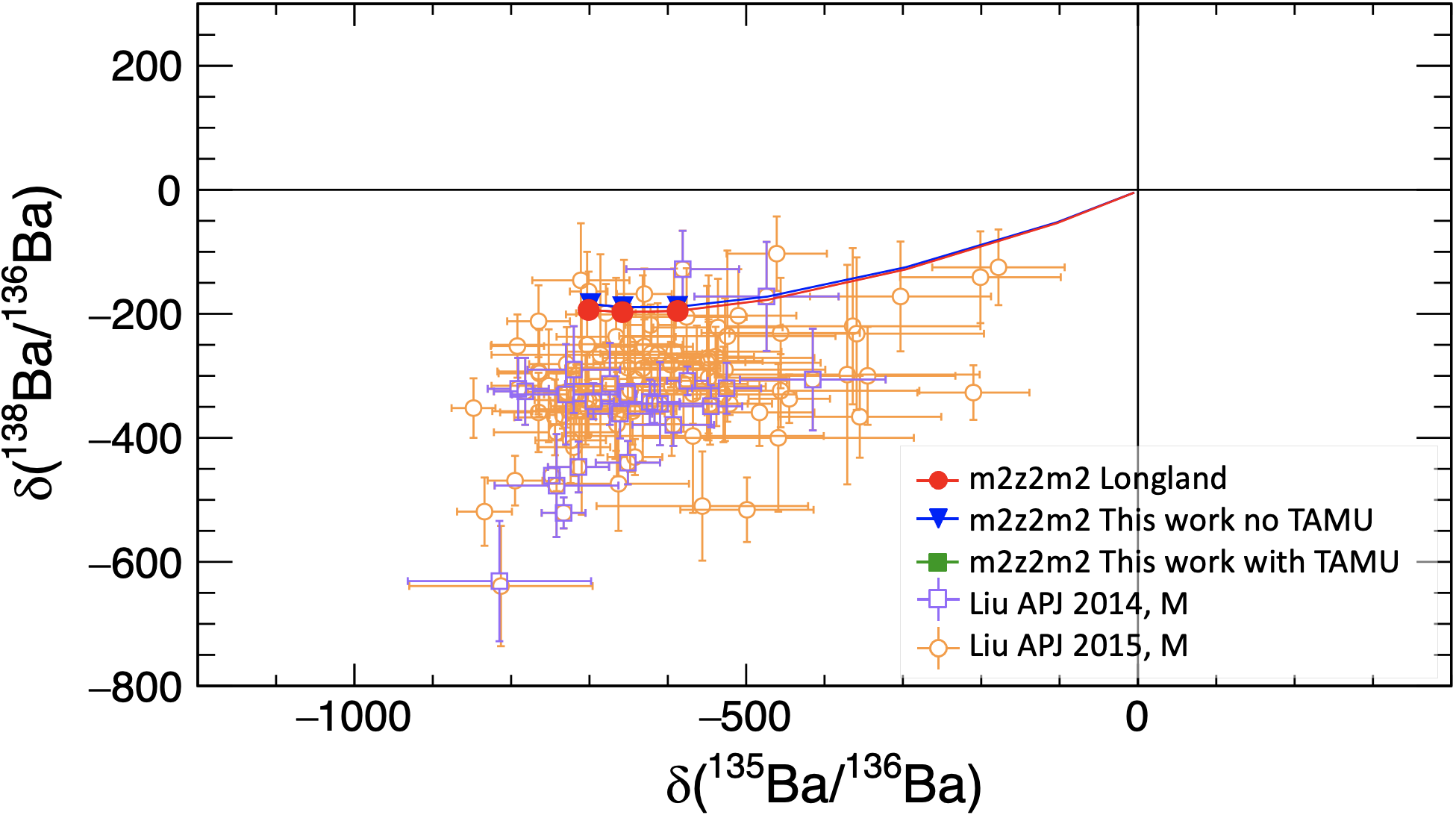}
    }
    \caption{As in Figure \ref{fig:Ba_isoratios}, but showing the theoretical results obtained with the NEWTON code. Lower panel: 2$M_{\odot}$ model. Upper panel: 3$M_{\odot}$ model.}
    \label{fig:Ba_isoratios_sara}
\end{figure}

The ${^{22}}$Ne($\alpha,n$)$^{25}$Mg reaction does not dominate the bulk of $s$-process production in low-mass AGB stars but it still leaves its fingerprint on isotopic ratios involving branching-points and neutron-magic nuclei. Figures \ref{fig:Ba_isoratios} shows the MESA prediction of barium isotopic ratios compared to laboratory measurement of Silicon-Carbide (SiC) grains  \citep{liu:14,liu:15}, which condensed in the ejected atmosphere of AGB stars, polluting the pristine solar nebula \citep{lugaro:03a,zinner:14}. The impact of the rate evaluated without the Texas A\&M results is visible but is smaller than the observational uncertainties.  In the rate using the Texas A\&M results a small yet visible impact is present shifting the theoretical tracks towards the observational data, favouring a better agreement. In Figure \ref{fig:Ba_isoratios_sara} the NEWTON theoretical predictions reach similar values for the ${^{138}}$Ba/${^{136}}$Ba ratio even adopting a lower metallicity (Z=0.02) because the ${^{13}}$C-pocket of the NEWTON code, which is assumed to form because of the stellar magnetic field, is larger (4.8$\times 10^{-3}$M$_{\odot}$) and poor in ${^{14}}$N (see Ref. \cite{trippella} for more details).

The main differences between the results shown in Figures \ref{fig:rb_cstars}/\ref{fig:Rb_cstars_sara} and \ref{fig:Ba_isoratios}/\ref{fig:Ba_isoratios_sara} are due to the stellar nucleosynthesis models employed for calculation and to the different initial metal content. The theoretical tracks in Figure \ref{fig:Ba_isoratios} cover a good fraction of the values in ${^{138}}$Ba/${^{136}}$Ba because of the high metallicity adopted (Z=0.03), which favour the production of first-peak elements (Sr,Y and Zr) over second peak ones (Ba and La) due to the higher neutrons over Fe seeds ratio \citep{lugaro:18}.

\begin{figure}[htbp]
    \centering
    \includegraphics[width=0.95\columnwidth]{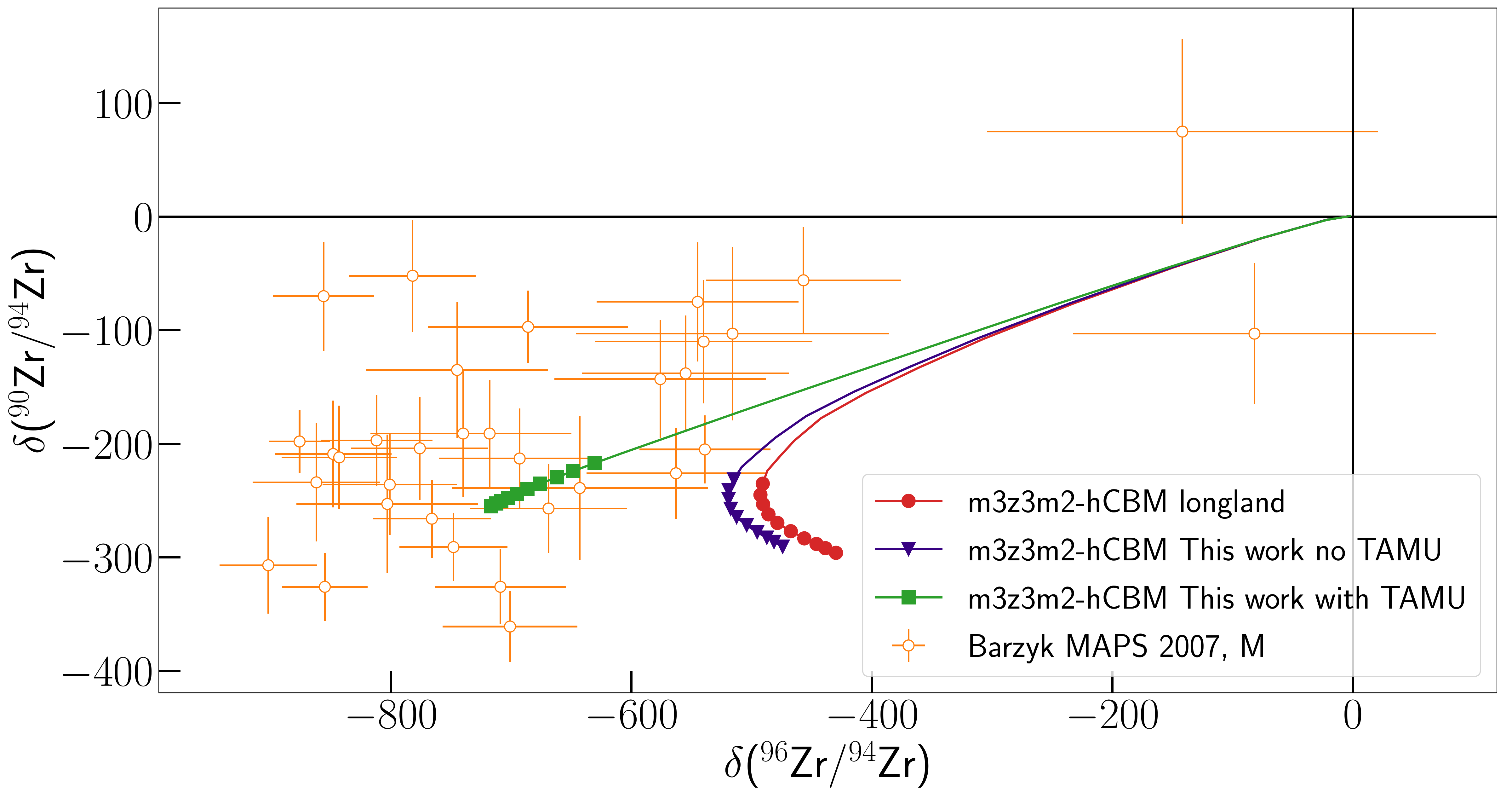}
    \caption{Comparison of MESA stellar models results with measured Zr isotopic ratios from presolar SiC grains. The theoretical tracks show the results obtained with the ${^{22}}$Ne($\alpha$,$n$)${^{25}}$Mg and ${^{22}}$Ne($\alpha$,$\gamma$)${^{26}}$Mg evaluated with and without the recent nuclear data inputs of the Texas A\&M measurements  \cite{OTA2020135256,JAYATISSA2020135267} and the rates of Longland {\it et al.} \cite{PhysRevC.85.065809}.}
    \label{fig:Zr_isoratios}
\end{figure}

Figure \ref{fig:Zr_isoratios} shows the impact on the zirconium isotopic ratios of our newly evaluated rates for ${^{22}}$Ne($\alpha$,$n$)${^{25}}$Mg and ${^{22}}$Ne($\alpha$,$\gamma$)${^{26}}$Mg with MESA code. Including the Texas A\&M results, the difference between the theoretical predictions is much larger, mainly due to a factor three reduction in the ${^{22}}$Ne($\alpha$,$n$)${^{25}}$Mg rate at $T=0.3$ GK when our rate is compared to that of Longland {\it et al.} \cite{PhysRevC.85.065809}, in particular bringing the theoretical predictions in agreement with measured barium isotopic ratio from presolar grains. The lower ${^{22}}$Ne($\alpha$,$n$)${^{25}}$Mg rate directly affects the $s$-process branching at ${^{95}}$Zr, impacting the production of ${^{96}}$Zr and lowering the predicted ${^{96}}$Zr/${^{94}}$Zr ratio; the comparison with measured ratios is greatly improved.

\begin{figure}[htbp]
    \centering
    \subfigure{
    \includegraphics[width=0.95\columnwidth]{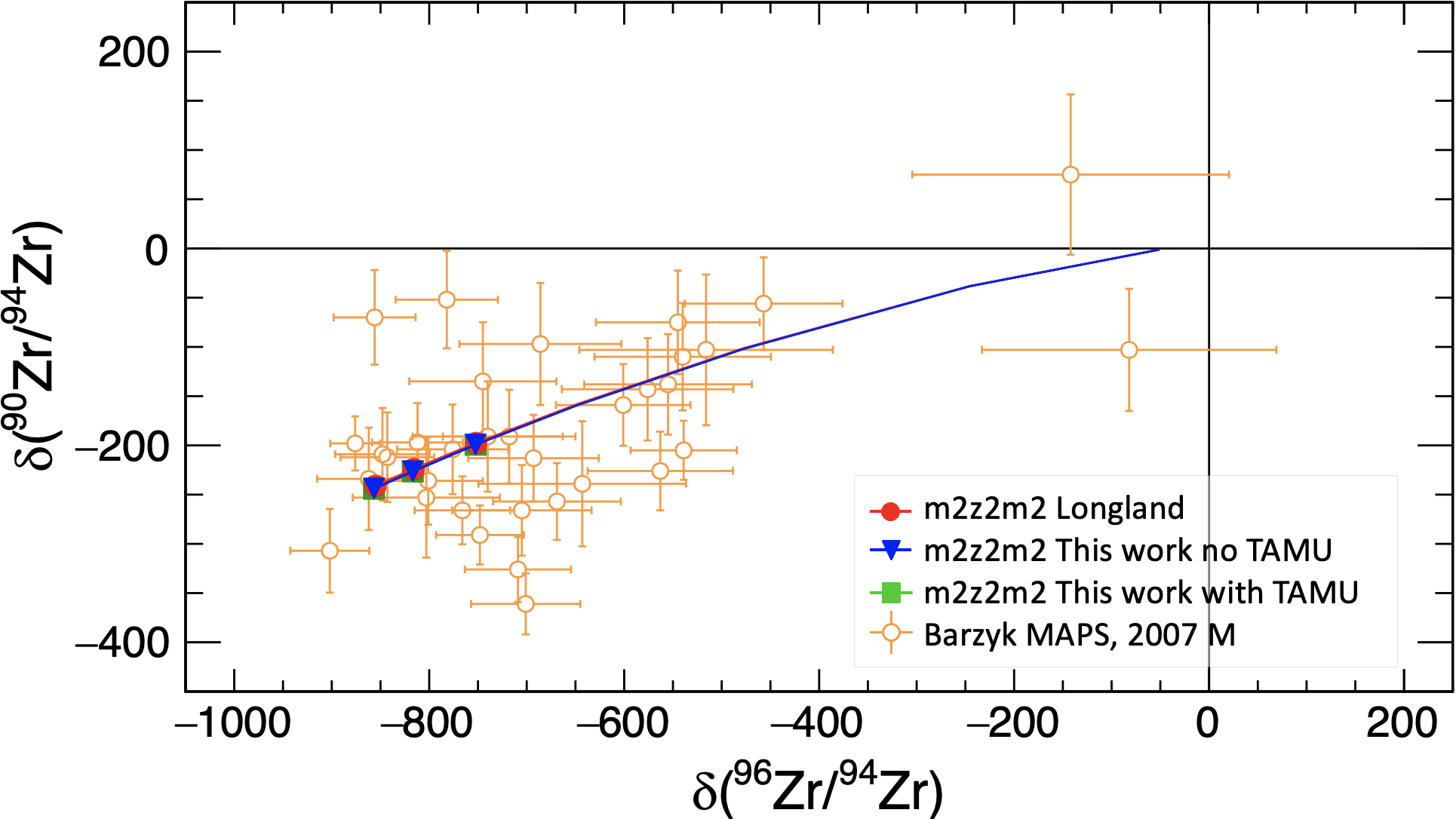}}
    \subfigure{
    \includegraphics[width=0.95\columnwidth]{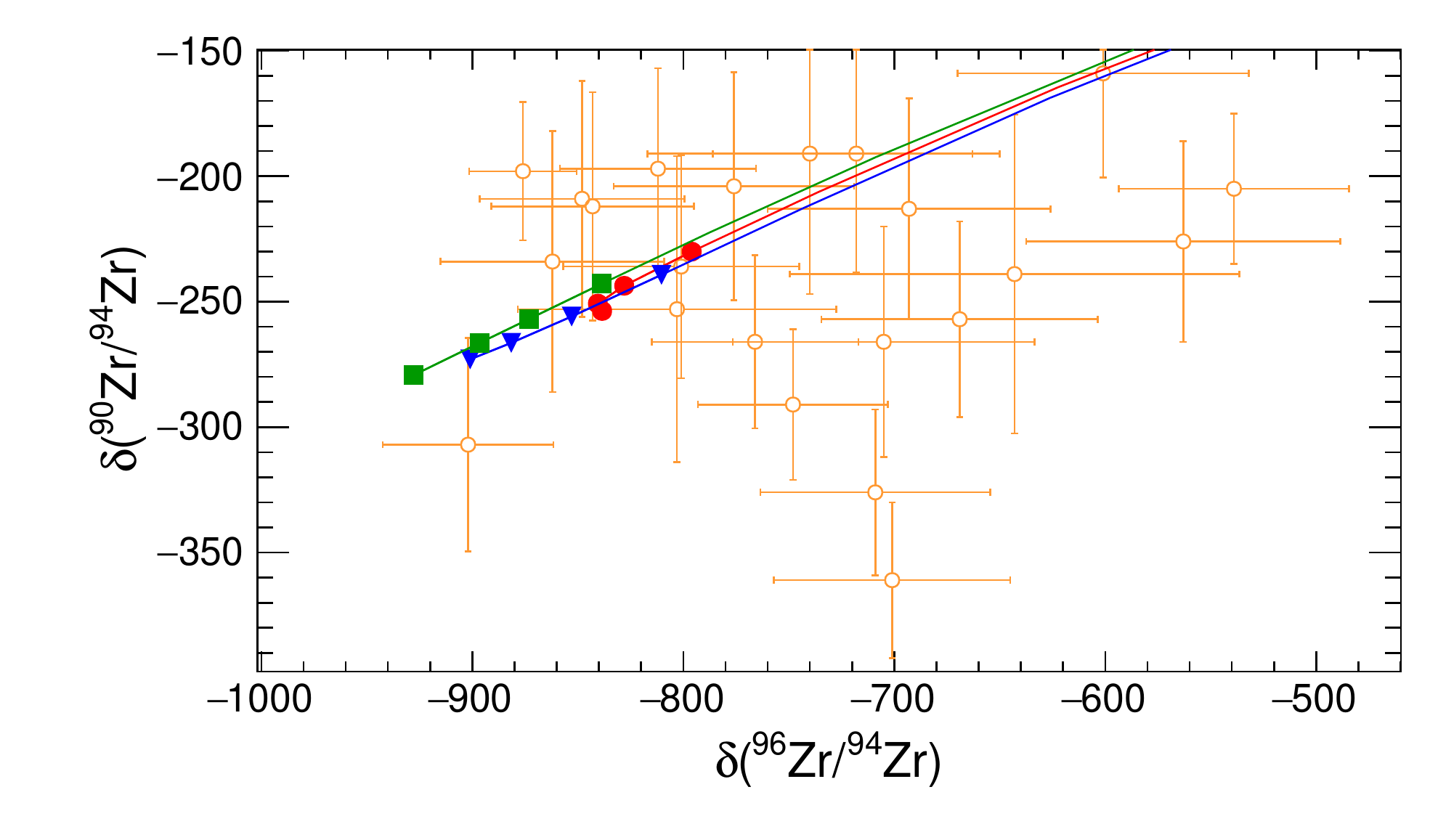}}
    \caption{As in Figure \ref{fig:Zr_isoratios} but showing the theoretical results obtained with the NEWTON code. The 3 theoretical tracks show the results obtained with ${^{22}}$Ne$+\alpha$ reaction rates of Longland {\it et al.} \cite{PhysRevC.85.065809} (in green) and the one suggested in this paper evaluated using the Texas A\&M measurements   \cite{OTA2020135256,JAYATISSA2020135267} (in red) and without this last input (in blue). The lower panel is an enhanced region of the upper panel.}
    \label{fig:Zr_isoratios_sara}
\end{figure}

In Figure \ref{fig:Zr_isoratios_sara} the Zr isotopic mixture of the sample of grains in Figure \ref{fig:Zr_isoratios} is compared with the predictions of the $s$-process using the NEWTON code for a 3$M_{\odot}$ star, this model shows a reduction on the ${^{96}}$Zr$/{^{94}}$Zr ratio too, but the sensitivity to the $^{22}$Ne$+\alpha$ reaction rates is less pronounced, as the ${^{96}}$Zr/${^{94}}$Zr ratio is more efficiently reduced because of the extended $^{13}$C pocket.
The results of the MESA and NEWTON calculations are consistent in showing that the ${^{96}}$Zr/${^{94}}$Zr ratio is a factor of two lower when computed by using the recent nuclear data inputs of the Texas A\&M measurements \cite{OTA2020135256,JAYATISSA2020135267}. The shift between the model curves is less visible in Figure \ref{fig:Zr_isoratios_sara} because the values of the ${^{96}}$Zr/${^{94}}$Zr ratio are smaller than in Figure \ref{fig:Zr_isoratios} and the $\delta$ notation reduces the visible size of this effect: the $\delta_i = 1000(\frac{(^iX/^jX)_\mathrm{grain}}{(^iX/^jX)_\mathrm{standard}}-1)$. An expanded plot of the $\delta(^{96}$Zr/$^{94}$Zr$)$ values is shown in the lower panel of Figure \ref{fig:Zr_isoratios_sara}.

The differing predictions of the zirconium isotopic ratios from MESA and NEWTON are a consequence of considering two different mixing processes at the convective boundaries, resulting in greater $^{94}$Zr production in the NEWTON models compared to the MESA models, and a concomitant lower $^{96}$Zr/$^{94}$Zr ratio. The nuclear reaction rates can be a source of uncertainty comparable to the model uncertainties (see Figs. \ref{fig:Zr_isoratios} and \ref{fig:Zr_isoratios_sara}), and minimising the nuclear-physics uncertainties is a necessary step before the impact of the convective boundary mixing schemes on $s$-process nucleosynthesis in evolved stars may be understood.

\subsection{Weak $s$-process from massive star}

The weak $s$-process component (producing solar system $s$-process abundances between iron and zirconium; e.g. \cite{travaglio:04}) is produced during convective core helium burning and convective shell carbon-burning stages in massive stars (M$>$10M$_\odot$).
The ${^{22}}$Ne($\alpha$,$n$)${^{25}}$Mg reaction is the principal neutron source for the weak $s$-process component. An important role is also played by the ${^{22}}$Ne($\alpha$,$\gamma$)${^{26}}$Mg reaction, which competes with the ${^{22}}$Ne($\alpha$,$n$)${^{25}}$Mg reaction in typical weak $s$-process conditions, lowering the number of neutrons available for the $s$-process. In the following section, we first discuss the $s$-process production without the inclusion of the Texas A\&M results, and then with those results.

\begin{figure}[htbp]
    \centering
    \subfigure{
    \includegraphics[width=0.95\columnwidth]{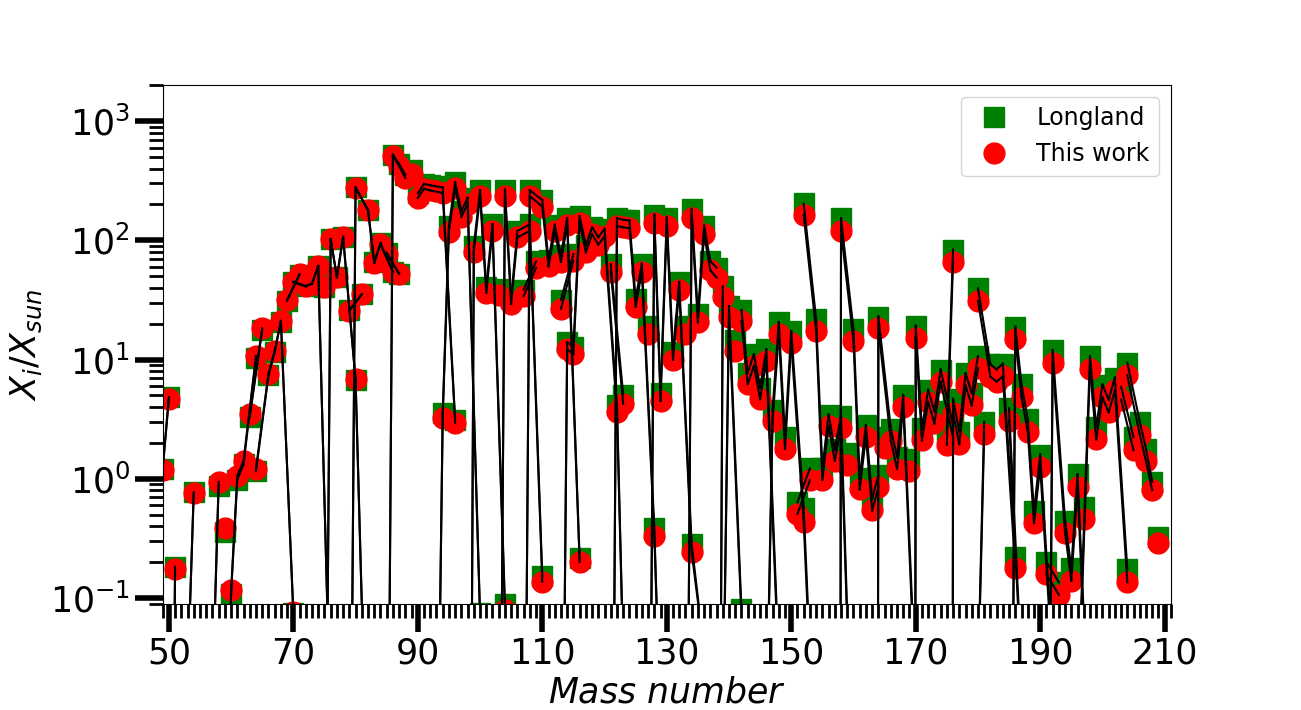}}
    \subfigure{
    \includegraphics[width=0.95\columnwidth]{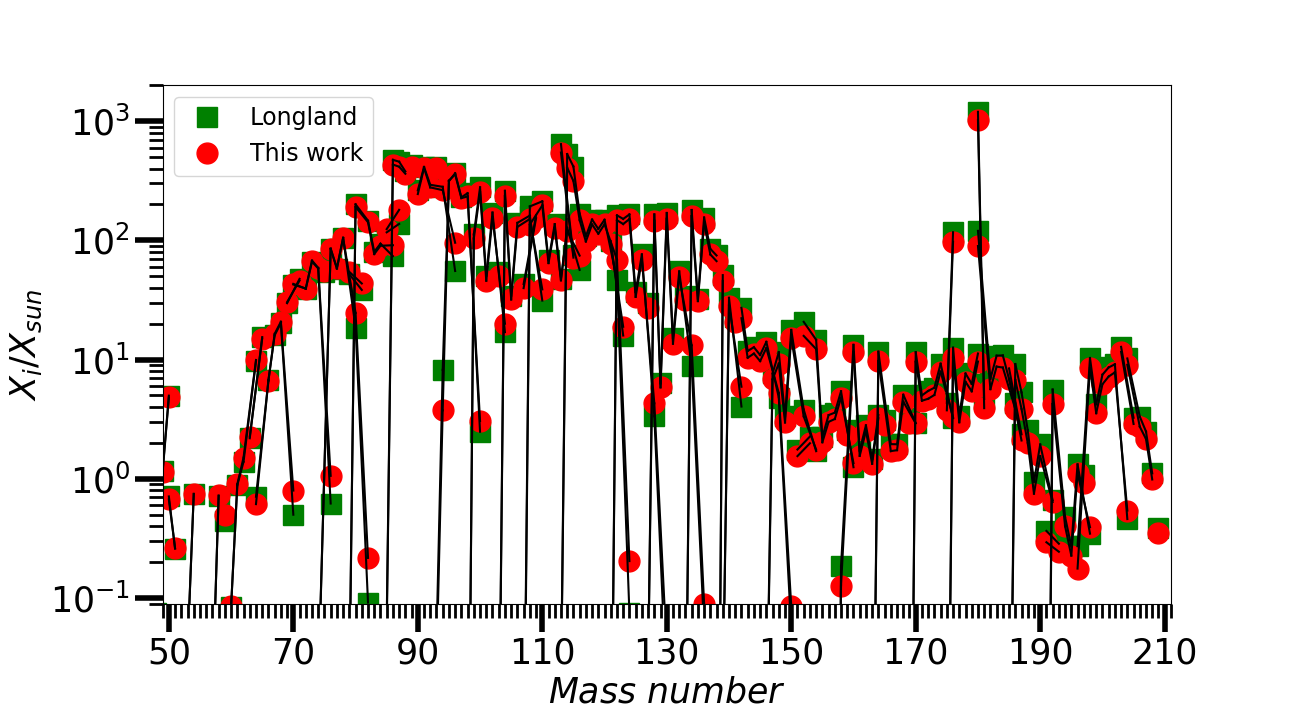}}
    \caption{Production factors of heavy isotopes between the iron and lead peak from the 25 M${_\odot}$, Z=0.006 model of Ref. \cite{ritter:18}. Top and bottom panels show the isotopic distribution at the end of core helium burning and shell carbon burning respectively. Lagrangian mass and time coordinate selected to extract the predictions here presented, are specified in figure \ref{fig:MS_kippe}. For each isotope, two different values are presented, corresponding to the adoption of the ${^{22}}$Ne+$\alpha$ rates without the inclusion of the Texas A\&M results which are presented in this work and from Longland {\it et al.} \cite{PhysRevC.85.065809}.}
    \label{fig:MS_pf}
\end{figure}

\begin{figure}[htbp]
    \centering
    \subfigure{
    \includegraphics[width=\columnwidth]{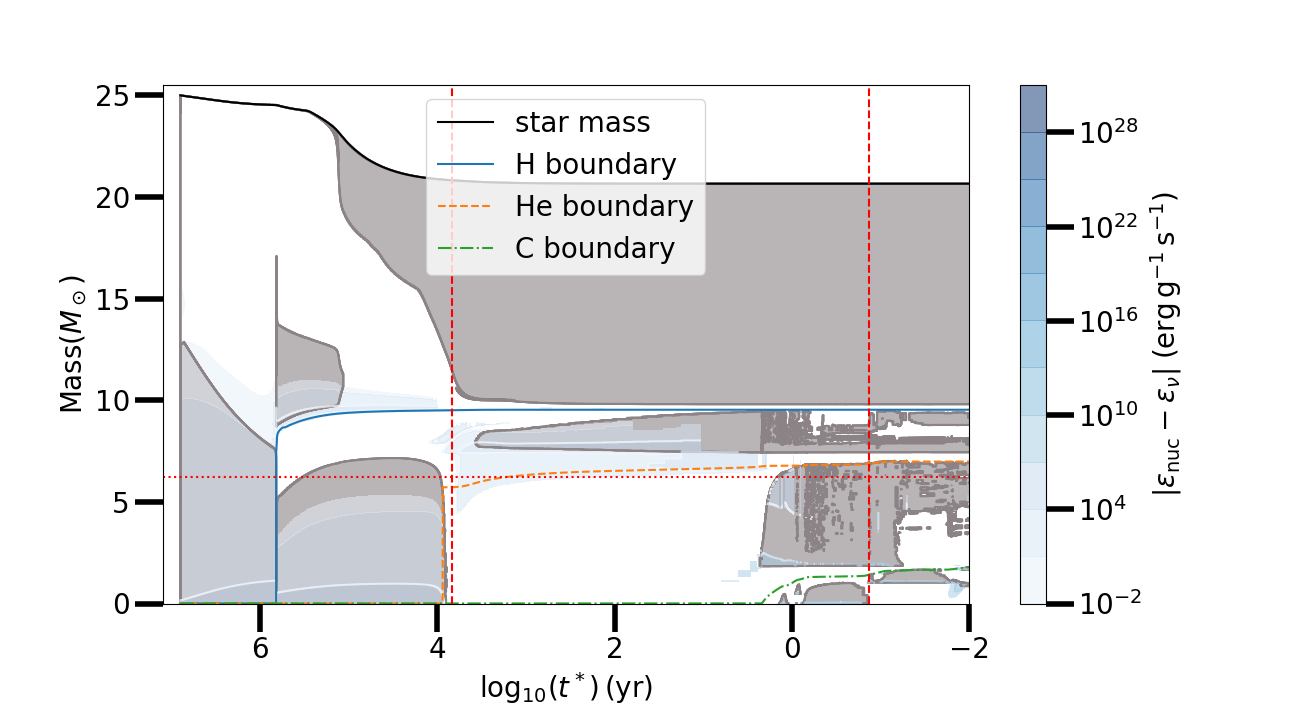}}
    \caption{Kippenhahn diagrams of our adopted  25 M${_\odot}$, Z=0.006 model. The convective zones are marked in gray. Lagrangian mass-coordinate is shown on the Y-axis, while the logarithm of the time remaining before the end of the star's life is given on the X-axis. Vertical red dashed lines mark the time selected to extract the abundance distributions presented in our plot, at the mass-coordinate specified by the red horizontal dotted line.}
    \label{fig:MS_kippe}
\end{figure}

In Figure \ref{fig:MS_pf} we present the impact on the weak $s$-process nucleosynthesis of our new rates computed using the MESA code without the inclusion of the Texas A\&M results. We post-processed the stellar structure model of a 25 M$_{\odot}$ star with initial metal content Z=0.006 from Ref. \cite{ritter:18}. The Top and bottom panels of Figure \ref{fig:MS_pf} show the isotopic distribution at the end of core helium burning and shell carbon-burning respectively, extracted at Lagrangian mass and time coordinate specified in Figure \ref{fig:MS_kippe}. Overall, the impact is not large. Elements between the first-peak elements (Sr, Y and Zr) and A=130 are all overproduced by two order of magnitudes compared to their initial abundances, and they are all reduced when our rates are adopted instead of the rates from Ref. \cite{PhysRevC.85.065809}, but always by less than a factor of two.

\begin{figure}[htbp]
    \centering
    \subfigure{
    \includegraphics[width=0.95\columnwidth]{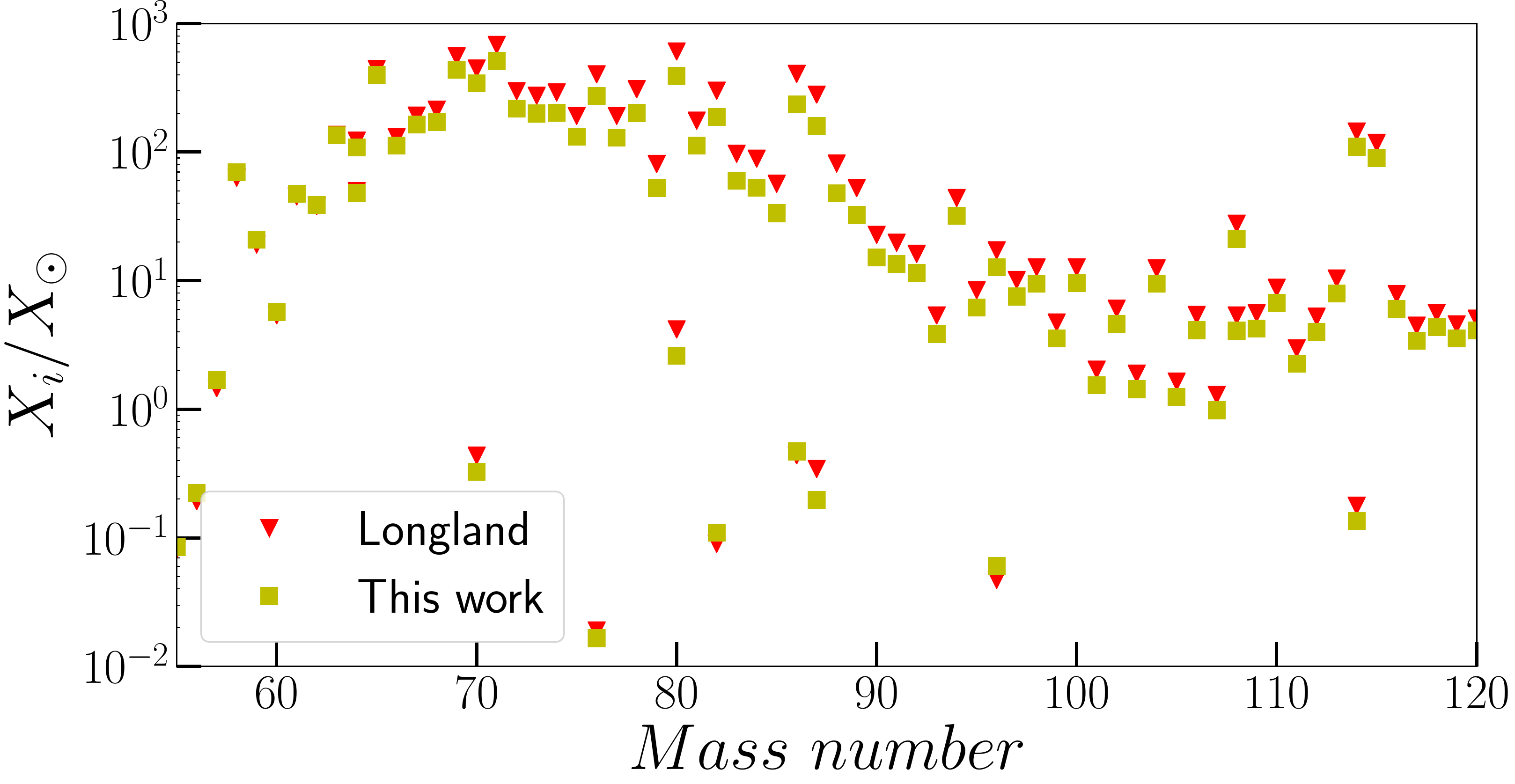}}
    
    \subfigure{
    \includegraphics[width=0.95\columnwidth]{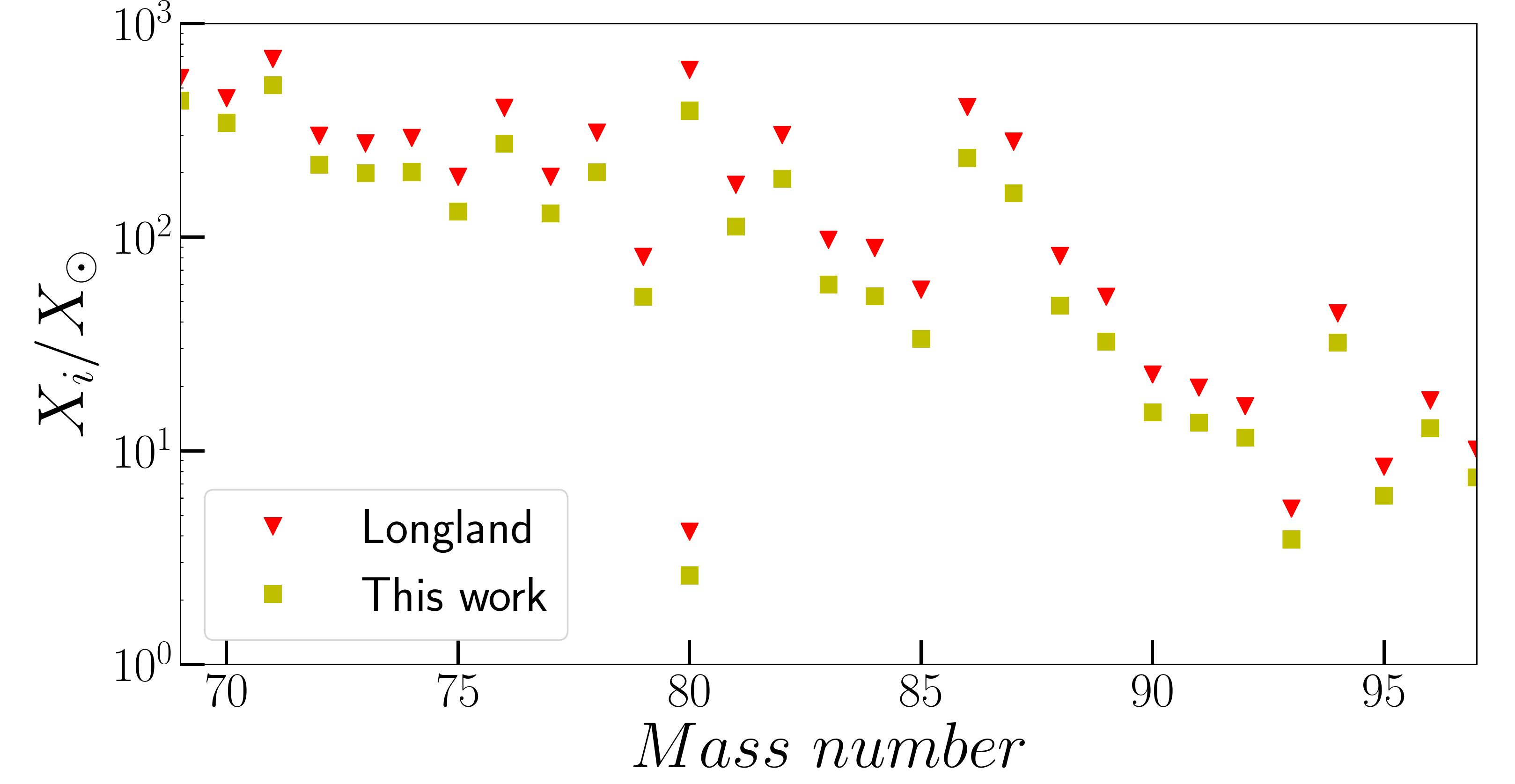}}
    
    \caption{Same test as presented in Figure \ref{fig:MS_pf}, but performed on the rotating massive star model with 25 M${_\odot}$ star, Z=0.00001 from Ref. \cite{hirschi:04}}
    \label{fig:MS_pf2}
\end{figure}

In Figure \ref{fig:MS_pf2}, we repeat the same test in a massive rotating star model (25 M$_{\odot}$ star, Z=0.0001) from \cite{hirschi:04}. 
Rotation has indeed an important role in the weak $s$-process in massive star, as it allows the primary production of ${^{14}}$N, which is later converted into ${^{22}}$Ne by $\alpha$-particle captures. As a consequence, a larger quantity of neutrons are released by ${^{22}}$Ne($\alpha,n$)${^{25}}$Mg reactions, globally increasing the $s$-process efficiency. In this case, we again confirm the low impact of the new rates without the Texas A\&M results compared to Longland {\it et al.} \cite{PhysRevC.85.065809}, as the variations we obtained are all smaller than a factor of two.

\begin{figure}[htbp]
    \centering
    \subfigure{
    \includegraphics[width=0.95\columnwidth]{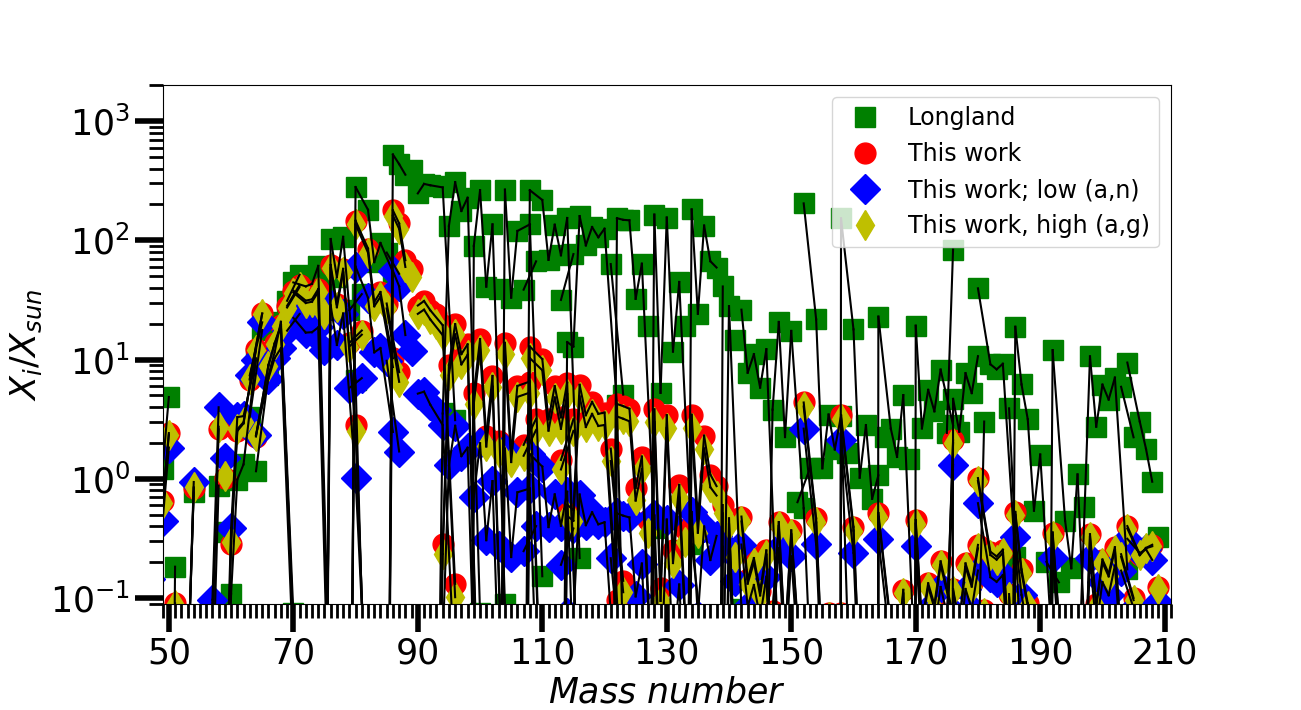}}
    \subfigure{
    \includegraphics[width=0.95\columnwidth]{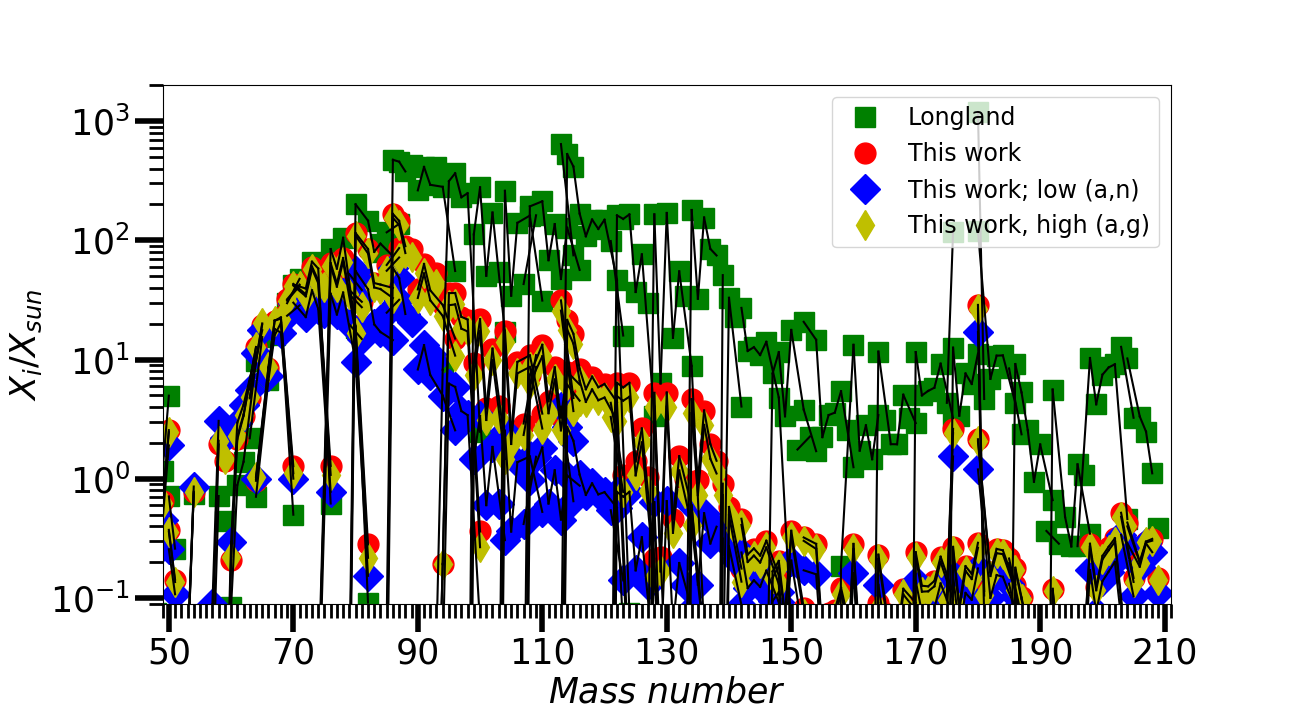}}
    \caption{Same as in Figure \ref{fig:MS_pf}, but showing our theoretical results obtained adopting the recommended ${^{22}}$Ne($\alpha$,$n$)${^{25}}$Mg and ${^{22}}$Ne($\alpha$,$\gamma$)${^{26}}$Mg reaction rates evaluated using the recent input nuclear data from Texas A\&M \cite{OTA2020135256,JAYATISSA2020135267}.}
    \label{fig:MS_pf_tamu}
\end{figure}

In the same manner, in Figure \ref{fig:MS_pf_tamu} we show the impact of our new recommended reaction rates for ${^{22}}$Ne($\alpha$,$n$)${^{25}}$Mg and ${^{22}}$Ne($\alpha$,$\gamma$)${^{26}}$Mg using the input nuclear data from the Texas A\&M experiments \cite{OTA2020135256,JAYATISSA2020135267}. As already seen for the main $s$-process from AGB stars, the difference from the theoretical predictions adopting the Longland {\it et al.} \cite{PhysRevC.85.065809} reaction rates is large. In particular, the production of Sr, Y and Zr drops by about a factor of five, and by one order of magnitude for A$>$100, or even two if the lower limit of the ${^{22}}$Ne($\alpha,n$)${^{25}}$Mg reaction rate is adopted.

\section{Conclusions}

The $^{22}$Ne($\alpha,\gamma$)$^{26}$Mg and $^{22}$Ne($\alpha,n$)$^{25}$Mg reactions control the production of neutrons for the weak $s$-process in evolved massive stars and in the helium-flash in low-mass AGB stars. In this paper we have critically re-evaluated 
the available nuclear data on the states in $^{26}$Mg which govern these reaction rates and have re-analysed these rates. We find that the rates are approximately unchanged from the evaluation of Longland {\it et al.} \cite{PhysRevC.85.065809}, in contrast to the evaluation of Talwar {\it et al.} \cite{PhysRevC.93.055803} which found a greatly increased $^{22}$Ne($\alpha,\gamma$)$^{26}$Mg reaction rate, and the evaluation of Massimi {\it et al.} \cite{Massimi20171} which found a greatly reduced $^{22}$Ne($\alpha,\gamma$)$^{26}$Mg reaction rate. Our disagreement with the results of Massimi {\it et al.} \cite{Massimi20171} may be attributed to the different methodologies for calculating the reaction rates as the inclusion of their nuclear data has only a small impact on the current evaluation.

The primary source of uncertainty in the $^{22}$Ne($\alpha,\gamma$)$^{26}$Mg reaction rate is whether a strong $\alpha$-cluster state exists at around $E_x = 11.17$ MeV. Evidence from two different measurements of the $^{22}$Ne($^6$Li,$d$)$^{26}$Mg $\alpha$-particle transfer reaction is contradictory, with experiments which have approximately the same excitation-energy resolution both observing two strong $\alpha$-cluster resonances separated by a similar energy gap but with inconsistent excitation energies. Resolution of this disagreement is an high priority for future experimental studies. Additional information, in particular estimates of resonance strengths and/or $\alpha$-particle partial widths of the states at $E_x = 10.949$, $11.112$, $11.163$ and $11.171$ MeV which are expected to control the behaviour of the reaction rates below the $E_r = 706$-keV resonance, and on the interference pattern between higher-energy resonances are also required.

In this work we have evaluated the reaction rates twice, once without the inclusion of the new results from two experiments performed at Texas A\&M and once with the inclusion of those results. This latter evaluation, with the inclusion of the Texas A\&M results, is our recommended rate.

The evaluations of the rates without the inclusion of the new Texas A\&M results produces no substantial change in the predictions of $s$-process nucleosynthesis. The inclusion of the new resonance strength for the $E_r = 706$-keV resonance derived from the Texas A\&M measurements results in predicted barium and zirconium isotopic ratios which better match the measured ratios from presolar SiC grains.

\section*{Acknowledgements}

This paper is based upon work from the \textquoteleft ChETEC\textquoteright\ COST Action (CA16117), supported by COST (European Cooperation in Science and Technology). PA thanks the Claude Leon Foundation for a fellowship under which part of this work was performed. UB and AB acknowledge support from the European Research Council (Grant agreements ERC-2015-STG No. 677497 and ERC-2019-STG No. 852016, respectively). RH acknowledges support from the World Premier International Research Centre Initiative (WPI Initiative).

\bibliography{Ne22ReactionRates}

\end{document}